\documentclass[prb,twocolumn,floatfix,longbibliography,superscriptaddress]{revtex4-2}
\usepackage[english]{babel}
\usepackage[utf8]{inputenc}
\usepackage[T1]{fontenc}
\usepackage{amsmath}
\usepackage{mathtools,amssymb}
\usepackage{graphicx}
\usepackage{bm}
\usepackage{gensymb}
\usepackage{color}
\usepackage{hyperref}
\usepackage{upgreek}
\usepackage{scalerel}
\usepackage{pgffor}
\usepackage{pdfpages}
\usepackage{float}
\usepackage{appendix}
\usepackage{physics} 
\usepackage{lscape}   
\usepackage{silence}
\usepackage{soul,xcolor} 
\setstcolor{red} 
\usepackage{comment}  
\makeatletter
\AtBeginDocument{\let\LS@rot\@undefined}
\makeatother

\begin{document}

\title{Superconducting phenomena in systems with unconventional magnets}

\author{Yuri Fukaya}
 \email[ ]{ yuri.fukaya@ec.okayama-u.ac.jp}
\affiliation{Faculty of Environmental Life, Natural Science and Technology, Okayama University, 700-8530 Okayama, Japan}

\author{Bo Lu}
 \email[ ]{ billmarx@tju.edu.cn}
\affiliation{Center for Joint Quantum Studies, Tianjin Key Laboratory of Low Dimensional Materials Physics and Preparing Technology, Department of Physics, Tianjin University, Tianjin 300354, China}

\author{Keiji Yada}
\email[]{ yada.keiji.b8@f.mail.nagoya-u.ac.jp}
\affiliation{Department of Applied Physics, Nagoya University, 464-8603 Nagoya, Japan}

\author{Yukio Tanaka}
 \email[ ]{ ytanaka@nuap.nagoya-u.ac.jp}
\affiliation{Department of Applied Physics, Nagoya University, 464-8603 Nagoya, Japan}
\affiliation{Research Center for Crystalline Materials Engineering, Nagoya University, 464-8603 Nagoya, Japan}

 \author{Jorge Cayao} 
 \email[Corresponding author: ]{ jorge.cayao@physics.uu.se}
\affiliation{Department of Physics and Astronomy, Uppsala University, Box 516, S-751 20 Uppsala, Sweden}

\date{\today}
\begin{abstract}
In this work we review the recent advances on superconducting phenomena in junctions formed by superconductors and unconventional magnets. Conventional magnets, such as ferromagnets and antiferromagnets, are characterized by broken time-reversal symmetry  but only ferromagnets produce a finite net magnetization due to parallel spin alignment and spin-split bands in momentum. Very recently, a new type of magnets has been reported and here we refer to them as  \emph{unconventional magnets} because they exhibit special properties of both ferromagnets and antiferromagnets:  they exhibit zero net magnetization (like antiferromagnets)  and  a nonrelativistic   spin splitting of energy bands (like ferromagnets), both leading  to anisotropic spin-polarized Fermi surfaces. An interesting property of unconventional magnets is that their magnetic order can be even or odd with respect to momentum, where \emph{$d$-wave altermagnets} and \emph{$p$-wave magnets} are the most representative examples. In this regard, $d$-wave altermagnets and $p$-wave magnets are seen as counterparts in magnetism of the unconventional $d$-  and $p$-wave superconducting states, respectively.  While the impact of   conventional magnetism on superconductivity has been largely studied, the combination of  unconventional magnets and superconductivity has only lately attracted considerably attention. This work provides a comprehensive review of  the recent progress on  the interplay between superconductivity and unconventional magnets. In particular, we focus on the fundamental emerging superconducting phenomena and also discuss the potential implications towards quantum applications.
\end{abstract}
  
\maketitle

\section{Introduction}
\label{section0}
Superconductivity and magnetism were initially saw to be incompatible  due to the destruction of spin-singlet Cooper pairs by the exchange field but their combination was later shown to be the key for novel superconducting phenomena and quantum applications \cite{BergeretReview,RevModPhys.77.935,eschrig2011spin,linder2015superconducting,Eschrig2015,yang2021boosting,mel2022superconducting,cai2023superconductor}. One salient example is the realization of spin-triplet Cooper pairs \cite{BergeretReview}, which not only characterize an emergent spin-triplet superconducting phase but also represent  the key for  superconducting spintronics \cite{eschrig2011spin,linder2015superconducting,Eschrig2015,yang2021boosting,mel2022superconducting,cai2023superconductor},  topological superconductivity and Majorana zero modes \cite{tanaka2011odd_review,sato2016,sato2017topological,Aguadoreview17,lutchyn2018majorana,frolov2019quest,Cayao2020odd,flensberg2021engineered,tanaka2024review}, and superconducting qubits \cite{sarma2015majorana,krantz2019quantum,benito2020hybrid,aguado2020majorana,beenakker2019search,kjaergaard2020superconducting,MarraJAP2022}.  The  role of magnetism for converting spin-singlet Cooper pairs into spin-triplet ones is crucial, an aspect that has been mostly studied in superconducting systems formed by  ferromagnets (FMs) and antiferromagnets (AFMs), see e. g., Refs.\,\cite{buzdin2006superconductorx,linder2015superconducting,Eschrig2015,jeon2021long,PhysRevB.109.184504,bobkova2024neel,PhysRevLett.131.076001,PhysRevB.108.054510,cai2023superconductor}. FMs and AFMs are examples of collinear magnetism  arising from the spontaneous arrangements of magnetic moments oriented parallel and antiparallel to each other, respectively \cite{RevModPhys.25.58,kaganov,nolting2008fundamentals,nolting2009quantum}; whenever necessary, FMs and AFMs will be referred to    as conventional magnets. In both types of these conventional magnets, time reversal symmetry is broken, but only FMs exhibit a finite net magnetization and spin-split electronic bands \cite{aharoni2000introduction,RevModPhys.77.1375,RevModPhys.91.035004,Baltz_RevModPhys_2018}. While both FMs and AFMs have been shown to promote novel superconducting phases, in terms of applications both posses considerable challenges. For instance, the finite net magnetization of FMs,   useful in some situations \cite{HIROHATA2020166711}, is detrimental for scalability and stability of superconductor-FM devices \cite{Newhorizons_spintronics,dal2024antiferromagnetic,jungwirth2018multiple}; moreover, AFMs are excellent in terms of scalability due to zero net magnetization \cite{jungwirth2016antiferromagnetic,Baltz_RevModPhys_2018,HIROHATA2020166711}, they do not host the useful spin-dependent  effects of FMs,  which difficulties reading data stored in AFMs \cite{Newhorizons_spintronics,dal2024antiferromagnetic,jungwirth2018multiple}. It would be  therefore desirable to realize magnetic phases containing the positive properties of both AFMs and FMs, such that, when combining with superconductivity, the emergent superconducting phase can be useful for realistic applications. 

Recently, a new family of magnetic materials with properties akin to both FMs and AFMs has been predicted \cite{noda2016momentum,Ahn2019,Hayami19,PhysRevB.101.220403,PhysRevB.102.014422,LiborSAv,MaNatcommun2021,jungwirth2024,jungwirth2024He3Ams}. These new magnets have been shown to have zero net magnetization (like AFMs) and a  spin-splitting of electronic bands (like FMs) that depends on momentum but of nonrelativistic origin \cite{landscape22,LiborPRX22,PhysRevB.109.L140402,hellenes2024P,brekke24} and hence distinct from common spin-orbit coupling \cite{galitski2013spin,manchon2015new}. For these reasons, the new magnets are known as \emph{unconventional magnets} \cite{MazinPRX22,jungwirth2024,Bai_review24,tamang2024,jungwirth2024He3Ams}. A particular consequence of the momentum dependent spin-split energy bands in  unconventional magnets is that they exhibit anisotropic spin-polarized Fermi surfaces in momentum space as well as a momentum dependent magnetic order that can have   even- \cite{landscape22,LiborPRX22}  and odd-parity \cite{PhysRevB.109.L140402,hellenes2024P,brekke24}, akin to what happens in unconventional superconductors \cite{RevModPhys.63.239,maki1998introduction}. The even-parity magnetic order can have $s$-, $d$-, $g$-, or $i$-wave symmetries, corresponding to the degree $0$, $2$, $4$, or $6$ of the exchange magnetic field polynomials with momenta space as variables. Similarly, the odd-parity magnetic order can have $p$- or $f$-wave symmetries, corresponding to the degree $1$ or $3$ of the exchange magnetic field in momentum. The simplest  unconventional magnets, where the even- and odd-parity exchange fields are  quadratic and linear in momenta,  have so far attracted an enormous attention and were coined $d$-wave altermagnets (AMs) \cite{landscape22,LiborPRX22} and $p$-wave   magnets \cite{PhysRevB.109.L140402,hellenes2024P,brekke24}, respectively. In what follows we will refer to all the even-parity  unconventional magnets   as to AMs, of course indicating always the type of parity symmetry.  Even though  the intriguing spin-momentum locking and  spin splitting in both types of  unconventional magnets have a nonrelativistic origin, they  originate from distinct symmetries that is different for even- and odd-parity  unconventional magnets. For instance, AMs are  collinear-compensated magnets in real space where time-reversal symmetry is broken but inversion symmetry is preserved \cite{landscape22,LiborPRX22}; atoms in AMs exhibit an alternating spin direction and spatial orientation on the neighbouring sites. Moreover, AMs break time-reversal symmetry but preserve inversion symmetry, while   opposite spins are connected by crystal rotation or mirror symmetries \cite{landscape22,LiborPRX22}. A recent study suggested that the concept of AMs can be extended to non-collinear spins as well \cite{cheong2024altermagnetism}, which certainly expands the materials search of altermagnetism. On the other hand,  $p$-wave   magnets are noncollinear and noncoplanar magnets,  protected by the combination of time-reversal and a translation of a half the unit cell, where  time-reversal symmetry is preserved but  inversion symmetry is broken \cite{PhysRevB.109.L140402,hellenes2024P,brekke24}. 

All the  intriguing properties of  unconventional magnets have inspired a large number of theoretical and experimental studies in the normal state, although mostly in AMs, see Refs.\, \cite{noda2016momentum,Ahn2019,Hayami19,PhysRevB.102.014422,LiborSAv,MaNatcommun2021,jungwirth2024,jungwirth2024He3Ams}. At the moment, there already are around 60 materials predicted to host altermagnetism \cite{landscape22,LiborPRX22,guo2023spin};  among the predicted materials, CrSb and MnTe are believed to be $g$-wave AMs, while RuO$_{2}$, Mn$_{5}$Si$_{3}$, MnF$_{2}$ and MnO$_2$ are expected to be $d$-wave AMs. Several of these materials have already been measured to be altermagnetic, such as  V$_2$Se$_2$O and  
V$_2$Te$_2$O~\cite{MaNatcommun2021,fzhang2024,jiang2024,hu2024},   RuO$_{2}$~\cite{Ahn2019,fedchenko2024observation,lin2024observation}, Mn$_{5}$Si$_{3}$~\cite{reichlova2021macro,han2024SciAdv},  La$_{2}$CuO$_{4}$~\cite{Moreno16}, MnTe~\cite{Lee24,Osumi2024,KrempaskyNature2024}, MnTe$_{2}$ \cite{zhu2024observation}, Weyl semimetal CrSb \cite{reimers2024,lu2024}.  In this regard, multiple experiments 
have reported band splitting in AMs \cite{liu2023inverse,KrempaskyNature2024,lin2024observation,reimers2024,Lee24,Osumi2024,fedchenko2024observation,PhysRevLett.133.056701,chen2024altermSpinSPLITmagne} as well as anomalous and spin transport effects \cite{ghimire2018large,reichlova2021macro,Feng_2022,LiborSAv,Tschirner_2023,wang2023emergent,PhysRevLett.132.056701,Lee24,PhysRevB.109.224430,takagi2025spontaneous},   spin currents \cite{shao2021spin,Rafael21,PhysRevLett.128.197202,Bose_2022,PhysRevLett.129.137201,PhysRevLett.130.216701,PhysRevB.108.024410,li2024spinsplittingaltermruo2,pan2024unvei,chen2024altermSpinSPLITmagne}, magnetoptical effects \cite{PhysRevB.104.024401,PhysRevB.106.064435,gray2024timeres,PhysRevLett.132.176701,PhysRevB.109.094413,KIMEL2024172039}; see also Refs.\,\cite{NakaPRB2020,PhysRevB.106.195149,PhysRevB.107.155126,PhysRevB.107.L161109,PhysRevB.108.115138,NakaNatCommun2019,PhysRevB.103.125114,PhysRevB.108.L180401,PhysRevB.108.L140408,PhysRevB.109.174438,PhysRevB.110.054446,ezawa2024_B,lin2024coulombdragAM,chen2024electrSwitchAM,PhysRevB.111.035423,PhysRevLett.132.176702,PhysRevB.110.144412,CCLiu2,PhysRevB.111.L020412,Pupim2024,PhysRevB.111.L161109,PhysRevB.111.184427,Ghorashi2025Dynamical,Yarmohammadi2025Anisotropic,PhysRevB.111.174436,khodas2025AMstrain,PhysRevB.110.094425,sorn2025AnomHallEff,yang2025MagnetoResistanceAM,fu2025SpintronicsAM,karetta2025straincontrolledgdwave} for  more  theoretical studies on     the normal state  AMs and  Refs.\,\cite{Bai_review24,tamang2024,fender2025altermagnetism} for recent reviews on normal state AMs.  In relation to  unconventional magnets with odd-parity, 
some materials have been proposed, such as Mn$_{3}$GaN and CeNiAsO~\cite{hellenes2024P} for $p$-wave    magnets,  and, given the ongoing activities \cite{chakraborty16378}, experiments are very likely to be reported in the  near future; see also \cite{brekke24,PhysRevB.110.205114,ezawa2024_A,ezawa2024_B} for recent theoretical activity in the normal state of  $p$-wave   magnets.   It is thus evident that the peculiar nonrelativistic spin splitting and spin-momentum locking in  unconventional magnets make them promising for  spin-dependent effects, with strong implications  towards scalable spintronic devices in the normal state \cite{Bai_review24}. 

The properties of  unconventional magnets have also triggered a great deal of attention in the superconducting state, placing unconventional magnetism as the key mechanism to engineer novel superconducting phenomena while maintaining zero net magnetization. In this regard, emergent superconducting states have been explored as   intrinsic effects and also as induced phenomena at interfaces of hybrid junctions when coupling  unconventional magnets and superconductors. The first study on the intrinsic coexistence of altermagnetism and superconductivity as well as  the possibility of superconductivity  due to AM spin fluctuations was theoretically carried out    in 2022 \cite{mazin2022notesaSC}, almost immediately  after the prediction of AMs in the normal state. Later, it was shown that such intrinsic coexistence of altermagnetism and superconductivity  can promote finite momentum superconductivity without any net magnetization 
\cite{PhysRevResearch.5.043171,chakraborty2024_1,bose2024altermagn,sim2024,hong2024,mukasa2024}, and, to date, there exist   signs of this coexistence effect in  strained RuO$_2$ \cite{PhysRevLett.125.147001,ruf2021strain,PhysRevMaterials.6.084802} and in monolayer FeSe \cite{mazin2023inducedmoFeSe}; see also Ref.\,\cite{PhysRevB.89.165126}. As an intrinsic effect, altermagnetism was also predicted to appear in  the parent cuprate La$_2$CuO$_4$ of a high-temperature superconductor \cite{LiborPRX22} and recently also suggested to appear in the strongly correlated CuAg(SO$_4$)$_2$  \cite{jeschke2024highly}, see also \cite{domanski2023unique}.
Intrinsic   altermagnetism was also shown to be useful for topological superconductivity \cite{PhysRevB.108.184505},  spin-polarized  $p$-wave superconductivity \cite{Brekke23}, and strained-induced spin-triplet superconductivity \cite{khodas2025AMstrain}, while its emergence was recently discussed in Sr$_2$RuO$_4$ \cite{autieri2025,ramires2025pureMixedAM}.

On a parallel front, hybrid systems formed by coupling  unconventional magnets and superconductors have also received an enormous interest, originating a plethora of novel phenomena with fundamental and applied prospects. For instance, under generic grounds, it was shown that   unconventional magnets not only induce a spin-singlet to spin-triplet conversion but also transfer their parity, generating superconducting correlations with spin-triplet symmetry and higher angular momentum \cite{maeda2025classifi,fukaya2024x}; this unveils the fundamental role of unconventional magnetism in superconductors; see also Refs.\cite{Brekke23,chakraborty2024_2,sukhachov2025,chatterjee2025,sun2025pseudoIsingSCpWaveUM,yokoyama2025SOCSCMixed,khodas2025AMstrain}. Moreover, in the simplest superconducting junction, when an AM is coupled to  a superconductor, the reflection of an incident electron into a hole known as Andreev reflection (AR)   was shown to strongly depend on the crystal orientation and  the spin-splitting \cite{Sun23,Papaj23,maeda2024,IkegayaAltermagnet,Niu2024,PhysRevB.109.L201404,Bo2025}; crossed ARs have also been studied in AM-superconductor-AM junctions \cite{PhysRevB.109.245424,Ping_Niu_2024}. There also exist preliminary theoretical work on ARs in  $p$-wave magnet-superconductor junctions \cite{maeda2024}, with predictions of even zero-bias peaks \cite{fukaya2024x}. We note that the dependence of ARs on the transverse momentum originates the appearance of Andreev bound states (ABSs)  that strongly disperse with such momentum,  shown for superconducting junctions with $d$-wave AMs and $p_{y}$-wave    magnets \cite{fukaya2024x}. Very recently, it was reported a single    experimental study on the AR in AMs, which involved a planar junction between an indium electrode and MnTe   altermagnetic candidate \cite{Kazmin_2025}. Since a finite AR in the  unconventional magnet of a NS junction  is directly related to the induced superconducting correlations in the  unconventional magnets \cite{Cayao2020odd,fukaya2024x}, we conclude that unconventional magnetism strongly affects the proximity effect \cite{klapwijk2004proximity}. The role of potential impurities in AMs was recently also studied \cite{maiani2024}, reporting impurity-induced local sign change of the order parameter and highly tunable spin-dependent tunneling that can be useful for quantum information processing.

The impact of unconventional magnetism was also investigated in junctions formed by two superconductors coupled via an  unconventional magnet in the form of superconductor-unconventional magnet-superconductor Josephson junctions (JJs), thus allowing to explore the  Josephson effect \cite{Josephson62}. In   JJs, ARs at both interfaces lead to the formation of ABSs, which become dependent on the superconducting phase difference that allows them to carry a dissipationaless supercurrent   and determine the Josephson effect
\cite{RevModPhys.51.101,kashiwaya2000,RevModPhys.76.411,Beenakker:92,cayao2016hybrid,sauls2018andreev,eschrig2018theory,mizushima2018}; see also Refs.\,
\cite{PhysRevLett.10.486,kulik1975,kulik1978josephson,FurusakiPRB1991,furusaki1991dc,PhysRevB.45.10563,Furusaki_1999,PhysRevB.64.224515,PhysRevLett.96.097007,cayao2018andreev}. In Josephson junctions based on $d$-wave AMs and spin-singlet $s$-wave superconductors, it was shown that altermagnetism induces in the free energy   $0-\pi$ oscillations \cite{Ouassou23,Beenakker23,zhang2024,PhysRevLett.133.226002,sun2024,fukaya2024x,Bo2025,alipourzadeh2025ABSJE} and oscillations away from $0,\pi$ with double degeneracy \cite{PhysRevLett.133.226002,fukaya2024x}, which result from the strong influence  of altermagnetism on the emergent ABSs. The $0$-$\pi$ transitions were also   predicted in altermagnet-based Josephson junctions formed by spin-singlet and spin-triplet superconductors~\cite{Cheng24}. For $0-\pi$ transitions in Josephson junctions without AMs, see Refs.~\cite{tanaka961,tanaka971,Buzdin03,Goldovin,RevModPhys.76.411}.  The nontrivial features induced by altermagnetism in the free energy and ABSs induce anomalous features in the current-phase curves, involving the contribution from multiple harmonics of the superconducting phase difference \cite{PhysRevLett.133.226002,fukaya2024x}, presence of multiple nodes \cite{PhysRevLett.133.226002}, tunable skewness \cite{sun2024}, and unconventional Fraunhofer patterns \cite{zhang2024}. 

Besides the intriguing Josephson phenomena, altermagnetism was also shown to be useful for realizing other emergent superconducting phenomena. For instance, it was predicted that altermagnetism can be used to engineer topological superconductivity \cite{PhysRevB.108.184505,PhysRevB.109.224502,PhysRevLett.133.106601,CCLiu1,chatterjee2025,pal2025JosephFloAM,hodge2025AM}, where Majorana states are found to appear with a zero net magnetization; similarly, Ref.\,\cite{sun2025pseudoIsingSCpWaveUM} recently reported the realization of topological superconductivity  by using $p$-wave magnetism, while Ref.\,\cite{nagae2025Majo} suggested the design of Majorana flat bands.   As in regular semiconductor-superconductor Majorana devices, AM-based heterostructures might be also susceptible to the formation to topologically trivial zero-energy ABSs \cite{Mondal24}, which can, however, be distinguished e. g., by Josephson measurements \cite{PhysRevLett.123.117001,PhysRevB.104.L020501,baldo2023zero,PhysRevB.105.054504}. For trivial ABSs in superconducting junctions, including Josephson junctions, we refer to Refs.\, \cite{prada2019andreev} and also to Refs.\,
\cite{Bagrets:PRL12,Pikulin2012A,PhysRevB.86.100503,PhysRevB.91.024514,PhysRevB.86.180503,PhysRevB.98.245407,dominguez2017zero,PhysRevLett.123.117001,PhysRevB.102.245431,DasSarma2021Disorder,PhysRevB.105.144509,PhysRevB.104.L020501,PhysRevB.104.134507,marra2022majorana,PhysRevB.107.184509,baldo2023zero,PhysRevB.107.184519,PhysRevB.110.165404,PhysRevB.110.085414,PhysRevB.110.224510,ahmed2024oddfreABS}. Altermagnetism was also proposed as a key ingredient for   bulk \cite{BanerjeePRB24,sim2024,chakrabortyd2024} and Josephson superconducting diode effects \cite{Qiang24,jiang2025josephson,sharma2025tunableJDE,boruah2025fieldfreeJD}, where bulk diodes can even achieve perfect efficiencies due to finite momentum states \cite{chakrabortyd2024}. It was also uncovered that  the perfect diode effect in \cite{chakrabortyd2024} is related to   topology, opening the possibility for  realizing topologically protected superconducting diodes, before suggested in superconductor-semiconductor Majorana systems \cite{PhysRevB.109.L081405,mondal2025JDE}; see also \cite{Tanakadiode1,Tanakadiode2,fu2023fieldeffect} for the diode effect in topological insulators and 
  Refs.\,\cite{FAndoNature2020,MisakiPRB2021,davydova2022universal,DaidoPRL2022,he2022phenomenological,pal2022josephson,mazur2022gatetunable,trahms2023diode,PhysRevLett.130.266003,PhysRevB.108.054522,nadeem2023superconducting,10.1063/5.0210660,PhysRevB.111.L020507} for  other studies on superconducting diodes.   It is worth noting that the peculiar spin-momentum locking of unconventional magnets enables a nontrivial light-matter coupling \cite{hu2025spinSong,Peihao2025,Ghorashi2025Dynamical} that is able to induce spin-triplet states, e.g., by means of Floquet engineering \cite{Peihao2025,yokoyama2025FloquetSCAM}, and control  the spin magnetization \cite{hu2025spinSong}.  We lastly note that AMs have recently been explored for realizing the magnetoelectric effect \cite{zyuzin2024,JXhu2024}, where a spin response is generated by an applied current and   is relevant for   superconducting spintronics.   Thus, despite the recent studies on the interplay between  unconventional magnets and superconductivity, the field has already seen an enormous and rapid progress, which reflects the importance of the field and believe  that the advances will motivate experiments soon. 
   
 In this work, we aim at reviewing the recent progress on superconducting phenomena in systems formed by   unconventional magnets and superconductors, with a particular attention on the  low-energy models and their theoretical predictions. We first in Section  \ref{section1}  introduce unconventional magnetism, including even- and odd-parity magnets,   how they are often modelled using minimal models, and their main properties. We then in Section \ref{section2} discuss the impact of superconductivity and unconventional magnetism, highlighting the possible superconducting symmetries that can emerge due to their combination in a hybrid system. In this part, we also address the Andreev reflection, proximity effect, and inverse proximity effect. In Section  \ref{section3}, we focus on the Josephson effect and comment on how to model Josephson junctions, the formation of ABSs, the current-phase curves, induced correlations, and Josephson diode effects. In Section \ref{section4}, we discuss the recent experimental advanced in unconventional magnets in the normal state and the single experiment involving superconductors. Our conclusions are presented in  \ref{section5}, where we also remark possible future research  directions driven by the interplay between unconventional magnetism and superconductivity.

This review does not intend to cover all the important advances in the field of unconventional magnetism and superconductivity. There are excellent  reviews addressing distinct aspects of these areas and we therefore   refer to those works. For reviews on altermagnetism in the normal state, we refer to Refs.\,\cite{Bai_review24,tamang2024}, while for readers interested in ferromagnets and antiferromagnets we refer to Ref.\,\cite{BergeretReview,RevModPhys.77.935,RevModPhys.91.035004}. Moreover, for reviews on anomalous Hall antiferromagnets, we refer to Ref.\cite{smejkal2021ano},  for antiferromagnetic spintronics to Refs.\,\cite{jungwirth2016antiferromagnetic,Baltz_RevModPhys_2018,10.1063/5.0023614,10.1063/5.0009482,jungwirth2018multiple,dal2024antiferromagnetic}, and for spintronics to Refs.\,\cite{wolf2001spintronics,Sarma_RevModPhys_2004,chappert2007emergence,bader2010spintronics,HIROHATA2020166711,Newhorizons_spintronics}. 
Furthermore, for readers interested in superconductor-ferromagnet heterostructures,  superconducting spintronics, and  topological superconductivity, we refer to  Refs.\,\cite{BergeretReview,RevModPhys.77.935},  Refs.\,\cite{eschrig2011spin,linder2015superconducting,Eschrig2015,yang2021boosting,mel2022superconducting,cai2023superconductor},
and Refs.\,\cite{tanaka2011odd_review,sato2016,Aguadoreview17,sato2017topological,lutchyn2018majorana,zhang2019next,beenakker2019search,frolov2019quest,prada2019andreev,aguado2020majorana,Cayao2020odd,flensberg2021engineered,MarraJAP2022,tanaka2024review}, respectively.
 

\begin{figure}[t!]
    \centering
    \includegraphics[width=8.5cm]{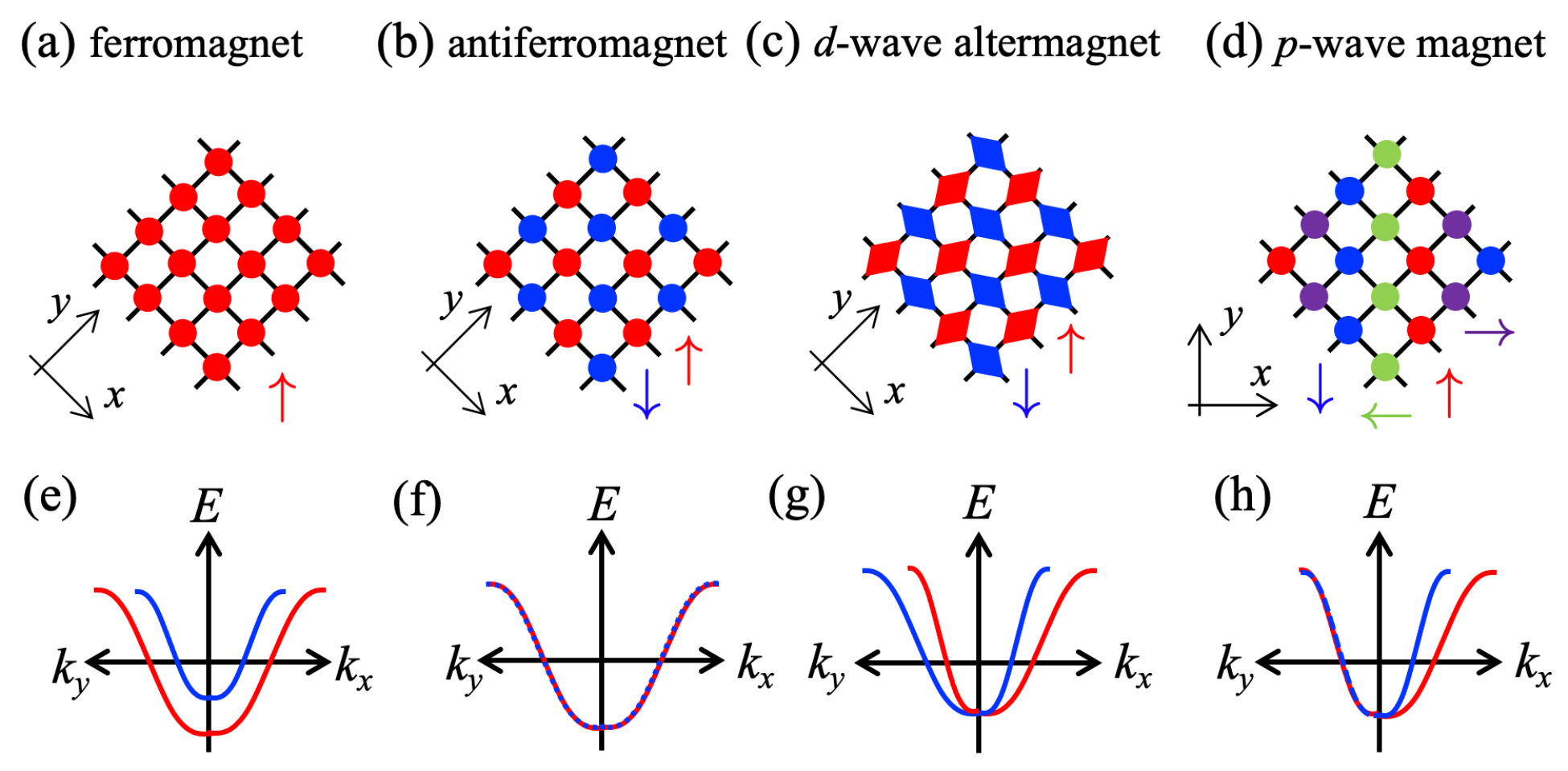}
    \caption{(a-d) Schematic illustration of the lattices in a ferromagnet  (a), antiferromagnet (b), altermagnet (c)~\cite{LiborPRX22}, and unconventional $p$-wave magnet (d)~\cite{brekke24};  (d) takes into account the time-reversal symmetry breaking term due to the isotropic sd coupling between  localized spins and itinerant electrons as in Ref.\,\cite{brekke24}. Up and down arrows colored with blue and red represent spin down and up, respectively. In panel (d), blue, red, green and purple arrows represent the noncollinear spin structure. (e-h) Energy dispersions along $k_x$ and $k_y$ directions for the corresponding magnetic systems of panels (a-d), where the blue and red curves indicate opposite spins.  We note that (h) does not include the time-reversal symmetry breaking term discussed in Ref.\,\cite{brekke24}.}
    \label{demoofAM}
\end{figure}

\section{Unconventional magnetism in the normal state}
\label{section1}
We start by  discussing   unconventional magnetism with even- and odd-parity magnetic orders which characterize  e.g., $d$-wave altermagnets (AMs) and $p$-wave   magnets. Altermagnetism is believed to be a third class of collinear magnetic phases, in addition to the well-known ferromagnets and antiferromagnets \cite{NakaNatCommun2019,Hayami19,Ahn2019,NakaPRB2020,LiborSAv,Hayami20,PhysRevB.102.014422,MazinPNAS,LiborPRX22,landscape22,BhowalPRX24,Bai_review24,hu2024,jungwirth2024,tamang2024,cheong2024altermagnetism}.  The term `altermagnet' was named after the anisotropic magnetization that `alternates' direction in momentum space. The study on this state can be traced back to a series of discoveries of unconventional magnetism before the name was given, such as in the momentum-dependent spin split bands with antiferromagnetic order \cite{NakaNatCommun2019,Hayami19,NakaPRB2020,PhysRevB.102.014422,Rafael21}, the crystal Hall effect \cite{LiborSAv}, the $C$-paired spin-valley locking in antiferromagnets \cite{MaNatcommun2021}, or in the ferromagnetic response in antiferromagnets \cite{MazinPNAS}. 

\begin{figure}[t!]
    \centering
    \includegraphics[width=8cm]{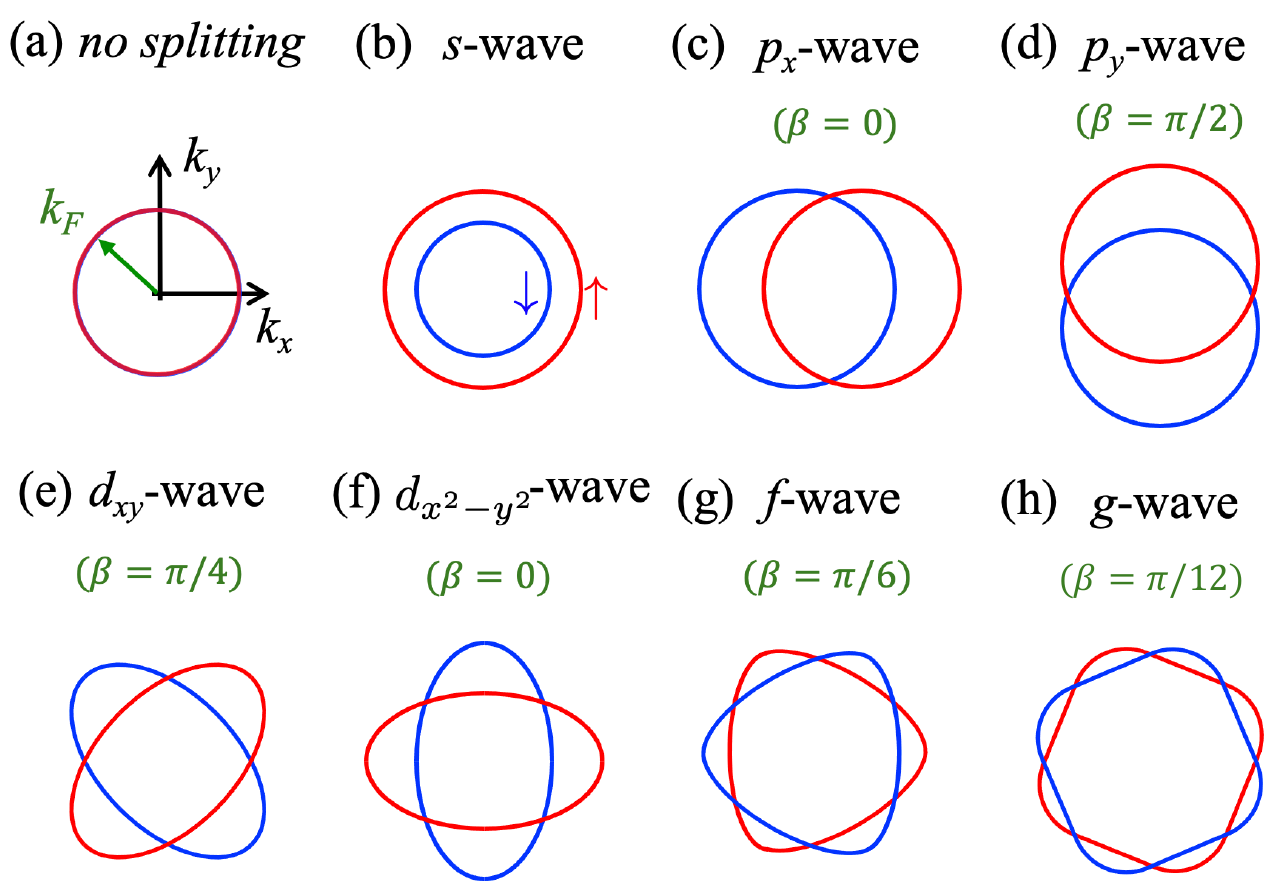}
    \caption{Sketch  of the Fermi surfaces for up (red circle) and down (blue circle) spins in distinct normal state systems, including   (a) a normal metal, (b) a ferromagnet ($s$-wave magnetic order), (c) a $p_{x}$-wave  magnet, (d) a $p_{y}$-wave  magnet, (e) a $d_{xy}$-wave  altermagnet (AM), (f) a $d_{x^{2}-y^{2}}$-wave AM, (g) a $f$-wave magnet, and (h)  $g$-wave AM.    The green arrow in (a) indicates the Fermi wavevector and $\beta$ indicates the direction of the nodes where the exchange field  becomes zero.}
    \label{FS_UM}
\end{figure}

To understand the properties of AMs, we first review some characteristics of ferromagnets and antiferromagnets since AMs share common features with them,  see Figs.\,\ref{demoofAM} and \ref{FS_UM}. For ferromagnets, the exchange field induces a majority spin state and a minority spin state for all lattice sites [Fig.~\ref{demoofAM}(a)], making the majority spin state more populated at the Fermi surface [Fig.~\ref{demoofAM}(e) and Fig.~\ref{FS_UM}(b)]; see also  Fig.~\ref{FS_UM}(a), which shows the Fermi surface of a normal metal without spin splitting is shown for comparison. In some half-metal materials, the minority spin state can disappear at the Fermi surface which results in full polarization. Thus, ferromagnets possess a non-vanishing net macroscopic magnet moment \cite{kaganov,nolting2009quantum}. In Landau's broken symmetry theory, ferromagnets can be described by a ferromagnetic order parameter: the exchange field \cite{kaganov}. Such an exchange field is isotropic in ferromagnets since the ferromagnetic state is invariant under a unitary lattice symmetry. Another type of collinear magnets are antiferromagnets \cite{kaganov,nolting2009quantum}, which has an alternating spin polarization in real space with neighboring sites possessing different spin Fig.~\ref{demoofAM}(b). Thus, antiferromagnets are invariant under a unitary lattice translation or inversion symmetries \cite{RevModPhys.25.58,kaganov,nolting2008fundamentals,nolting2009quantum}. The spin degeneracy of an antiferromagnet is not lifted as dictated by parity-time-inversion ($\mathcal{P}\mathcal{T}$) symmetry [Fig.~\ref{demoofAM}(f)].

Based on the above discussion, AMs are similar to ferromagnets  because both of them have a spin splitting of energy bands and spin-polarized Fermi surfaces,  see Figs.\,\ref{demoofAM}(a,c) and \ref{FS_UM}(b,e,f,h). However, the spin states of AMs compensate each other and result in zero net magnetization, akin to what occurs in antiferromagnets but distinct from ferromagnets. As already pointed out, AMs exhibit an exchange field that is even in momentum (even-parity), which can posses   $d$-, $g$-, $i$-wave symmetries [Fig.\,\ref{FS_UM}(b,e,f,h)].  Unlike antiferromagnets, the Kramers degeneracy in AMs is lifted  because the opposite-spin sublattices can not map onto each other by a translation or inversion but are connected by a real-space rotation transformation \cite{landscape22}, as shown in Fig. \ref{demoofAM}(c,g). In AMs, $\mathcal{PT}$ is broken by the combined lattice geometry and spin ordering \cite{landscape22}. Furthermore, an important property of AMs is that the spin-splitting field  does not originate from
relativistic effects, which is distinct from relativistically spin-orbit coupling  \cite{galitski2013spin,manchon2015new}.For this reason, the spin-splitting field  in AMs acquire   larger values, e. g., reaching the order of $1$\,eV \cite{landscape22}.  
In relativistically spin-orbit coupling systems, such as semiconductors with Rashba spin-orbit couplings, it requires heavy element to produce large spin-splitting but their values are not very large \cite{manchon2015new,galitski2013spin,RevModPhys.96.021003}. In contrast, AMs   have advantages in providing significant spin-splitting field in lighter compounds based on e.g., Fe and Mn. Even though the field is still developing, there have already been a large number of theoretical and experimental studies  addressing AMs and their properties; see e. g., Refs.\,\cite{KrempaskyNature2024,NakaNatCommun2019,Ahn2019,NakaPRB2020,LiborSAv,LiborPRX22,fedchenko2024observation,jiang2024,fzhang2024} for some previous experiments and Section \ref{section4} for more details on the experimental advances. In terms of theoretical developments in the normal state, AMs have been proposed to realize ultrafast THz switching devices with no stray fields and with low damping spin currents \cite{Magnon1,Baltz2024,weber2024} and they have also were  suggested to generate a giant tunneling magnetoresistance effect \cite{Libor011028}.  Moreover, AMs were shown to be useful for   anomalous Hall effects \cite{LiborSAv,Rafael21,Feng_2022,Lee24,PhysRevB.106.195149,PhysRevB.110.094425}, including a skyrmion Hall effect~\cite{jin2024skyrmion}, as well as for promoting  high harmonic generation \cite{werner2024}. Other studies of AMs involved impurity-induced Friedel oscillations in the local density of states~\cite{PhysRevB.110.205114}, strain-induced spin splitting \cite{belashchenko2024,karetta2025straincontrolledgdwave}, spin splitting Nernst effect~\cite{PhysRevB.108.L180401,PhysRevB.111.035423}, higher-order topological state~\cite{CCLiu2}, field-sensitive dislocation bound states~\cite{zhu2024}, parity anomaly in weak topological insulators~\cite{PhysRevB.111.045407}, bilayer stacking ferromagnets with antiferromagnet coupling \cite{Zeng_2024theory}, in heavy fermions systems \cite{zhao2024ss}, a    Landau theory of altermagnetism~\cite{PhysRevLett.132.176702}, and   spin currents \cite{shao2021spin,ezawa2024_B,yang2025MagnetoResistanceAM,fu2025SpintronicsAM,belashchenko2024,shao2021spin,golub2025spincurrent}; see also Refs.\,\cite{Bai_review24,tamang2024,jungwirth2024He3Ams,jungwirth2024} for a more detailed discussion of the advances on AMs in the normal state. It is clear that the number of candidate altermagnetic materials is now rapidly growing \cite{landscape22,LiborPRX22,Bai_review24,tamang2024,hu2024,hu2024,fzhang2024,lu2024,PhysRevB.111.014434,meier2025,yang2025}, and all the advances support the benefits of AMs for potential spintronic applications. 

After the prediction and discovery of altermagnetism, the field moved quickly to the study of  odd-parity unconventional magnets, which can exhibit a magnetic order with $p$ and $f$-wave symmetries \cite{PhysRevB.101.220403}. In the case of $p$-wave unconventional magnets, they are both  noncollinear and noncoplanar magnets with preserved time-reversal symmetry  and broken inversion symmetry \cite{brekke24,hellenes2024P}.  More precisely,   $p$-wave unconventional magnets are protected by the combination of the time-reversal operation ($\mathcal{T}$) and a translation of half the unit cell ($\tau$)~\cite{hellenes2024P}.   To visualize some preliminary properties of $p$-wave magnets without delving into the specific models, in Figs.~\ref{demoofAM}(d) we show an schematic illustration of the $p$-wave magnet lattice based on the effective model of Ref.\,\cite{brekke24}, induced by   the spin-rotational structure along one direction, while  Figs.~\ref{demoofAM}(h) shows the energy dispersion for the same model   but without the isotropic sd  coupling between   between a localized spins and itinerant electrons discussed in \cite{brekke24}; later on we will deepen the discussion of models and their properties. In Figs.~\ref{FS_UM}(c,d,g), we also depict the Fermi surfaces of $p$-wave magnets, while in Figs.~\ref{FS_UM}(g) of a $f$-wave magnet. In terms of materials for odd-parity magnets, Ref.\ \cite{hellenes2024P} predicted Mn$_3$GaN and CeNiAsO as candidate materials for   $p$-wave magnetism but their experimental verification is still an open task.   The interest in $p$-wave magnets has recently also attracted more attention, with studies   addressing   Friedel oscillations in the local density of states~\cite{PhysRevB.110.205114},  transverse spin currents ~\cite{PhysRevB.111.035404}, linear and nonlinear conductivities 
 \cite{ezawa2024_A},   spin-current diode effect \cite{ezawa2024_B}, 
 and highly efficient non-relativistic Edelstein effect~\cite{chakraborty2024Edelstein}.

Most of the theoretical advances on both AMs and  $p$-wave magnets have employed minimal models that capture the main properties of both  unconventional magnets. Due to the relevance of these models for the advance of the field and for understanding most of the phenomena in unconventional magnets,  we discuss them  in the next subsection.

 \begin{figure*}[t!]
    \centering
    \includegraphics[width=17.5cm]{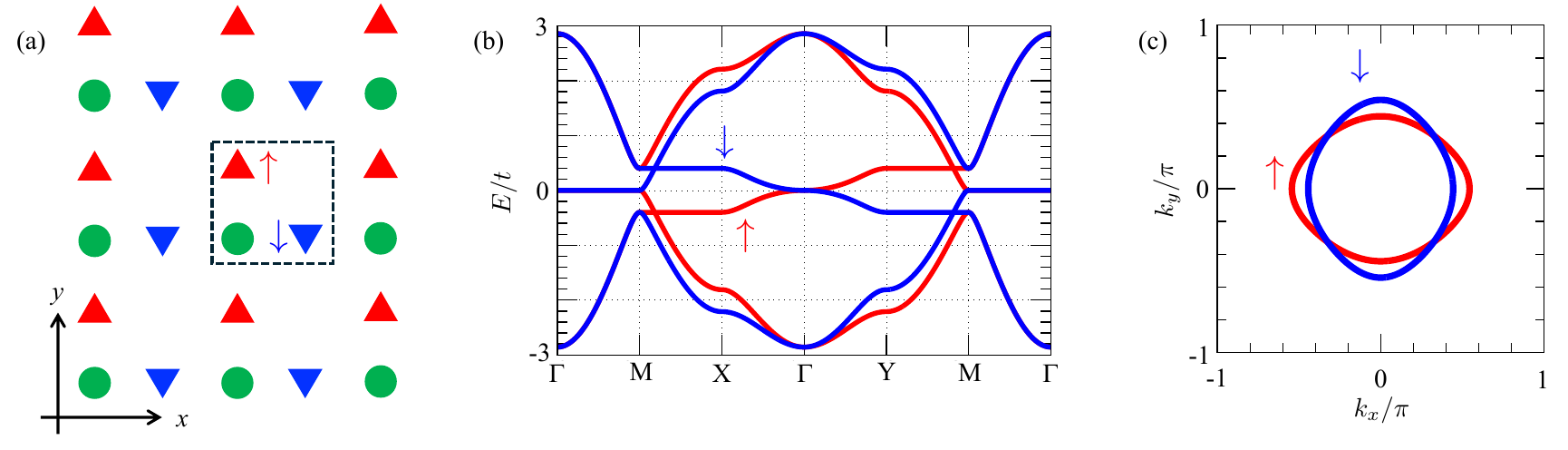}
    \caption{(a) Schematic illustration  of the lattice in a $d$-wave altermagnet, adapted from Ref.\ \cite{Brekke23}.    Red and blue triangles indicate the magnetic sites with spin-up and down, respectively.  A green circle is a nonmagnetic site.
    The black-dotted line means the unit cell.
    (b) Energy dispersion and (c) Fermi surface of (a).
    The red and blue lines correspond to the up and down-spin, respectively. 
    We set the magnetic order as $M=0.4t$.    In (c), we choose the chemical potential as $\mu=2.5t$. 
    }
    \label{mini_AM}
\end{figure*}
 
\subsection{Minimal models for understanding unconventional magnetism}
\label{section2a}
 In this part, we discuss the minimal models to describe unconventional
magnetism, with a particular focus on $d$-wave AMs and $p$-wave  magnets. Within a 
simple, but proven useful, description of the itinerant electrons in  unconventional magnets is given by the phenomenological exchange field with momentum dependence $M_{\bm{k}}$ \cite{landscape22}. From a microscopic point of view, the tight-binding models for  unconventional magnets are proposed based on fundamental symmetries \cite{Brekke23,PhysRevB.110.144412,hellenes2024P,brekke24}. In what follows, we summarize the   minimal models for  unconventional magnets and their consequent electronic band structures.

To model unconventional magnets, effective two-band models   describing momentum-dependent spin-splitting bands in the continuum model were adopted and they are given by \cite{landscape22}
\begin{equation}
\label{eq:AM_H}
\hat{H}(\bm{k})=\frac{\hbar^2\bm{k}^2}{2m}-\mu-M_{\bm{k}}\bm{n}_{\bm{k}}\cdot{\bm{\sigma}}\,,
\end{equation}
where $\mu$ is the chemical potential, $M_{\bm{k}}$ is the momentum-dependent exchange field characterizing the unconventional magnetic order,   $\bm{n}_{\bm{k}}$ a unit vector that dictates the direction of the spin polarization and ${\bm{\sigma}}=[{\sigma}_{x},{\sigma}_y,{\sigma}_{z}]$ is  the vector of Pauli matrices in spin space. The energy dispersion of Eq.\ (\ref{eq:AM_H}) is given by 
\begin{equation}
    \varepsilon_{\bm k}^\pm=\hbar^2{\bm k}^2/2m-\mu\pm M_{\bm k}\,,
\end{equation}
which clearly reveals that $M_{\bm k}\neq0$ induces a momentum-dependent splitting of spin bands.  One of the simplest situations of the above Hamiltonian is a two-dimensional model with  a collinear spin configuration along the direction $\bm{n}_{\bm{k}}=(0, 0, 1)$, which is the focus of our discussion in what follows.

The symmetry of  $M_{\bm k}$ can be classified in the same way as that of the superconducting pair potential. Thus, the conventional ferromagnetic state is classified as $s$-wave magnetic order with $M^{s}_{\bm k}=J$, which does not exhibit any momentum dependence. Similarly, the magnetic order of $p$-wave magnets  can be modelled by \cite{hellenes2024P,brekke24}
\begin{equation}
\label{eq:mp}
    M^{p}_{\bm{k}}=\frac{J}{k_\mathrm{F}}[k_x\cos\beta  +k_y\sin\beta],
\end{equation}%
while the magnetic order of $d$-wave AMs by \cite{landscape22}
\begin{equation}
\label{eq:am}
    M^{d}_{\bm{k}}=\frac{J}{k^2_\mathrm{F}}[2k_x k_y\sin 2\beta +(k_x^2-k_y^2)\cos2\beta]\,, 
\end{equation}%
where $k_{\rm F}=\sqrt{2m\mu/\hbar^{2}}$ is the Fermi wavevector as shown in Fig.~\ref{FS_UM}, while $\beta$ represents the angle between the $x$-axis and the nodes where the exchange field becomes zero. By setting $\beta=0$ or $\beta=\pi/2$, $ M^{p}_{\bm{k}}$ models a  $p_{x}$- or a $p_{y}$-wave  magnet.  At this point, we note that the minimal model for the $p$-wave magnets given by Eq.\,(\ref{eq:mp}) does not take into account the sd coupling between  localized spins and itinerant electrons depicted in Figs.~\ref{demoofAM}(d) and thus cannot spin-flip processes intrinsically present in noncollinear magnets \cite{hellenes2024P}. However, in spite of this seemingly issue with the model in Eq.\,(\ref{eq:mp}), such a model interestingly captures the inherent $p$-wave parity of the spin polarization \cite{brekke24}, see Fig.~\ref{demoofAM}(h) and Figs.\,\ref{FS_UM}(c,d). That is why  Eq.\,(\ref{eq:mp}) is useful and can help us understand some key properties of $p$-wave magnets.   In relation to AMs, we also find two types at given angles, while a mixture of them at other angles. For instance, for  $ M^{d}_{\bm{k}}$ at $\beta=0$ models $d_{x^{2}-y^{2}}$-wave AM, while $ M^{p}_{\bm{k}}$ at $\beta=\pi/4$ models a $d_{xy}$-wave AM. Note that we use $J$ to label the strength of magnetic order, irrespective of the type of magnet. Following a similar discussion, the magnetic order of $f$-, $g$-, and $i$-wave magnets \cite{ezawa2024_B,maeda2025classifi}
\begin{equation}
\begin{split}
    M^{f}_{\bm{k}}&=\frac{J}{k^3_\mathrm{F}}[k_x (k_x^2-3k_y^2)\cos3\beta \\
    &+k_y (3k_x^2-k_y^2)\sin3\beta]\,,\\
        M^{g}_{\bm{k}}&=\frac{J}{k^4_\mathrm{F}}\Big[(k_{x}^{4}-6k_{x}^{2}k_{y}^{2}+k_{y}^{4}){\cos}4\beta\\
    &+4k_xk_y(k^2_x-k^2_y){\rm sin}4\beta\Big]\,,\\
   M^{i}_{\bm{k}}&=\frac{J}{k_{\rm F}^{6}}\Big[(k^{6}_{x}-15k^{4}_{x}k^{2}_{y}+15k^{2}_{x}k^{4}_{y}-k^{6}_{y})\cos{6\beta}\\
   &+2k_{y}k_{y}(3k_{x}^{2}-k_{y}^{2})(k_{x}^{2}-3k_{y}^{2})\sin{6\beta}\Big].
    \end{split}
\end{equation}%
These continuum models can be generalized to three dimensions as well, see Ref.\ \cite{ezawa2024_B}; and also  Ref.\ \cite{PhysRevB.110.144412} for higher order even-parity  unconventional magnets.  Note that the symmetry of the Hamiltonian in Eq.\ (\ref{eq:AM_H}) for the even-parity  unconventional magnets ($s$-wave, $d$-wave, $g$-wave, $i$-wave) and odd-parity  unconventional magnets ($p$-wave, $f$-wave) are completely different in terms of the time-reversal symmetry. While the odd-parity  unconventional magnets have time-reversal symmetry as given by
\begin{equation}
\label{eq:TRS}
\mathcal{T}\hat{H}(\bm{k})\mathcal{T}^{-1}=\hat{H}(-\bm{k})\,,
\end{equation} 
the even-parity  unconventional magnets break this symmetry, where $\mathcal{T}=-i{\sigma}_y \mathcal{K}$ is the time-reversal operator. Thus, according to the common view that a magnet  should break time-reversal symmetry \cite{kaganov,nolting2008fundamentals,nolting2009quantum}, $p$-wave magnets  described by Eq.\,(\ref{eq:mp}) cannot be regarded as magnets but instead as another kind of magnetic phase. This is of course in contrast to the spin structure given in Fig.\,\ref{demoofAM}(d) which clearly breaks time-reversal symmetry. In this regard, as already noted above, Eq.\,(\ref{eq:mp}) does not completely reproduce the noncollinear magnetic structure and thus cannot describe unavoidable spin-flip processes. This issue can be resolved by incorporating into Eq.\,(\ref{eq:mp}) a term due to the isotropic sd coupling between localized spins and itinerant electrons proportional to $\sigma_{x}$ as done in Ref.\,\cite{brekke24}; see also Ref.\,\cite{hellenes2024P}. Nevertheless, we emphasize that both Eq.\,(\ref{eq:mp}) and the more elaborated model taking into account the noncollinear spin structure produce very similar   Fermi surfaces and spin expectation values $S_{z}$. For this reason,   Eq.\,(\ref{eq:mp})can be seen as useful to describe certain properties of   $p$-wave magnets. Thus,  we will always refer to them as $p$-wave magnets as already adopted in the literature, see e. g.,\,Ref.\, \cite{hellenes2024P,brekke24,ezawa2024_B,maeda2025classifi}.

\begin{figure*}[t!]
    \centering
    \includegraphics[width=17.5cm]{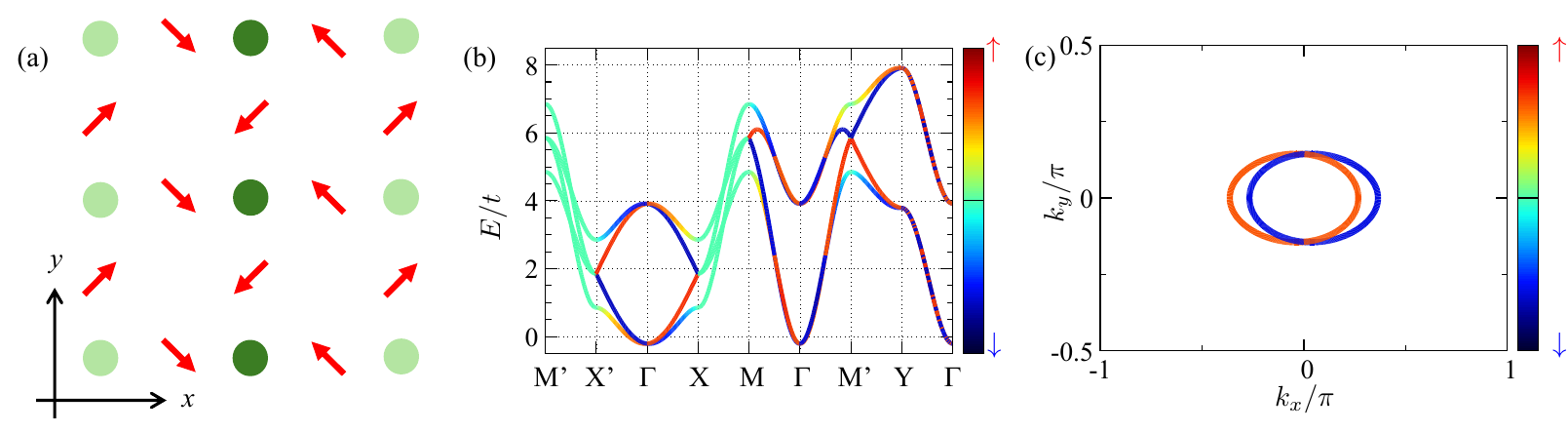}
    \caption{(a) Schematic illustration of the lattice in a  $p$-wave magnet,  adapted  from   Ref.\ \cite{hellenes2024P}.     Dark and light green colors denotes nonmagnetic atoms and red color denotes the spins of magnetic atoms in the $xy$-plane.    This sketch   corresponds to Eq.\ (\ref{eq:dAMH3}).     (b) Energy dispersion  and (c) Fermi surface of (a) with $t_J=0.25 t$ and the chemical potential $\mu=-3.85 t$.
     }
    \label{mini_UPM}
\end{figure*}

Before going further, we note that, under certain circumstances,  it is also important to consider the lattice models  of  unconventional magnets, where one can also incorporate point group symmetries and other properties of actual crystals. Broadly speaking, when considering a Hamiltonian in a crystal, e.g.\, given by Eq.\,(\ref{eq:AM_H}) , a lattice regularization is performed on said Hamiltonian. Then, the Hamiltonian must have the property that $\hat{H}(\bm{k})=\hat{H}(\bm{k}+\bm{G})$, where $\bm{G}$ is the reciprocal lattice vector. In the case of a tetragonal crystal system, $\bm{G}=(2\pi/a, 0)$ and $(0, 2\pi/a)$. Thus, $\hat{H}(\bm{k})$ is expressed by   sine and cosine functions, $\sin(nk_{x,y}a)$ and $\cos(nk_{x,y}a)$, and their products, where $n$ and $a$ are arbitrary integer and the lattice constant;  we set $a=1$ unless otherwise specified. Here, we can simply do the  following replacements: $k_{x,y}\rightarrow\sin{k_{x,y}}$ and $k_{x,y}^2\rightarrow2-2\cos k_{x,y}$. Then, Eq.(\ref{eq:AM_H}) for an  unconventional magnet in a square lattice  is given by 
\begin{equation}
\label{eq:dAM_H}
\hat{H}_1(\bm{k})=-2t(\cos k_x+\cos k_y) -\mu+\hat{H}^{j}_{\rm M}(\bm{k})\,,
\end{equation}
where the first term is the kinetic energy with $t$ being the hopping parameter, while the second term is the exchange field $\hat{H}^{j}_{\rm M}(\bm{k})$ describing the type of magnetic order in a lattice.  Taking as a starting point the continuum models given by Eq.\,(\ref{eq:am}) and Eq.\,(\ref{eq:mp}), for $s$-wave magnets (ferromagnet), $p$-wave  magnets, and $d$-wave AMs,  $\hat{H}^{j}_{\rm M}(\bm{k})$ are given by
\begin{equation}
\label{MTB}
\begin{split}
    \hat{H}^{s}_\mathrm{M}(\bm{k})&=m_{z}{\sigma}_z\,,\\
        \hat{H}^{p}_\mathrm{M}(\bm{k})&=[t_{x}\sin{k_x}+t_{y}\sin{k_y}]{\sigma}_z\,,\\
             \hat{H}^{d}_\mathrm{M}(\bm{k})&=[2t_{xy}\sin{k_x}\sin{k_y}\\
             &+t_{x^2-y^2}(\cos{k_x}-\cos{k_y})]{\sigma}_z\,,
 \end{split}
\end{equation}
where $m_{z}$   is the amplitude of the ferromagnetic order, $t_{x}$ and $t_{y}$ are the strength of unconventional $p_x$ and $p_y$-wave magnetic order, while   $t_{xy}$ and $t_{x^2-^y2}$ are strengths  of $d_{xy}$ and $d_{x^2-y^2}$-wave altermagnetic order. The second and third expressions of Eqs.\,(\ref{MTB}) characterize spin-dependent hopping integrals in  unconventional magnets, incorporated here within a two-band model description although their origin requires a more elaborated microscopic modelling, as discussed below.

To understand the microscopic origin of the spin-dependent hopping integrals in  unconventional magnets beyond two-band models,    minimal models that incorporate the sublattice degree of freedom and local magnetic spins have been proposed for  unconventional magnets \cite{landscape22,hellenes2024P,brekke24}. To be more precise, in the case of $d$-wave AMs \cite{landscape22}, the minimal model consists of two magnetic sites with distinct spin and one nonmagnetic site, as schematically shown in Fig.\,\ref{mini_AM}(a). Now, to comprehend  the spin dependent hopping integrals, it is important to consider the   hopping processes between the neighboring nonmagnetic sites, denoted by green filled circles in the Fig.\,\ref{mini_AM}(a). The first process is the direct hopping terms, which is included in the first term of Eq.\,(\ref{eq:dAM_H}) and characterized by $t$. Moreover, in addition to the direct hopping,  there are also virtual processes through the magnetic sites, denoted by blue (and red) triangles in Fig.\,\ref{mini_AM}(a). These virtual processes cause the spin-dependent hopping integrals in the third term of Eq.\,(\ref{eq:dAM_H}), see also second and third expressions  of Eqs.\ (\ref{MTB}). Thus, in the case when the magnetic moment between the nearest neighbor magnetic sites along the $x$-direction and the $y$-direction are opposite, the correspondent effective hopping integrals become those given by the third expression of the Hamiltonian in Eq.\ (\ref{eq:dAM_H}) for  $d$-wave AMs.  The minimal model discussed here can be formulated in terms of a Hamiltonian by considering a sublattice structure composed of a nonmagnetic site and two magnetic sites with positive and negative magnetic moments shown in  Fig.~\ref{mini_AM}(a)~\cite{Brekke23}.
Thus, the minimal model Hamiltonian for $d$-wave AMs can be written as
\begin{equation}
\label{eq:dAM_H2}
\hat{H}_2(\bm{k})=
\begin{pmatrix}
-\mu&-2t\cos{k_x/2}&-2t\cos{k_y/2}\\
-2t\cos{k_x/2}&M\sigma_z+V-\mu&0\\
-2t\cos{k_y/2}&0&-M\sigma_z+V-\mu\\
\end{pmatrix}
\end{equation}%
where $t$ is the hopping between the non-magnetic and two magnetic sites, while
$M$ and $V$ are the magnetic moments and the on-site potential of the magnetic sites, respectively. When the Fermi energy is close to the energy levels of the magnetic sites, we can trace out the nonmagnetic site and an effective $4\times4$ Hamiltonian in Ref.\,\cite{chakraborty2024_2} can be obtained. This two-dimensional tight-binding Hamiltonian can capture the characteristics of AMs where the opposite-spin arrangement can not be obtained from the lattice translation but from the four-fold rotation of the system. In Fig.~\ref{mini_AM}(b,c), the energy dispersion at $M=0.4t$ and the Fermi surface at $\mu=2.5t$ for the $d_{x^2-y^2}$-wave altermagnetic order are obtained by solving the tight-binding Hamiltonian.

Having understood the spin-dependent hopping integrals in AMs in terms of a minimal model, we now focus on   $p$-wave magnets. As we already know, the effective model for $p$-wave magnets given by Eq.\,(\ref{eq:mp}) does not break   time-reversal symmetry because $M_{\bm k}$ is an odd function in ${\bm k}$-space and Eq.\,(\ref{eq:TRS}) is satisfied.  We also discussed that a $p$-wave magnet that breaks time-reversal symmetry can be realized by considering a noncollinear magnetic structure \cite{hellenes2024P,brekke24}. One of the  minimal models beyond Eq.\,(\ref{eq:mp}) to describe  $p$-wave magnet was presented in Ref.\, \cite{hellenes2024P}, where the unit cell consists of two nonmagnetic sites and four magnetic sites with noncollinear spins, see Fig.~\ref{mini_UPM}; see also  Refs.\ \cite{brekke24,PhysRevB.110.205114}. In this case, the spin-dependent hopping integrals arise from the noncollinear magnetic structure between nonmagnetic sites \cite{hellenes2024P}. Following the spirit of Ref.\,\cite{hellenes2024P}, the  effective minimal Hamiltonian for a $p$-wave magnet can be written as \cite{hellenes2024P}
\begin{equation}
\label{eq:dAMH3}
\begin{split}
    \hat{H}_{3}(\bm{k})&=-\{2t[\cos(k_x/2)\tau_x+\cos(k_y)]+\mu\}\sigma_0 \\
    &+2t_J[\sin(k_x/2)\sigma_x\tau_y+\cos(k_y)\sigma_y\tau_z]\,,
\end{split}
\end{equation}
where ${\sigma}_{i}$ and ${\tau}_{i}$ are the $i$-th Pauli matrix in spin and sublattice spaces, respectively. In this model, the real-space magnetic structure is noncollinear [Fig.~\ref{mini_UPM} (a)], and the spin along the $(x+y)$- and $(x-y)$- direction couples to $\sigma_y$ and $\sigma_x$, respectively.
Intuitively, the direction of the spin corresponds to the direction of the couplings, while there is a mismatch between the spin-directions of Fig.\ \ref{mini_UPM}(a) and the coupling in Eq.\ (\ref{eq:dAMH3}).
In the proposed material in Ref.~\cite{hellenes2024P}, the $z$-components of atomic coordinates of magnetic sites and non-magnetic sites are different, which could be the cause for this mismatch between models.  The spin expectation values in this model have  $S_x$ and $S_y$ parts and also a $S_z$  term \cite{hellenes2024P}. In Fig.~\ref{mini_UPM}(b), the energy dispersion  is shown from the tight-binding Hamiltonian in Eq.\ (\ref{eq:dAMH3}), color coded by the  the spin expectation values along $z$. We see that the there is a splitting of  momentum dependent bands having opposite values of  $S_z$  at low energies   around the $\Gamma$ point. This spin-splitting is further confirmed by looking at the two spin-split Fermi surfaces shown in Fig.~\ref{mini_UPM}(c).   Having a spin polarization that is opposite for positive and negative momenta unveils its odd-parity nature, which can be also proved to be linear in momentum, and hence of $p$-wave nature. 

The minimal model discussed here is    similar to the models discussed in Refs.\ \cite{brekke24,PhysRevB.110.205114} for noncollinear spin structures along one direction in real space, see Fig.\,\ref{demoofAM}(d). In particular, Refs.\,\cite{brekke24} suggest a minimal model for a $p$-wave magnet given by $ \hat{H}_{4}(\bm{k})=-\{2t[{\rm cos}(k_{x}a)+{\rm cos}(k_{y}a)+\mu]\}\sigma_{0}\tau_{0}+[t_{x}{\rm sin}(k_{x}a)+t_{y}{\rm sin}(k_{x}a)]\sigma_{z'}\tau_{0}+J_{sd}\sigma_{x'}\tau_{z}$, where the prime spin coordinates in $\sigma_{i'}$ indicate that they are in principle decoupled from the crystal coordinates; in the simplest case, the spin coordinates are the same as the crystal coordinates, but distinct magnetic textures require appropriate rotation of the primed spin coordinates, see Ref.\,\cite{brekke24}. Moreover, in the model of \cite{brekke24}, the $J_{sd}$ characterizes the sd coupling which arises due to the interaction between itinerant electrons and a localized spin with a noncollinear magnetic structure. Thus, $p$-wave magnets can be modelled by either $\hat{H}_{3}(\bm{k})$ or $\hat{H}_{4}(\bm{k})$. Nevertheless, we stress   that, while $\hat{H}_{3(4)}(\bm{k})$ describe the noncollinear nature of $p$-wave magnets, Eq.\,(\ref{eq:mp}) might be enough if one is interested in the physics surrounded the $p$-wave parity. Furthermore, the spin-dependent hopping in Eq.\,(\ref{eq:dAMH3}) and the   spin-splitting   in Eq.\,(\ref{eq:mp}) originate from the magnetic structure of $p$-wave magnet which is different from   the relativistic spin-orbit coupling. Besides this, the momentum dependence of the exchange field in unconventional $p$-wave magnets [Eq.\,(\ref{eq:mp})] is equivalent to the momentum configuration in a persistent helix, see e. g.\,, Refs.\,\cite{hellenes2024P,PhysRevLett.97.236601,PhysRevB.92.024510,PhysRevB.95.075304,PhysRevB.101.155123,PhysRevB.103.L060503,PhysRevB.103.104509,PhysRevB.104.L020502}. This spin-helix phase is accidentally realized by the combination of Rashba- and Dresselhaus-type spin-orbit coupling when their strength are of the same order~\cite{PhysRevLett.97.236601,PhysRevB.74.235322,RevModPhys.89.011001,kohda2017physics} and therefore, it has a relativistic origin. It is worth noting  that  Fermi surfaces similar to the one discussed for $p$-wave magnets appear in a layered ferrimagnet Mn$_3$Si$_{2}$Te$_6$ \cite{seo2021colossal}. In this material, the low energy effective Hamiltonian is composed of $p_{x}$ and $p_{y}$ orbitals, reflecting a layered structure. There are thus four degenerate states at the $\Gamma$ point in its nonmagnetic states. These fourfold degenerate states split into two spin-polarized doubly degenerate states by the magnetization. In this subspace, only the $L_{z}S_{z}$ term is relevant as an atomic spin-orbit coupling $\bm{L}{\cdot} \bm{S}$, and therefore, twofold degeneracy is not lifted by an in-plane magnetization. As a result, degenerate bands appear on the $k_{z}$-axis, implying that  the shape of the obtained Fermi surfaces is similar to that of the $p$-wave magnets discussed here. However, their spin-states are different because Mn$_3$Si$_{2}$Te$_6$ is ferrimagnet and spin is polarized on these two Fermi surfaces, making it different from $p$-wave magnets.

At this point, we comment on the proper use of these Hamiltonians to model  unconventional magnets. For $d$-wave AMs, the Hamiltonians given in Eq.(\ref{eq:dAM_H}) and (\ref{eq:dAM_H2}) provide similar bulk information, such as the dispersion and their magnetic splitting near the Fermi level. On the other hand,  the momentum dependent spin-splitting may affect transport in junctions or systems with edge states, and in this case one has to be careful when using the real space or continuum descriptions of unconventional magnets.  This problem becomes important when  addressing   magnetic properties of the edge states or the spin-transport in junctions   because local spin structure at the edge/interface may affect them. 
For instance, this is relevant when   considering the particular [100] edge, the magnetic moments at the edge depend on the termination of the system.   
To consider the local spin-structure in continuum model, magnetic boundary conditions are required, which in principle involve  distinct conditions for up- and the down-spins.  A similar discussion applies to the use of $p$-wave magnets. To summarize, we have reviewed the minimal models of unconventional magnets. The effective Hamiltonians in the continuum are given by Eqs.\ (\ref{eq:AM_H}) while in the tight-binding description is given by Eqs.\,(\ref{MTB}). While these models have been broadly used   for investigating    superconductivity involving unconventional magnets,   discussed in Sections \ref{section2} and \ref{section3}, we note that 
 it would be interesting to adopt the microscopic models in the form of Eqs.\ (\ref{eq:dAM_H2}) and (\ref{eq:dAMH3}) in   future studies of transport phenomena.

\subsection{Topological phases in unconventional magnets}
The past two decades have witnessed significant progress in topological materials, with an enormous focus on systems with spin-orbit coupling. In this regard, unconventional magnets provide an alternative ground to realize functional topological materials. Since the field is still developing \cite{landscape22,PhysRevLett.133.106601}, most of the studies are theoretical but experimental activity will very likely be reported soon.
By now, it has already been predicted that it is possible to realize topological nontrivial  unconventional magnets \cite{PhysRevLett.133.106601,Junxiang23,antonenko2024weyl,parshukov2024,dasroy2024,Yufei24,PhysRevB.109.024404,brekke24,li2024crsb,zhu2024,JinFeng24,jungwirth2024}, efforts that also suggest to engineer topological states by combing unconventional magnets with topological materials. 

For intrinsic altermagnetic materials, relativistic spin-orbit coupling, which generally exists in actual crystal environment or induced by external inversion symmetry breaking  \cite{galitski2013spin,manchon2015new}, are crucial for producing topological phases. Several examples are related to the spin-degenerate nodal lines in the Brillouin zone. In Ref.\ \cite{PhysRevB.109.024404,jungwirth2024}, the authors showed that the nodal lines enabled by spin-orbit coupling give rise to pinch points in altermagnetic metals. Such pinch points behave as single or double type-II Weyl nodes due to a nontrivial Berry curvature around them and thus can be viewed as topological objects. Consequently, a topological transition occurs from altermagnetic to ferromagnetic phases  driven by an external field. 
Another representative example is the altermagnetic Weyl semimetal CrSb \cite{li2024crsb}, where the splitting of the non-relativistic nodal plane by the relativistic spin-orbit coupling leads to the emergence of pairs of Weyl points.
For a two-dimensional AM, the band structure normally has spin-polarized Dirac points but the presence of spin-orbit coupling can gap out the Dirac points, resulting in the formation of the first-order topological insulator \cite{parshukov2024,antonenko2024weyl}. The topological phases in AMs can even be tuned by the strength and directions of an applied external magnetic field combined with spin-orbit coupling, see Ref.\,\cite{Rao2024}. Moreover, when AMs are combined with first-order topological insulators, Refs.\,\cite{PhysRevLett.133.106601,CCLiu2} showed that it is possible to induce high-order topological insulator phases. Apart from these studies, it is important to remark that among the most interesting phenomena associated with the topological nature of AMs are the anomalous Hall effect and its thermal counterpart (known as anomalous Nernst effect) reported in Refs.\,\cite{LiborSAv,PhysRevLett.130.036702,Feng_2022,wang2023emergent,han2024ObsNernst}. The authors showed that these Hall phases occur due to  nontrivial berry curvatures induced by broken $\mathcal{T}\tau$ and $\mathcal{P}\mathcal{T}$ symmetries in AMs; here, $\mathcal{T}$, $\mathcal{P}$ and $\tau$ represent the time-reversal, space inversion, and translation symmetries. The intriguing role of the Berry curvature was recently taken further in Ref.\,\cite{Fang106701},  which showed that quantum geometry, involving the Berry curvature and quantum metric \cite{yu2024Qgeometry}, can induce a nonlinear third-order response in planar AMs.  Quantum geometry has  recently been generalized to    multiband systems \cite{bouhon2023QuantumGeom}, which was already applied to explore normal state systems \cite{jain2024AnonGeoEul} and also superconductors \cite{chau2025OptEul}. Given that AMs are in general multiband [Eq.\,(\ref{eq:dAM_H2})], we anticipate that investigating quantum geometry in AMs will uncover new physics at the intersection of geometry, topology, and altermagnetism.

 In relation to  topological phases based on $p$-wave magnets, less number of works have been reported in the normal state \cite{Ezawa24prb,brekke24} since the field is still in its infancy. Ref.\,\cite{Ezawa24prb} shows that a   hybrid system formed by a $p$-wave magnet and a metal with orbital degree of freedom can realize a topological insulator without   spin-orbit interaction. Furthermore, by considering momentum-dependent sd-coupling in $p$-wave magnets, Ref.\,\cite{brekke24} demonstrated  that it is possible to induce nontrivial topology. We hope that these studies will encourage the community to further expand the topology of $p$-wave magnets.

By now, it is well accepted that unconventional magnets provide promising opportunities for engineering topological phases without relaying on relativistic spin-orbit coupling in contrast to the  arena offered by semiconductors \cite{galitski2013spin,manchon2015new}.  
 Of interest is also that unconventional magnetism can also enhance the often small relativistic spin-orbit coupling, thus allowing the possibility to boost  topological order.  Another important route is to combine the characteristic magnetic order of unconventional magnets with other topological phases, e. g., in the form of heterostructures, a situation that can lead to induced emergent phases with entirely new functionalities. Motivated by these  possibilities, in the next section, we focus on the interplay between unconventional magnetism and superconductivity, with a broader goal to identify novel emergent phenomena.

\section{Superconductivity and unconventional magnetism}
\label{section2}
After discussing the properties of unconventional magnets in the normal state, here we focus on their interplay with superconductivity. As we have already pointed out in the introduction, superconductivity in unconventional magnets has been studied as an intrinsic effect in AMs \cite{PhysRevB.89.165126,PhysRevResearch.5.043171,mazin2022notesaSC,PhysRevB.108.184505,Brekke23,PhysRevB.109.134515,chakraborty2024_1,bose2024altermagn,sim2024,Carvalho24,chakraborty2024_2,hong2024,mukasa2024,chakrabortyd2024} and $p$-wave magnets \cite{sukhachov2025} but also as a proximity-induced effect in hybrid junctions formed by superconductors and AMs \cite{Papaj23,Sun23,IkegayaAltermagnet,zhang2024,PhysRevB.109.L201404,PhysRevB.109.245424,Niu2024,zyuzin2024,chatterjee2025,Bo2025,JXhu2024,chourasia2024,PhysRevB.110.094508,PhysRevB.110.054446,BanerjeePRB24,PhysRevLett.133.106601,Cheng24,Qiang24,sun2024,PhysRevB.109.L201404,fukaya2024x,PhysRevB.110.L140506,PhysRevLett.133.226002,maeda2025classifi,Peihao2025,Mondal24} and  $p$-wave magnets  \cite{maeda2024,sukhachov2025,fukaya2024x,PhysRevB.111.035404,maeda2025classifi,kokkeler2024,Peihao2025}. As we can indeed see, superconductivity in AMs and in hybrid junctions has attracted the majority of the attention. 

To understand how superconductivity has been addressed in the  systems mentioned above, we start by providing a basic introduction to the topic.  Since superconductivity is an ordered state, it is characterized by an order parameter often known as pair potential. This pair potential represents the macroscopic wavefunction of Cooper pairs and its functionalities determine the type of superconducting state. Under generic circumstances, the pair potential is obtained by a mean-field decomposition of an interacting Hamiltonian and taking into account anomalous averages of two annihilation (or creation) operators. Thus, for an homogeneous system in space, the pair potential can be   written as
\begin{equation}
\label{OrderPar}
\Delta_{\sigma\sigma'}(\bm{k})=\frac{1}{\beta N}\sum_{\bm{k}'}\sum_{n}V^{\sigma\sigma'}_{\bm{k},\bm{k}'}F_{\sigma\sigma'}(\bm{k},\bm{k}',\omega_{n})\,,
\end{equation}
where $N$ is the volume of the system, $\beta=1/(\kappa_{B}T)$, with $\kappa_{\rm B}$ being the Boltzmann constant, $T$ the temperature, $(\bm{k},\bm{k}')$ denotes the momenta,  and  $(\sigma,\sigma')$ represents the spins,
while $V_{\bm{k},\bm{k}'}^{\sigma\sigma'}$ is an attractive electron interaction and $F_{\sigma\sigma'}(\bm{k},\bm{k}',\omega_{n})$ is the  the anomalous Green's function defined as
 \begin{equation}
 \label{Fdefin}
 F_{\sigma\sigma'}(\bm{k},\bm{k}',\omega_{n})=-\int_{0}^{\beta}d\tau {e}^{i\omega_{n}\tau}\langle c_{\bm{k}\sigma}(\tau)c_{\bm{k}'\sigma'}(0)\rangle\,.
 \end{equation}
 Here, $\omega_{n}=\pi\kappa_{\rm B}(2n+1)T$ is the Matsubara frequency, with $T$ being the temperature, $c_{\bm{k}\sigma}(\tau)=e^{\tau H_{\rm MF}}c_{k\sigma}{e}^{-\tau H_{\rm MF}}$, with $H_{\rm MF}$ the mean field Hamiltonian describing the system under study. It is worth noting that $F_{\sigma\sigma'}(\bm{k},\bm{k}',\omega_{n})$ represents the thermodynamic expectation value of two annihilation operators, sometimes referred to as Cooper pair wavefunction or simply as \emph{pair amplitude} \cite{RevModPhys.63.239}. Since to find $c_{\bm{k}\sigma}(\tau)$ one needs the mean field Hamiltonian $H_{\rm MF}$,  to determine the pair potential $\Delta_{\sigma\sigma'}(\bm{k})$, Eq.\,(\ref{OrderPar}) needs to be solved within a self-consistent approach. In this case, it is common to write the mean field Hamiltonian in Nambu space where electrons and holes are taken at the same footing, which gives us the so-called Bogolyubov-de Gennes (BdG) Hamiltonian
\cite{zagoskin,TextTanaka2021}
\begin{equation}
\label{BdGH}
H_{\rm BdG}(\bm{k})=
\begin{pmatrix}
\hat{H}({\bm{k}})&\hat{\Delta}(\bm{k})\\
\hat{\Delta}^{\dagger}(\bm{k})&-\hat{H}^{*}(-{\bm{k}})
\end{pmatrix}\,,
\end{equation}
 where $\hat{H}({\bm{k}})$ and $-\hat{H}^{*}(-{\bm{k}})$ are matrices in spin space describing electrons and holes, respectively, while $\hat{\Delta}(\bm{k})$ is a matrix containing all the pair potential elements given by   Eq.\,(\ref{OrderPar}). Thus,   the pair potential, being a function of momentum $\bm{k}$ and spins $(\sigma,\sigma')$, are written as a matrix in spin space
\begin{equation}
\label{matrixPP}
    \hat{\Delta}(\bm{k})=
    \begin{pmatrix}
        \Delta_{\uparrow\uparrow}(\bm{k}) & \Delta_{\uparrow\downarrow}(\bm{k})  \\
        \Delta_{\downarrow\uparrow}(\bm{k}) & \Delta_{\downarrow\downarrow}(\bm{k})
    \end{pmatrix},
\end{equation}%
where $\Delta_{\sigma\sigma'}(\bm{k})$ is given by Eq.\,(\ref{OrderPar}). 
\begin{figure}[t!]
    \centering
    \includegraphics[width=8.5cm]{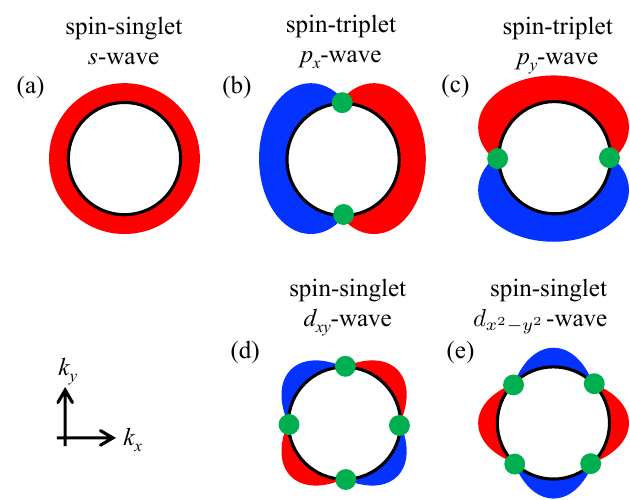}
    \caption{(a-e)  Schematic representation of the pair potentials with  spin-singlet $s$-wave,   spin-triplet $p_{x}$-wave, spin-triplet $p_{y}$-wave,     spin-singlet $d_{xy}$-wave, and  spin-singlet $d_{x^{2}-y^{2}}$-wave symmetries. The red and blue colors denote the sign of the pair potentials, while the green circle indicates their nodes.}
    \label{PPs}
\end{figure}%

To further understand the type of pair potential, and hence of superconducting state, the symmetries of $\Delta_{\uparrow\uparrow}(\bm{k})$ need to be inspected. For this purpose, we note that the pair potential is composed of an  orbital part and a spin part, with the former determined by the momentum dependence. Since $\Delta_{\sigma\sigma'}(\bm{k})$ describes the Cooper pair wavefunction, the orbital and spin dependences are linked by the antisymmetry condition imposed by the fermionic nature of the paired electrons. In a two-electron system, the possible spin states are spin singlet and three spin triplet states, which are \emph{odd} and \emph{even} functions under the exchange of spins, respectively. Thus, a spin singlet pair potential must be even in momentum (also known as parity) which leads to $\hat{\Delta}^{s}(\bm{k})=i\sigma_{y}d_{0}(\bm{k})$: here $d_{0}(\bm{k})=[\Delta_{\uparrow\downarrow}(\bm{k})-\Delta_{\downarrow\uparrow}(\bm{k})]/2$, $d_{0}(\bm{k})=d_{0}(-\bm{k})$,   $\Delta_{\uparrow\downarrow}(\bm{k})=-\Delta_{\downarrow\uparrow}(\bm{k})$, $\Delta_{\sigma\sigma}(\bm{k})=0$. In Fig.\,\ref{PPs}(a,d,e), we schematically show two examples of spin-singlet even-parity pair potentials, which include   isotropic $s$-wave  and anisotropic $d$-wave symmetries; these two cases are examples of conventional and unconventional superconducting states respectively, and can be described by the following pair potentials: $d_{0}^{s}(\bm{k})=\Delta$ and $d_{0}^{d}(\bm{k})=k_{x}k_{y}\Delta/k_{\rm F}^{2}$ or  $d_{0}^{d}(\bm{k})=\Delta(k_{x}^{2}-k_{y}^{2})/k_{\rm F}^{2}$, with $k_{\rm F}$ the Fermi wavevector and the $s(d)$ superscript  symbol denoting the $s(d)$-wave parity, see also  Tab.\,\ref{tab:PPs}. 
Similarly, a spin triplet pair potential must be an \emph{odd} function of momentum and can be written as $\hat{\Delta}^{t}(\bm{k})=[\bm{d}(\bm{k}){\cdot} \bm{\sigma}](i\sigma_{y})$, where $\bm{d}(\bm{k})=(d_{x}(\bm{k}),d_{y}(\bm{k}),d_{z}(\bm{k}))$ with
$\bm{d}(\bm{k})=-\bm{d}(-\bm{k})$,  $d_{x}(\bm{k})=[\Delta_{\downarrow\downarrow}(\bm{k})-\Delta_{\uparrow\uparrow}(\bm{k})]/2$,  $d_{y}(\bm{k})=[\Delta_{\downarrow\downarrow}(\bm{k})+\Delta_{\uparrow\uparrow}(\bm{k})]/(2i)$, and  $d_{z}(\bm{k})=[\Delta_{\uparrow\downarrow}(\bm{k})+\Delta_{\downarrow\uparrow}(\bm{k})]/2$. In Fig.\,\ref{PPs}(b,c), we schematically show an example of spin-triplet odd-parity pair potential with a $p$-wave symmetry; this is also an example of unconventional superconductivity and with a pair potential $\bm{d}(\bm{k})=\Delta(k_{x}+i k_{y})\hat{\bm{z}}/k_{\rm F}$, see also Tab.\,\ref{tab:PPs}. It is well-understood in the community that, while the isotropic spin-singlet $s$-wave pair potential describes conventional superconductors, the anisotropic spin-singlet $d$-wave and spin-triplet $p$-wave pair potentials describe unconventional superconductors \cite{RevModPhys.63.239}. 

\begin{table}\label{tab:PPs}
  \centering
  \begin{tabular}{|c|c|}
    \hline
{\bf   Type of superconductor}   & {\bf Pair potential} \\
    \hline\hline
    spin-singlet $s$-wave & $d^{s}_0(\bm{k})=\Delta$ \\
    spin-singlet $d$-wave & $d^{d}_0(\bm{k})=\Delta[(k_x^2-k_y^2)\cos2\beta$ \\
    &\quad $+2{k_xk_y}\sin2\beta]/k_\mathrm{F}^2$\\
    spin-triplet chiral $p$-wave & $\bm{d}(\bm{k})=\Delta(k_x+ik_y)\hat{\bm{z}}/k_\mathrm{F}$ \\
    spin-triplet helical $p$-wave & $\bm{d}(\bm{k})=i\Delta(k_x\hat{\bm{y}}+k_y\hat{\bm{x}})/k_\mathrm{F}$ \\
    \hline
  \end{tabular}
    \caption{Expressions of the $d$ vectors for distinct types of  superconductors. Here, $d_{0}(\bm{k})$  describes   the spin-singlet $s$- and $d$-wave superconductors,  while $\bm{d}(\bm{k})$ describes the spin-triplet chiral and helical $p$-wave superconductors. The spin-triplet chiral (helical)  $p$-wave pair potential corresponds to the mixed spin-triplet (equal spin-triplet) superconductor \cite{bernevig2013topological}. Here,   $\Delta$ is a constant quantity representing the amplitude of the pair potential, while    $\beta$ depicts the angle between the $x$-axis and the lobe of the $d$-wave superconductor \cite{TK95,TK96,tanaka971,kashiwaya2000,Lofwander2001}, see  Fig.~\ref{PPs}.  
  Moreover, $k_{\rm F}$ is the Fermi wavevector and $(\hat{\bm{x}},\hat{\bm{y}},\hat{\bm{z}})$ is a unit vector in three dimensions.}
\end{table}

Taking into account the previous discussion, the pair potential given by Eq.\,(\ref{matrixPP}) can be parametrized as 
\begin{equation}
\label{matrixPP2}
 \hat{\Delta}(\bm{k})=[d_{0}(\bm{k})+\bm{d}(\bm{k}){\cdot}\bm{\sigma}](i\sigma_{y})
\end{equation}
 An interesting point about the   matrix representation of the pair potential is that it becomes simpler to identify which ones are unitary and nonunitary, given that a unitary matrix $R$ requires that the product with its Hermitian conjugate is proportional to the unit matrix, namely, $R R^{\dagger}\propto \sigma_{0}$. In this regard, we find that the singlet pair potential is unitary since $\hat{\Delta}^{s}(\bm{k})[\hat{\Delta}^{s}(\bm{k})]^{\dagger}=|d_{0}(\bm{k})|^{2} \sigma_{0}$, while the triplet pair potential can develop both unitary and  nonunitary components   as, 
 \begin{equation}
 \label{DinP}
\hat{\Delta}^{t}(\bm{k})[\hat{\Delta}^{t}(\bm{k})]^{\dagger}=|\bm{d}(\bm{k})|^{2}\sigma_{0}+\bm{P}{\cdot}\bm{\sigma}\,,
 \end{equation}
where $\bm{P}=i\bm{d}{\times}\bm{d}^{*}$ characterizes the spin polarization of Cooper pairs and its finite value reflects time-reversal symmetry breaking \cite{RevModPhys.63.239}. Following the discussion presented in this part, it is possible to explore superconductivity, which can emerge intrinsically due to e. g., electron interactions as in Eq.\,(\ref{OrderPar}) and, depending on the type of pairing channel, spin-singlet and spin-triplet pair potential can be studied. It is worth noting that superconducting states can be also induced by means of the so-called proximity effect when placing   superconductors in contact with normal state materials, thereby enabling the possibility for superconducting states with properties coming from the normal materials. In what follows we will discuss intrinsic superconductivity in unconventional magnets and later will focus on hybrid systems formed by superconductors and unconventional magnets.

\subsection{Intrinsic superconductivity in unconventional magnets}
Most of the studies that addressed intrinsic superconductivity in  unconventional magnets focused on  the coexistence of superconductivity and altermagnetism in a single material as well as the possible generation of superconductivity by spin fluctuations in AMs. One of the initial works investigated the formation of superconductivity in a $d_{x^{2}-y^{2}}$-wave AM under the presence of Rashba spin-orbit coupling (SOC) \cite{PhysRevB.108.184505} since it naturally appears during the growing process of AMs on substrates \cite{bychkov1984properties}. In particular, the authors consider an extended attractive Hubbard interaction allowing for both spin-singlet $s$- and equal spin-triplet $p$-wave pairing channels, where their respective pair potentials are found  within a self-consistent approach by an expression similar to Eq.\,(\ref{OrderPar}) in the $d_{x^{2}-y^{2}}$-wave AM modelled by the third expression of Eqs.\,(\ref{MTB}) under Rashba SOC. The system was then shown to  host a mixture of spin-singlet $s$-wave and spin-triplet $p$-wave superconductivity due to the simultaneous effect of breaking time-reversal symmetry by altermagnetism and inversion symmetry by SOC \cite{PhysRevB.108.184505}. Another key aspect of this study is that it reports   dominant $p$-wave superconductivity when the combined action of time-reversal   and four-fold rotational around the $z$ axis symmetries is preserved \cite{PhysRevB.108.184505}; this symmetry combination is also preserved in AMs and uncovers the intriguing   role of $d_{x^{2}-y^{2}}$-wave altermagnetism on the  type of emergent superconductivity. In an almost parallel study, superconductivity and altermagnetism  was addressed in a system with itinerant electrons and magnetic sites but without SOC \cite{Brekke23},  where the onsite exchange interaction between itinerant and localized spins induces a spin splitting, expected in AMs, and is also responsible for an electron-magnon coupling due to fluctuations in the localized spins. Interestingly, Ref.\,\cite{Brekke23} found that, while nonmagnetic sites are   essential  for $d$-wave altermagnetism, a symmetry also exhibited by the splitting of magnon bands, the   electron-magnon coupling  gives rise to electron-electron interactions (as  the  one leading to   Eq.\,(\ref{OrderPar})) but mediated by two magnons;   this then produces a dominant equal spin-triplet $p$-wave  superconductivity with sizeable spin polarization determined by $\bm{P}$ in Eq.\,(\ref{DinP}), perhaps already anticipated in Ref.\,\cite{mazin2022notesaSC}. This electron-magnon induced superconductivity was then also confirmed under many-body effects \cite{PhysRevB.109.134515}, carried out within the strong-coupling approach of the Eliashberg theory \cite{eliashberg1960interactions,eliashberg1960temperature}.  Within the weak coupling limit,  the spin-triplet odd-momentum superconducting state can be dominant by the electron-phonon coupling~\cite{leraand2025}.

In a subsequent study, the role of magnetic exchange interactions in AMs was further   explored  as a mechanism to drive superconductivity \cite{bose2024altermagn}, where the chosen system involved a multiorbital $t{-}J$ model on a square-octagon lattice. By allowing all (charge, spin, and pairing) mean-field channels \cite{bose2024altermagn},  distinct  phases appear, such as the coexistence of altermagnetism and superconductivity having  a mixture of spin-singlet $d$-wave and mixed spin-triplet $s$-wave as well as a  $d$-type pair-density wave due to the Fermi surface nesting, thus reflecting the intriguing role of altermagnetism for realizing nontrivial states. Moreover, in a study earlier than Refs.\,\cite{PhysRevB.109.134515,bose2024altermagn}, it was shown that finite momentum superconductivity can naturally appear in $d_{x^{2}-y^{2}}$-wave AMs with spin-singlet $d_{{x}^{2}-y^{2}}$-wave superconductivity even in the absence of net magnetization, as long as their nodes coincide; here,  intrinsic superconductivity is due to an effective nearest-neighbor attraction for spin-singlet  pairing, where the pair potential is found from an equation similar to Eq.\,(\ref{OrderPar}) but with the momentum decomposition allowing for a finite center-of-mass momentum of Cooper pairs; see also Refs.\,\cite{sim2024,hong2024,chakrabortyd2024,mukasa2024,chakraborty2024_2} for other theoretical studies supporting the appearance of finite momentum superconductivity in AMs. Another studied aspect of altermagnetism is its potential for modifying the parity symmetry, which was shown in Ref.\,\cite{Carvalho24} with the appearance of spin-triplet $f$-wave superconductivity in $d$-wave AMs with spin-singlet and spin-triplet pair potentials. Besides AMs, superconductivity was very recently   studied in $p$-wave magnets taking into account a pair potential from the simplest electron-phonon mediated interaction similar to Eq.\,(\ref{OrderPar}), which then was shown to produce spin-singlet $s$-wave and equal spin-triplet  $p$-wave superconductivity \cite{sukhachov2025}.  Moreover, Ref.\,\cite{PhysRevB.111.054520} recently reported a systematic approach to construct superconducting order parameters within the framework of spin space groups, which describe the symmetry of  unconventional magnets. Taking altogether, the theoretical advances on intrinsic superconductivity in  unconventional magnetism uncovered the potential of  unconventional magnets for hosting novel superconducting phases.

 \begin{table*}
\centering
 \begin{tabular}{||c||c| c| c| c||} 
 \hline
{\bf  Magnet\textbackslash superconductor} & {\bf singlet $s$-wave} & {\bf singlet $d$-wave} & {\bf triplet chiral $p$-wave}&{\bf triplet helical $p$-wave} \\ [0.5ex] 
 \hline\hline
 {\bf $d$-wave altermagnet}&
 \begin{tabular}{c}
$s$-wave ESE \\ 
$d$-wave OTE($\parallel\hat{\bm{z}}$) 
\end{tabular} &
 \begin{tabular}{c}
$d$-wave ESE \\ 
$g$-wave OTE($\parallel\hat{\bm{z}}$) 
\end{tabular}  & 
 \begin{tabular}{c}
$p$-wave ETO($\parallel\bm{d}$)  \\ 
$f$-wave OSO 
\end{tabular}  & 
 \begin{tabular}{c}
$p$-wave ETO($\parallel\bm{d}$) \\ 
$f$-wave ETO($\parallel\hat{\bm{z}}\times\bm{d}$) 
\end{tabular}  \\ 
 \hline
  {\bf $p$-wave magnet}& 
  \begin{tabular}{c}
$s$-wave ESE \\ 
$p$-wave ETO($\parallel\hat{\bm{z}}$)  
\end{tabular}  & 
 \begin{tabular}{c}
$d$-wave ESE \\ 
$f$-wave ETO($\parallel\hat{\bm{z}}$)  
\end{tabular}  &  
\begin{tabular}{c}
$p$-wave ETO($\parallel\bm{d}$) \\ 
$d$-wave ESE 
\end{tabular}  & 
 \begin{tabular}{c}
$p$-wave ETO($\parallel\bm{d}$) \\ 
$d$-wave OTE($\parallel\hat{\bm{z}}\times\bm{d}$) 
\end{tabular}  \\ [1ex] 
 \hline
\end{tabular}
 \caption{
List of   pair symmetries due to the combination of  superconductivity and unconventional magnetism, characterized by altermagnets and  $p$-wave magnets as two representative examples of even- and odd-parity unconventional magnets. The  top row and leftmost column show the type of superconductor and  unconventional magnet, respectively. The symmetries of the induced pair amplitudes are shown  starting from the second column and second row. As an example, the cell  at the intersection between   the second column and second row indicates that combining spin-singlet $s$-wave superconductivity with $d$-wave altermagnetism gives rise to  two pair symmetry classes:  $s$-wave ESE and $d$-wave OTE, with a $d$-vector parallel to the $z$-axis.  Here, ESE (OSO) stands for even-frequency (odd-frequency) spin-singlet   even-parity (odd-parity), while OTE (ETO) for odd-frequency (even-frequency) spin-triplet even-parity (odd-parity). We note that the symbols ``$s, p, d, f, g,\dots$-wave''   represent, respectively, the degree $0,1,2,3, 4,\dots$ of the homogeneous polynomials of momenta given by the pair amplitudes. Taken from Ref.\,\cite{maeda2025classifi}.
    }
\label{tab:pair_amp_symmetry}
 \end{table*}
 
\subsection{Possible superconducting pair correlations due to unconventional magnetism}
\label{subsection32}
In order to further uncover the role of unconventional magnetism for allowing novel superconducting states, and given that the pair potential is intimately   related to the pair amplitude $F_{\sigma\sigma'}(\bm{k},\bm{k}',\omega_{n})$ via Eq.\,(\ref{OrderPar}), we now analyze the symmetries of $F_{\sigma\sigma'}(\bm{k},\bm{k}',\omega_{n})$. With this, we can identify  the emergent superconducting correlations in unconventional magnets with superconductivity. To carry out this task, it is necessary to remind that the pair amplitude represents a two-electron wavefunction and, therefore, fulfills the antisymmetry condition imposed by the fermionic nature of paired electrons under the total exchange of involved quantum numbers as
\begin{equation}
\label{ASC}
F_{\sigma\sigma'}(\bm{k},\bm{k}',\omega_{n})=-F_{\sigma'\sigma}(\bm{k}',\bm{k},\omega_{n})\,.
\end{equation}
This condition here involves momenta, spins, and Matsubara frequencies but, in general, the pair amplitude can depend on other quantum numbers such as bands, valley, layer, etc., see Refs.\,\cite{BergeretReview,tanaka2011odd_review,Cayao2020odd,RevModPhys.91.045005,triola2020role}. Interestingly, the antisymmetry condition given by Eq.\,(\ref{ASC}) implies that the  parity of the momentum part, the spin configuration, and the frequency dependence are linked. As such, under the individual exchange of such quantum numbers, the pair amplitude can be either \emph{even} (E) or \emph{odd} (O), provided the total exchange of quantum numbers makes the pair amplitude antisymmetric. At this point, it is also useful to decompose the spin symmetry and write the pair amplitudes as a $2\times2$ matrix in spin space as for the pair potential in Eq.\,(\ref{matrixPP}) and Eq.\,(\ref{matrixPP2}),
\begin{equation}
 \hat{F}(\bm{k},\omega_{n})=[F_{s}(\bm{k},\omega_{n})+\bm{F}_{t}(\bm{k},\omega_{n}){\cdot}\bm{\sigma}](i\sigma_{y})\,,
\end{equation}
where $F_{s}(\bm{k},\omega_{n})$ is the spin-singlet pair amplitude, while  $\bm{F}_{t}(\bm{k},\omega_{n})$ is the vector of   spin-triplet  pair components $\bm{F}_{t}(\bm{k},\omega_{n})=(F_{x}(\bm{k},\omega_{n}), F_{y}(\bm{k},\omega_{n}), F_{z}(\bm{k},\omega_{n}))$. The singlet and  triplet parts are given by: $F_{s}(\bm{k},\omega_{n})=[F_{\uparrow\downarrow}(\bm{k},\omega_{n})-F_{\downarrow\uparrow}(\bm{k},\omega_{n})]/2$,  $F_{x}(\bm{k},\omega_{n})=[-F_{\uparrow\uparrow}(\bm{k},\omega_{n})+F_{\downarrow\downarrow}(\bm{k},\omega_{n})]/2$, $F_{y}(\bm{k},\omega_{n})=[F_{\uparrow\uparrow}(\bm{k},\omega_{n})+F_{\downarrow\downarrow}(\bm{k},\omega_{n})]/(2i)$, $F_{z}(\bm{k},\omega_{n})=[F_{\uparrow\downarrow}(\bm{k},\omega_{n})+F_{\downarrow\uparrow}(\bm{k},\omega_{n})]/2$; details on the decomposition of pair symmetries noted here can be found in Refs.\,\cite{BergeretReview,Eschrig2007,tanaka2011odd_review,Cayao2020odd}. Moreover, it can be also defined a  spin polarization of Cooper pairs associated to the pair amplitudes as $\bar{\bm{P}}=i\bm{F}_{t}{\times}\bm{F}_{t}^{*}$, which is akin to $\bm{P}$ in Eq.\,(\ref{DinP}). The expressions for the pair amplitudes given here are similar to those for the singlet and triplet parts of the pair potential given in the paragraph below Eq.\,(\ref{matrixPP2}). 
 
Now, going back to identifying the symmetry of the pair amplitudes, we realize that, due to the antisymmetry condition, the spin-singlet pair amplitude can have either even-frequency even-parity  symmetry or  odd-frequency odd-parity symmetry. By a similar analysis, we find that the spin-triplet pair amplitudes can exhibit either even-frequency odd-parity or odd-frequency even-parity. As a result, taking into account frequencies, spins, and parity, 
 there exist four pair symmetry classes that respect the antisymmetry condition: i) even-frequency spin-singlet even-parity (ESE),  ii) even-frequency spin-triplet odd-parity (ETO), iii) odd-frequency spin-singlet odd-parity (OSO), and iv) odd-frequency spin-triplet even-parity (OTE); see more details about the OTE class in  Refs.\,\cite{Berezinskii74,Kirkpatrick91,Belitz99,odd1,odd3b,Eschrig2007,tanaka2011odd_review,Asano2013,PhysRevB.90.220501,Fominov,PhysRevB.92.205424,PhysRevB.95.174516,PhysRevB.96.155426,tanaka2018surface,PhysRevB.97.134523,PhysRevB.98.161408,Tamura2019PRB,Spectralbulk,PhysRevB.100.115433,PhysRevB.101.214507,Takagi2020,Cayao2020odd,Takagi2022,PhysRevB.106.L100502,PhysRevResearch.2.022019,PhysRevB.110.125408,ahmed2024oddfreABS}. It is worth noting that, while there exists a direct relation between pair amplitudes and pair potential via Eq.\,(\ref{PPs}),  the pair potential in Eq.\,(\ref{PPs}) is not frequency dependent, and thus symmetric, but the pair amplitude given by Eq.\,(\ref{Fdefin}) depends on frequency. A generalization of the pair potentials discussed here, such that they become frequency dependent, can be done within Eliashberg theory \cite{eliashberg1960interactions,eliashberg1960temperature,Vojta,Shigeta1,Shigeta2,Shigeta3,Fuseya,Kusunose}, see also
 Refs.\,\cite{PhysRevB.92.054516,PhysRevB.104.174518}. We will however concentrate on the four allowed pair symmetries, whose emergence, in spite of being allowed, depends on the unconventional magnetic system under investigation \cite{maeda2025classifi}, with a particular relation to the breaking of the underlying symmetries of the system \cite{RevModPhys.63.239}  (time translation, spatial translation, spin-rotation, inversion, etc.).   Even though the resulting four pair symmetry classes have already been studied before  in systems with relativistic SOC \cite{tanaka2011odd_review,Asano2013,PhysRevB.96.155426,PhysRevB.98.075425,Tamura2019PRB,Spectralbulk,Cayao2020odd,ahmed2024oddfreABS}, the full classification in even- and odd-parity  unconventional magnets with superconductivity was only recently reported \cite{maeda2025classifi}, see also Refs.\,\cite{PhysRevB.109.L201404,chakraborty2024_2,fukaya2024x,sukhachov2025}.

In order to unveil the role of unconventional magnetism for inducing novel superconducting correlations, let us first consider a $d$-wave AM with  spin-singlet superconductivity. This is  modelled by a BdG Hamiltonian given by Eq.\,(\ref{BdGH}) but with  $\hat{H}({\bm{k}})=\xi_{k}\sigma_{0}+M^{d}_{\bm{k}}\sigma_{z}$ characterizing the $d$-wave AM, with the exchange field  $M^{d}_{\bm{k}}$ given by Eq.\,(\ref{eq:am})  and  $\xi_{\bm{k}}$ the kinetic term, while
 $\hat{\Delta}(\bm{k})=i\sigma_{y}d_{0}(\bm{k})$ describes the spin-singlet pair potential discussed in the previous subsection. Then, by using Eq.\,(\ref{Fdefin}) for the anomalous Green's function, which is the same as obtaining the off-diagonal component of the Nambu Green's function in Matsubara representation $\hat{G}(\bm{k},i\omega_{n})=(i\omega_{n}-H_{\rm BdG}(\bm{k}))^{-1}$, the spin-singlet and spin-triplet pair amplitudes can be obtained as \cite{maeda2025classifi}
\begin{equation}
\label{Fstsingletdwave}
\begin{split}
  F_{s}(\bm{k},i\omega_n)&=
  -\frac{d_0(\bm{k})Q_{\bm{k}}(\omega_{n})}
  {[Q_{\bm{k}}(\omega_{n})]^2+4[M_{\bm{k}}^{d}]^2\omega_n^2}\,,\\
  \bm{F}_{t}(\bm{k},i\omega_n)&=
  \frac{2i\omega_nd_0(\bm{k}) M^{d}_{\bm{k}}\hat{\bm{z}}}
  {[Q_{\bm{k}}(\omega_{n})]^2+4[M_{\bm{k}}^{d}]^2\omega_n^2}\,,
  \end{split}
\end{equation}
where $Q_{\bm{k}}(\omega_{n})=\omega_n^2+\xi_{\bm{k}}^2 - [M^{d}_{\bm{k}}]^2+|d_0(\bm{k})|^2$.   From Eqs.\,(\ref{Fstsingletdwave}), we can identify the symmetries of superconducting correlations that emerge when combining $d$-wave altermagnetism with spin-singlet superconductors. The first property to note is that spin-singlet and spin-triplet pair amplitudes emerge, with the triplet parallel to $\hat{\bm{z}}$, which reflects the role of $d$-wave altermagnetism as a mechanism to convert spin-singlet into spin-triplet. The spin-singlet pair amplitude, first  expression in Eqs.\,(\ref{Fstsingletdwave}),  is even in both frequency and momentum given that $Q_{\bm{k}}(\omega_{n})=Q_{-\bm{k}}(\omega_{n})$, $Q_{\bm{k}}(\omega_{n})=Q_{\bm{k}}(-\omega_{n})$, and $d_{0}(\bm{k})=d_{0}(-\bm{k})$. As a result, $F_{s}(\bm{k},i\omega_n)$ has a pair symmetry that corresponds to the ESE class discussed at the beginning of this subsection. Depending on whether $d_{0}(\bm{k})$ describes a spin-singlet $s$-wave or spin-singlet  $d$-wave pair potential,  the spin-singlet pair amplitude $F_{s}(\bm{k},i\omega_n)$ would acquire the $s$- or $d$-wave parity of the parent superconductor. 
 \begin{figure}
    \centering
    \includegraphics[width=0.95\linewidth]{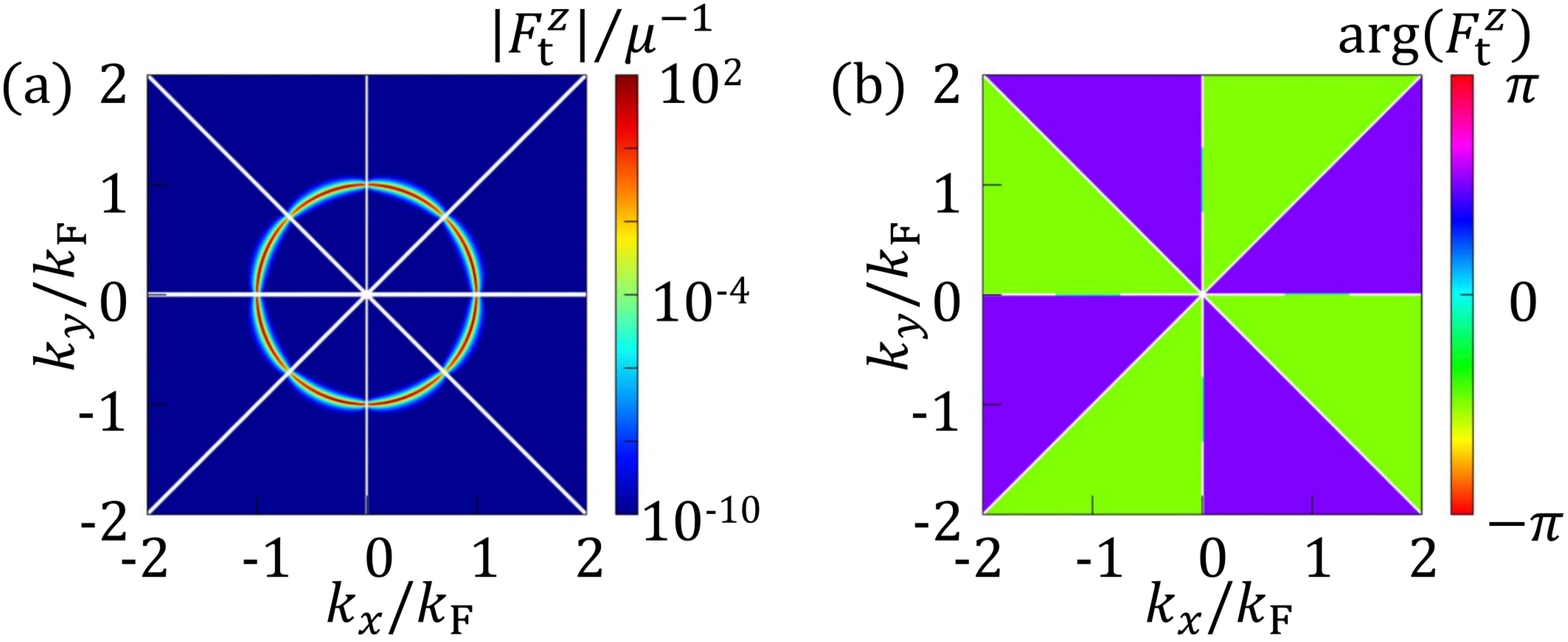}
    \caption{Emergent pair amplitudes  in a  $d_{x^{2}-y^{2}}$-wave   altermagnet with spin-singlet $d_{xy}$-wave superconductivity.  (a) Absolute value of the  odd-frequency spin-triplet even-parity (OTE)   pair amplitude given by Eq.\,(\ref{Fstsingletdwave})  as a function of momenta.  The diagonal  lines indicate the nodes of the pair amplitude. (b) Argument of the OTE pair amplitude as a function of  momenta, where  the purple and green colors indicate $\pi/2$ and $-\pi/2$, respectively.   Parameters: $\Delta=0.01\mu$, $\omega_n=0.5\Delta$ and $J=0.3\Delta$. Adapted from Ref.\,\cite{maeda2025classifi}.}
    \label{FigFdxySCdx2y2AM}
\end{figure}

  \begin{figure*}
    \centering
    \includegraphics[width=0.98\linewidth]{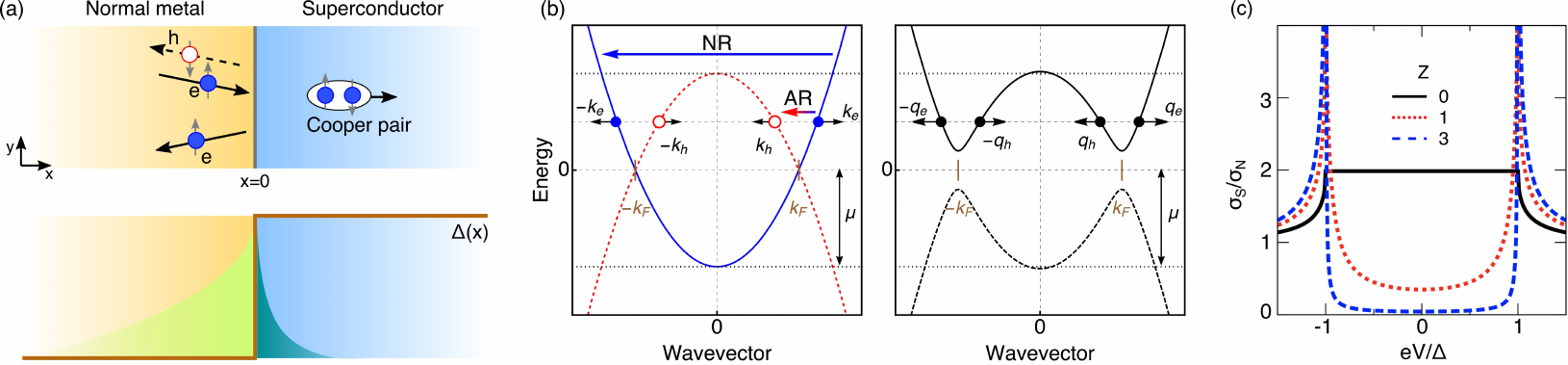}
    \caption{(a) Top: Andreev reflection   in a normal metal-superconductor (NS) junction formed by a semi-infinite normal metal (N, yellow) and a semi-infinite conventional superconductor (S, blue). An incident electron (filled blue circle) at the NS interface  located at $x=0$ (gray vertical line) can be reflected back either as an electron or as a hole (empty red circle).  The reflection of an electron   into an electron   is known as normal reflection (NR). The reflection of an electron into a hole is known as Andreev reflection  (AR), a process that also allows the transfer of a Cooper pair (white ellipse with two filled circles) into the superconductor. The Andreev reflection discussed here implies that the reflection angle is the same  as the angle of incidence (retroreflection).   Bottom: The Andreev reflection carries superconducting correlations from the superconductor with pair potential $\Delta(x)$ (brown curve) into the normal metal, known as the proximity effect (light green region), and also modifies the superconducting correlations in the superconductor, known as the inverse proximity effect (blue-green region).  (b) Energy versus momentum in the normal metal (left) and superconductor (right), indicating the right moving electrons and holes in the former as well as quasielectrons and quasiholes in the latter. (c) Tunneling conductance in the NS junction of (a) as a function of energy for distinct values of the interface barrier strength $Z$.}
    \label{NSjunction_1}
\end{figure*}

The situation becomes more interesting for the spin-triplet pair amplitude $\bm{F}_{t}(\bm{k},i\omega_n)$ given by the second expression in Eqs.\,(\ref{Fstsingletdwave}). In fact, while such a  pair component has a mixed spin-triplet symmetry, it develops and odd frequency dependence ($\sim \omega_{n}$ in the numerator) and a parity symmetry that results from the interplay between superconductivity (via $d_{0}(\bm{k})$) and $d$-wave altermagnetism (via $M_{\bm{k}}^{d}$). This implies that $\bm{F}_{t}(\bm{k},i\omega_n)$ has an odd-frequency spin-triplet even-parity and belongs to the OTE class. For a spin-singlet $s$-wave superconductor ($d_{0}(\bm{k})=\Delta$), the even-parity of $\bm{F}_{t}(\bm{k},i\omega_n)$ corresponds to $d$-wave symmetry entirely  determined by $d$-wave altermagnetism via $M_{\bm{k}}^{d}$; hence the parity of the AM is transferred to the emergent pair amplitude. For a spin-singlet $d$-wave superconductor, with a $d$ vector as in Tab.\,\ref{tab:PPs}, the even-parity of $\bm{F}_{t}(\bm{k},i\omega_n)$ has a $g$-wave symmetry due to the combined action of the $d$-wave AM and $d$-wave superconductor; the amplitude and argument of the OTE pair amplitude are shown in Fig.\,\ref{FigFdxySCdx2y2AM}, with a clear node structure that reflects its parity. The ESE and OTE pair symmetry classes due to $d$-wave altermagnetism and spin-singlet superconductivity are indicated in Tab.\,\ref{tab:pair_amp_symmetry}.  Ref.\,\cite{maeda2025classifi} also addressed the role of AMs with spin-triplet superconductivity as well as other $p$-wave magnets with distinct types of superconductivity, which allowed to establish unconventional magnetism as a mechanism that not only mixes spins, thereby leading to a spin-singlet to spin-triplet conversion, but also transfers the parity symmetry to the emergent superconducting state, see Tab.\,\ref{tab:pair_amp_symmetry}. Thus, it is possible to induce superconducting states with a higher parity \cite{maeda2025classifi}, such as the $f$-wave pairing state in AMs with spin-triplet $p$-wave superconductivity as shown in Tab.\,\ref{tab:pair_amp_symmetry}, already predicted before   in Ref.\,\cite{Carvalho24} in the context of intrinsic superconductivity. Due to the direct relation between pair amplitude and pair potential, the classification carried out in Ref.\,\cite{maeda2025classifi} is expected to also hold for the pair potential under appropriate interactions, being thus useful to understand for appearance of intrinsic superconductivity in unconventional magnets. We can therefore conclude that unconventional magnetism is a promising ground  for novel superconducting states.

\subsection{Superconducting junctions formed by unconventional magnets and superconductors}
Having discussed the effect of unconventional magnetism on the   superconducting  pair symmetries, in this part we address the emergent phenomena when  unconventional magnets are placed in contact with a superconductor. In particular, we focus on NS hybrid junctions where N  represents a normal state material, which can be any normal metal or unconventional magnet, while S denotes a superconductor, as schematically shown in Fig.\,\ref{NSjunction_1}(a). NS junctions are the simplest superconducting junctions that permit  us to analyze fundamental superconducting phenomena due to the combination of unconventional magnetism, superconductivity, and the breaking of spatial translation at the junction's interface.

\subsubsection{Modelling NS junctions}
Before going further, it is important to remark that, when dealing with systems that are inhomogeneous in space,  the pair potential is written in terms of two spatial coordinates; thus, the  expression of the pair potential in spin space corresponding to Eq.\,(\ref{matrixPP}) is given by,
\begin{equation}
\hat{\Delta}(\bm{r},\bm{r}')=[d_{0}(\bm{r},\bm{r}')+\bm{d}(\bm{r},\bm{r}'){\cdot}\bm{\sigma}](i\sigma_{y})\,,
\end{equation}
where $d_{0}(\bm{r},\bm{r}')$ describes the spin-singlet  pair potential and $\bm{d}(\bm{r},\bm{r}')$ the vector of spin-triplet components. The individual components hold properties according to their symmetries, see more on Refs.\,\cite{TextTanaka2021,tanaka2024review}. Since the analysis of NS junctions often involves findings the spectrum $E$ and wavefunction $\Psi$, it is necessary to solve the eigenvalue problem associated to the BdG Hamiltonian, which is similar to the one given by Eq.\,(\ref{BdGH}) but for inhomogeneous systems. The eigenvalue problem in Nambu space is given by
\begin{equation}
\label{BdGEqs}
  \int d\bm{r}'
    \mathcal{H}_{\rm BdG}(\bm{r},\bm{r}')\Psi(\bm{r}')=E\Psi(\bm{r})\,,
\end{equation}
where $\Psi(\bm{r})=(u_{\uparrow}(\bm{r}),u_{\downarrow}(\bm{r}),v_{\uparrow}(\bm{r}),v_{\downarrow}(\bm{r}))^{\rm T}$ is the Nambu vector  with electron- ($u_{\sigma}$) and hole-like ($v_{\sigma}$) parts, while  $\mathcal{H}_{\rm BdG}(\bm{r},\bm{r}')$ is the BdG Hamiltonian written as
\begin{equation}
\mathcal{H}_{\rm BdG}(\bm{r},\bm{r}')=
\begin{pmatrix}
\delta(\bm{r}-\bm{r}')\hat{h}(\bm{r}') &\hat{\Delta}(\bm{r},\bm{r}')\\
\hat{\Delta}^{\dagger}(\bm{r},\bm{r}') &-\delta(\bm{r}-\bm{r}')\hat{h}(\bm{r}') 
\end{pmatrix}\,.
\end{equation}
Here, $\hat{h}(\bm{r})$ is the   Hamiltonian  describing electrons and $\hat{\Delta}(\bm{r},\bm{r})$ the pair potential, both matrices in spin space.  The eigenvalue problem given by Eq.\,(\ref{BdGEqs}) is also known as BdG equations. While $\hat{h}(\bm{r})$ is in general a $2\times2$ matrix in spin space with all its elements finite,  in the absence of   fields that couple spins $\hat{h}(\bm{r})$ is a diagonal matrix in spin space given by  $\hat{h}(\bm{r})=h(\bm{r})+M(\bm{r})\sigma_{z}$, with $h(\bm{r})=-(\hbar^{2}/2m)\nabla^{2}-\mu(\bm{r})+U(\bm{r})$, where $\mu(\bm{r})$ is the chemical potential that measures the filling of the band from the band bottom, $U(\bm{r})$ is a one-body potential characterizing the interface, and $M(\bm{r})$ describes the exchange field of an  unconventional magnet appropriately written from Eqs.\ (\ref{eq:mp}) and (\ref{eq:am}). Thus, when spins are not coupled, the BdG equations can be decoupled, leading to simpler expressions to solve, as discussed e. g., in Refs.\,\cite{TextTanaka2021,tanaka2024review}. This is the case, for instance, of  a spin-singlet $s$-wave pair potential and in the absence of spin mixing fields and $M(\bm{r})=0$, where the BdG equations simply become
\begin{equation}
\label{BdG_swave}
\begin{split}
\begin{pmatrix}
        h(\bm{r}) & \Delta(\bm{r}) \\
        \Delta^{*}(\bm{r}) & -h(\bm{r})
    \end{pmatrix}
 \begin{pmatrix}
        u_{\uparrow}(\bm{r})\\
        v_{\downarrow}(\bm{r}) 
    \end{pmatrix}
    &=E
  \begin{pmatrix}
        u_{\uparrow}(\bm{r})\\
        v_{\downarrow}(\bm{r}) 
    \end{pmatrix}\,,\\
    \begin{pmatrix}
        h(\bm{r}) & -\Delta(\bm{r})  \\
        -\Delta^{*}(\bm{r}) & -h(\bm{r})
    \end{pmatrix}
    \begin{pmatrix}
        u_{\downarrow}(\bm{r})\\
        v_{\uparrow}(\bm{r}) 
    \end{pmatrix}
    &=E
    \begin{pmatrix}
        u_{\downarrow}(\bm{r})\\
        v_{\uparrow}(\bm{r}) 
    \end{pmatrix}\,.
    \end{split}
\end{equation}%
While for conventional superconductors, the BdG equations are simpler,  under general circumstances, one always has to solve the Eq.\,(\ref{BdGEqs}), which is of course a challenging task. We also note that there are clever approximations that can reduce the complexity of the BdG equations, such as the so-called Andreev or semiclassical approximation where the chemical potential is the largest energy scale. The Andreev equations are simpler to solve than  the BdG equations because they involve first derivatives in the single particle Hamiltonian   in contrast to the second derivatives in $\hat{h}(\bm{r})$ of the BdG equations, see more in Refs.\,\cite{TextTanaka2021,tanaka2024review}.

To find the energies and wavefunctions of  a NS junction along $x$ with a spin-singlet $s$-wave pair potential $\Delta(\bm{r})=\theta(x)\Delta$, we can just solve the BdG equations Eqs.\,(\ref{BdG_swave}) in terms of plane waves for a given single particle Hamiltonian $\hat{h}(\bm{r})$. This approach assumes that N and S are semi-infinite systems along $x$ and that there is a pair potential only in the superconducting region S but  no pair potential in the normal region N exists. While this is a simplification of what happens at  interfaces between superconductors and normal state materials, it turns out that such an assumption already provides interesting insights on the emergent superconducting phenomena at such NS interfaces \cite{asano2021andreev}. Upon appropriate modifications of the profile of the pair potential, it is also possible to model finite N and S regions in superconducting junctions, see e. g., Refs.\,\cite{PhysRevB.96.155426,PhysRevB.98.075425,PhysRevB.100.115433,PhysRevB.104.134507,PhysRevB.106.L100502,Tanakadiode1,PhysRevResearch.6.L012062,Tanakadiode2,PhysRevLett.133.226002}. The total solutions would then need to comply with the continuity of wavefunctions at the respective interfaces, such as at $x=0$ in Fig.\,\ref{NSjunction_1}(a). Another possibility to model  NS junctions with finite N and S regions is to carry out a tight-binding discretization of the BdG Hamiltonian within a tunneling Hamiltonian approach \cite{PhysRevB.91.024514,cayao2018andreev,PhysRevB.110.085414,cayao2016hybrid,PhysRevB.107.184519,ahmed2024oddfreABS,tanaka2024review}, thus allowing to explore finite size effects. Whenever necessary, we will refer to the type of approach used to model superconducting junctions, see also Refs.\,\cite{zagoskin,mahan2013many,datta1997electronic,PhysRevB.54.7366}.

\subsubsection{Andreev reflections and Andreev bound states}
To understand the most fundamental effect in superconducting junctions between superconductors and  unconventional magnets, we here address charge transport in a NS junction  
modelled as discussed in the previous subsection. To start with, we first focus on a one-dimensional   NS junction located at $x=0$ when N is   a normal metal and S a conventional spin-singlet $s$-wave superconductor; this will help us unveil the role of unconventional magnetism later.  Then, to uncover the transport processes at the interface of NS junctions, we employ scattering states which are build by using the wavefunctions of the N and S regions \cite{asano2021andreev,TextTanaka2021}. 
Thus, by taking the BdG equations given by Eq.\,(\ref{BdG_swave}) and using plane waves $\Psi(x)=A{\rm e}^{ikx}$, the BdG Hamiltonian  gives the energies in the superconductor $E_{k}=\pm\sqrt{\xi_{k}^{2}+\Delta^{2}}$, where $\xi_{k}=\hbar^{2}k^{2}/2m-\mu$, while $E_{k}=\pm\xi_{k}$ are the normal energy bands in N. The associate wavevectors in N are $\pm k_{e(h)}$ with
\begin{equation}
\label{kehN}
k_{e(h)}=k_{\rm F}\sqrt{1\pm\frac{E}{\mu}}\,,
\end{equation}
while in S are $\pm q_{e(h)}$ with
\begin{equation}
\label{qehN}
q_{e(h)}=k_{\rm F}\sqrt{1\pm\frac{\Omega}{\mu}}\,,
\end{equation}
where $k_{\rm F}=\sqrt{2m\mu/\hbar^{2}}$ is the Fermi wavevector, and 
\begin{equation}
  \Omega=
  \begin{cases}
 \sqrt{E^2-|\Delta|^2}\,, & E\geq|\Delta| \\
        i\sqrt{|\Delta|^2-E^2}\,, & -|\Delta|\leq E \leq |\Delta| \\
        -\sqrt{E^2-|\Delta|^2}\,, & E\leq|\Delta|\,.
\end{cases}
\end{equation}
Thus, the wavefunctions in N are given by 
\begin{equation}
\begin{split}
\Psi_{\pm k_{e}}(x)&=e^{\pm ik_{e}x}
\begin{pmatrix}
1\\
0
\end{pmatrix}\,,\\
\Psi_{\pm k_{h}}(x)&=e^{\pm ik_{h}x}
\begin{pmatrix}
0\\
1
\end{pmatrix}\,,
\end{split}
\end{equation}
while in S by
\begin{equation}
\begin{split}
\Psi_{\pm q_{e}}(x)&=e^{\pm iq_{e}x}
\begin{pmatrix}
u\\
v
\end{pmatrix}\,,\\
\Psi_{\pm q_{h}}(x)&=e^{\pm iq_{h}x}
\begin{pmatrix}
v\\
u
\end{pmatrix}\,,
\end{split}
\end{equation}
where $u=\sqrt{1+\Omega/E}/\sqrt{2}$ and $v=\sqrt{1-\Omega/E}/\sqrt{2}$.
The energy versus momentum in N and S as well as their associated wavevectors are shown in Fig.\,\ref{NSjunction_1}(b).  To identify the scattering processes, we look at the direction of motion of the respective states at fixed wavevectors. The direction of motion of a state at the given wavevector can be obtained by looking at the group velocity
\begin{equation}
\label{groupvel}
v_{g}(k)=\frac{1}{\hbar}\frac{dE(k)}{dk}\,,
\end{equation}
where $E(k)$ is the energy dispersion discussed in the paragraph before Eq.\,(\ref{kehN}), see also  Fig.\,\ref{NSjunction_1}(b). This allows us to identify that electrons in N with $k_{e}$ and $-k_{e}$  move to the right and left directions, respectively. Moreover, holes in N with $k_{h}$ and $-k_{h}$ move to the left and right, respectively. In a very similar way, we can identify the direction of motion for the quasiparticles in S. The direction of motion in N and S is depicted by horizontal black arrows in Fig.\,\ref{NSjunction_1}(b). Having this allowed states in mind, we therefore find that a right moving electron $\Psi_{k_{e}}(x)$ from N     can be reflected back at the NS interface into N either as a left-moving electron with $\Psi_{-k_{e}}(x)$  or as a left-moving hole with $\Psi_{k_{h}}(x)$. The reflection into a particle of the same kind is known as normal reflection (NR), while the reflection of an electron into a hole or vice-versa is  known as Andreev reflection (AR) \cite{andreev1964thermal,andreev1965thermal}, see  Fig.\,\ref{NSjunction_1}(b); for more details on the AR, see Refs.\,\cite{asano2021andreev}. The right moving electron from N can be also transmitted into S as a quasielectron or quasihole, processes  known as normal transmissions and Andreev transmissions, respectively. When scattering occurs at energies within the superconducting gap, the AR involves   the transfer of a Cooper pair into S \cite{asano2021andreev}, sketched in Fig.\,\ref{NSjunction_1}(a). We can thus see that Andreev processes are unique to superconducting interfaces, which is why they represent   perhaps one of the most fundamental superconducting scattering effects, carrying information about the superconductor and the NS interface as well see next. The Andreev reflection is perhaps the most fundamental effect in superconducting junctions and does not have any analog in normal state heterostructures \cite{andreev1964thermal,andreev1965thermal,asano2021andreev}.  We can then quantify ARs and their effects in transport by   constructing scattering states, which involve the discussed processes for a given particle moving towards the NS interface. For instance, for a right-moving   electron towards the interface, we obtain in N and S,
\begin{equation}
\begin{split}
\Psi_{\rm N}(x)&=e^{ik_{e}x}
\begin{pmatrix}
1\\
0
\end{pmatrix}
+
a e^{ik_{h}x}
\begin{pmatrix}
0\\
1
\end{pmatrix}
+
b e^{-ik_{e}x}
\begin{pmatrix}
1\\
0
\end{pmatrix}\,,\\
\Psi_{\rm S}(x)&=ce^{iq_{e}x}
\begin{pmatrix}
u\\
v
\end{pmatrix}
+
d e^{-iq_{h}x}
\begin{pmatrix}
v\\
u
\end{pmatrix}\,.
\end{split}
\end{equation}
The coefficients in $\Psi_{\rm N,S}(x)$ are then determined by the boundary conditions at the interface, which, including a delta potential barrier $U(x)=\lambda\delta(x)$ of barrier strength $\lambda$, we find
\begin{equation}
\label{Wavematching}
\begin{split}
[\Psi_{\rm N}(x)]_{x=0}&=[\Psi_{\rm S}(x)]_{x=0}\,,\\
[\partial_{x}\Psi_{\rm S}(x)]_{x=0}-[\partial_{x}\Psi_{\rm N}(x)]_{x=0}&=\frac{2m\lambda}{\hbar^{2}} [\Psi_{\rm N}(x)]_{x=0}\,.
\end{split}
\end{equation}
Then, within the Andreev approximation where the chemical potential is the largest energy scale, we assume $k_{e(h)}\sim q_{e(h)}\sim k_{\rm F}$. In this regime, the coefficients of $\Psi_{\rm N,S}(x)$ take simple forms, and for the normal and Andreev reflections, one finds $a=\Gamma/[(1+Z^{2})-Z^{2}\Gamma^{2}]$ and $b=[iZ(1-iZ)(\Gamma^{2}-1)]/[(1+Z^{2})-Z^{2}\Gamma^{2}]$, with $Z=m\lambda/(\hbar^{2}k_{\rm F})$, $\Gamma=v/u$. We can then obtain the differential conductance at zero temperature $dI/dV=(2e^{2}/h)\sigma_{\rm S}(E=eV)$,
where \cite{PhysRevB.25.4515}
\begin{equation}
\label{GNS1}
{\displaystyle {\displaystyle \sigma_{\rm S}}=
\begin{cases}
\frac{\displaystyle 2\sigma_{\rm N}^{2}\Delta^{2}}{\displaystyle (\sigma_{\rm N}-2)^{2}\Delta^{2}-4(1-\sigma_{\rm N})E^{2}}\,,|E|<\Delta\,,\\
\frac{\displaystyle 2\sigma_{\rm N}E}{\displaystyle (2-\sigma_{\rm N})\Omega+\sigma_{\rm N}E}\,,|E|>\Delta\,,
\end{cases}}
\end{equation}
where $\sigma_{\rm N}=1/(1+Z^{2})$ is the transmissivity in the normal state and we have used that $\sigma_{\rm S}=(1-|b|^{2}+|a|^{2})$. The role of ARs is thus to increase $\sigma_{\rm S}$, while NRs tend to reduce it \cite{PhysRevB.25.4515,asano2021andreev}. The conductance $\sigma_{\rm S}$ as a function of energy is presented in  
Fig.\,\ref{NSjunction_1}(c) for distinct values of $Z$ which captures the barrier strength. Without any barrier $Z=0$, the junction is transparent and $\sigma_{\rm N}=1$, which then leads to a constant conductance $\sigma_{\rm S}=2$ within the superconducting gap due to perfect AR $|a|=1$ and $|b|=0$, see black curve in Fig.\,\ref{NSjunction_1}(c). The finite constant subgap conductance at $Z=0$ is therefore a signature of Cooper pair transport, which is remarkable given that the AR coefficient is obtained in N; the AR hence carries superconducting information deep into N.  At finite $Z$, normal reflections appear and reduce the subgap conductance, see dotted red curve in Fig.\,\ref{NSjunction_1}(c). At very large Z, the normal transmissivity is very small and leads to very small values of the subgap conductance, as indeed seen in the dashed blue curve in Fig.\,\ref{NSjunction_1}(c). Conductance can thus provide information about the superconducting interface, via ARs, and also about the superconductor. While the physics discussed so far in this part is well-known, it will help us understand how unconventional magnetism affects the Andreev transport in superconducting junctions.

\begin{figure*}
    \centering
    \includegraphics[width=0.98\linewidth]{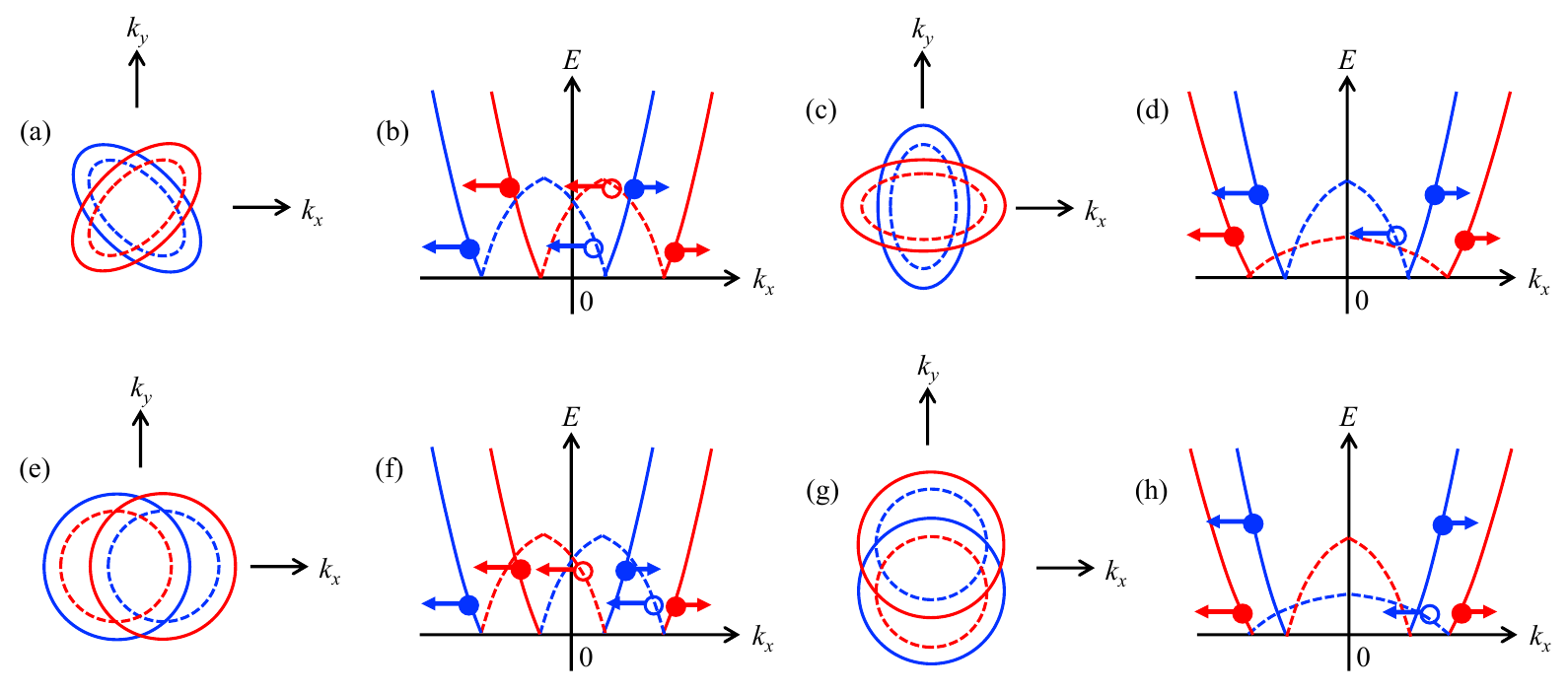}
    \caption{Fermi surfaces and energy dispersions in  unconventional magnets, including $d$-wave  altermagnets (AMs) (a-d) and $p$-wave magnets (e-h).  These figures were inspired by Ref.\,\cite{Papaj23}. More precisely, (a,c,e,g) shows the quasiparticle energy contour at $E>0$ for (a) $d_{xy}$ and (c) $d_{x^2-y^2}$-wave AMs, while (e,g) for $p_x$- and $p_y$-wave magnets. The solid red and blue lines represent the electron energy bands for up and down spins, respectively.
    Dotted red and blue lines stand for the hole energy bands with up and down spins, respectively.   (b,d,f,h) Sketch of scattering states due to right moving electrons with spin up and down at distinct energies, which involve  Andreev and normal reflections for  (e) $d_{x^2-y^2}$ and (d) $d_{xy}$-wave AMs, and (f) $p_x$ and (h) $p_y$-wave magnets.   Filled (empty) red and blue circles denote the electron (hole) with up and down spin at an energy $E$, and arrows represent the direction of the traveling wave of electrons and holes, respectively.
    }
    \label{Energydispersion_UMs}
\end{figure*}

We are now ready to analyse the scattering processes at the NS interface located at $x=0$ with N being an unconventional magnet and S a conventional superconductor. For unconventional magnets, we consider $d$-wave AMs and $p$-wave magnets discussed in Section \ref{section1}, and then follow the same steps as the ones we carried out before for finding the conductance [Eq.\,(\ref{GNS1})]  when N is a just a normal metal. We also assume that our system is infinite along $y$ so that $k_{y}$ is still a good quantum number.  Moreover, we have to distinguish between up and down spins, even though spins are not necessarily coupled since the exchange field in unconventional magnets we consider to be along $z$, see Section \ref{section1}. Hence, the $4\times4$ BdG matrix can be written in terms of two independent $2\times2$ blocks.  Without carrying out any calculation, it is important to have an intuition about the role of unconventional magnetism. For this reason, in Fig.\,\ref{Energydispersion_UMs} we show the Fermi surfaces and energy dispersions at positive energies of each unconventional magnet including $d_{xy}$- and $d_{x^{2}-y^{2}}$-wave AMs as well as $p_{x}$- and $p_{y}$-wave magnets. By inspecting the Fermi surface of $d_{xy}$-wave AMs and $p_{y}$-wave magnets [Fig.\,\ref{Energydispersion_UMs}(a,f)], we notice that opposite spin states have $k_{y}$ and $-k_{y}$, reflecting that ARs as traveling waves are always allowed but depend on the spin splitting of the energy dispersion. In the case of a $d_{x^{2}-y^{2}}$-wave AM and $p_{x}$-wave  unconventional magnet [Fig.\,\ref{Energydispersion_UMs}(b,e)], the states with $k_{y}$ and $-k_{y}$ have the same spins; here, NR is possible as a traveling wave but the AR becomes an evanescent wave. Further insights can be also obtained from Fig.\,\ref{Energydispersion_UMs}, where we show the energy dispersions and also indicate the  possible scattering states when incident electrons with spin up and down towards the interface. Keeping all the discussions in mind, and in order to quantify the contributions from NR and ARs to charge conductance, we now write down the scattering state in N ($x<0$) for an incident electron from N with spin up towards the NS interface,
\begin{equation}
\label{scatte1}
\begin{split}
\Psi_{\rm N}^{\uparrow}(x,k_{y})&=
{\rm e}^{ik_{y}y}
\left[
\Psi_{p^{+}_{e\uparrow}}(x)
+a_{\uparrow}\Psi_{p^{-}_{h\downarrow}}(x)
+b_{\uparrow}\Psi_{p^{-}_{e\uparrow}}(x)\right]\,,\\
\Psi_{\rm N}^{\downarrow}(x,k_{y})&=
{\rm e}^{ik_{y}y}
\left[
\Psi_{p^{+}_{e\downarrow}}(x)
+a_{\downarrow}\Psi_{p^{-}_{h\uparrow}}(x)
+b_{\downarrow}\Psi_{p^{-}_{e\downarrow}}(x)\right]\,,\\
\end{split}
\end{equation}
where $\Psi_{p^{+}_{e\uparrow}}(x)={\rm e}^{ip^{+}_{e{\uparrow}}x}(1,0,0,0)^{\rm T}$, $\Psi_{p^{-}_{h\downarrow}}(x)={\rm e}^{ip^{-}_{h{\downarrow}}x}(0,0,0,1)^{\rm T}$, $\Psi_{p^{-}_{e\uparrow}}(x)={\rm e}^{ip^{-}_{e{\uparrow}}x}(1,0,0,0)^{\rm T}$, $\Psi_{p^{+}_{e\downarrow}}(x)={\rm e}^{ip^{+}_{e{\downarrow}}x}(0,1,0,0)^{\rm T}$, $\Psi_{p^{-}_{h\uparrow}}(x)={\rm e}^{ip^{-}_{e{\uparrow}}x}(0,0,1,0)^{\rm T}$, $\Psi_{p^{-}_{e\downarrow}}(x)={\rm e}^{ip^{-}_{e{\downarrow}}x}(0,1,0,0)^{\rm T}$.
Moreover, the coefficients $a_{\sigma}$ and $b_{\sigma}$ represent the amplitudes of Andreev and normal reflections for an incident electron with spin $\sigma=\uparrow,\downarrow$. Moreover, $p_{e\sigma}^{\pm}$ and $p_{h\sigma}^{\pm}$ are  wavevectors in N and obtained from the energy dispersion for unconventional magnets. In the case of AMs, they are given by \textcolor{red}{\cite{Sun23}}
\begin{widetext}
\begin{equation}
\label{EqsqAM}
\begin{split}
p_{e\sigma}^{\pm}&=-\frac{2\sigma mJ\sin{2\beta}}{\hbar^{2}k^2_\mathrm{F}+2\sigma mJ\cos{2\beta}}k_y\pm\frac{\hbar k^2_\mathrm{F}}{\hbar^2k^2_\mathrm{F}+{2\sigma mJ\cos{2\beta}}}\sqrt{2m\left(1+\frac{2\sigma mJ\cos{2\beta}}{\hbar^{2}k^2_\mathrm{F}}\right)(\mu+E)-\hbar^{2}k^2_y+\frac{4m^{2}J^2}{\hbar^{2}k^4_\mathrm{F}}k^2_y}\,,\\
p_{h\sigma}^{\pm}&=\frac{2\sigma mJ\sin{2\beta}}{\hbar^{2}k^2_\mathrm{F}+2\sigma mJ\cos{2\beta}} k_y\mp\frac{\hbar k^2_\mathrm{F}}{\hbar^2k^2_\mathrm{F}+2\sigma mJ\cos{2\beta}}\sqrt{2m\left(1+\frac{2\sigma mJ\cos{2\beta}}{\hbar^{2}k^2_\mathrm{F}}\right)(\mu-E)-\hbar^{2}k^2_y+\frac{4m^{2}J^2}{\hbar^{2}k^4_\mathrm{F}}k^2_y}\,,\\
\end{split}
\end{equation}%
while for  $p$-wave magnets we have
\begin{equation}
\label{EqsqUM}
\begin{split}
p_{e\sigma}^{\pm}&=-\frac{\sigma mJ\cos\beta}{\hbar^{2}k_\mathrm{F}}\pm\sqrt{\frac{2m}{\hbar^{2}}(\mu+E)-\left(k_y+\frac{\sigma mJ\sin\beta}{\hbar^{2}k_\mathrm{F}}\right)^2+\frac{m^{2}J^2}{\hbar^{4}k^2_\mathrm{F}}}\,,\\
p_{h\sigma}^{\pm}&=\frac{\sigma mJ\cos\beta}{\hbar^{2}k_\mathrm{F}}\mp\sqrt{\frac{2m}{\hbar^{2}k_\mathrm{F}}(\mu-E)-\left(k_y-\frac{\sigma mJ\sin\beta}{\hbar^{2}k_\mathrm{F}}\right)^2+\frac{m^{2}J^2}{\hbar^{4}k^2_\mathrm{F}}}\,.
\end{split}
\end{equation}%
\end{widetext}%

We also note that the scattering states given by Eqs.\,(\ref{scatte1})  also hold when replacing N by a ferromagnet with magnetization along $z$, but the wavevectors are then given by \cite{PhysRevLett.74.1657}  
\begin{equation}
\begin{split}
p_{e\sigma}^{\pm}&=\pm\sqrt{\frac{2m}{\hbar^2}(\mu-\sigma J+E)-k^2_y}\,,\\
p_{h\sigma}^{\pm}&=\mp\sqrt{\frac{2m}{\hbar^2}(\mu+\sigma J-E)-k^2_y}\,.
\end{split}
\end{equation}%
The expressions for the wavevectors in AMs written in Eqs.\,(\ref{EqsqAM}) are   those reported in Refs.\,\cite{Sun23,PhysRevLett.133.226002} and used to study ARs in $d$-wave AMs. The expressions we write in Eqs.\,(\ref{EqsqUM}) slightly differ from those employed in \cite{maeda2024} to study ARs in $p$-wave magnets due to the definition of the $p$-wave magnetic order. In the S region, the wavevectors are still the same as those given by Eqs.\,(\ref{qehN}) but the wavefunctions need to be written taking into account up and down spins, see e. g., Refs.\,\cite{PhysRevB.92.134512,PhysRevB.98.075425}. Having the scattering states in N [Eq.\,(\ref{scatte1})], which now represents an unconventional magnet, and S, we can use Eqs.\,(\ref{Wavematching}) upon appropriate modifications, to find the coefficients $a_{\sigma}$ and $b_{\sigma}$ representing ARs and NRs. It is worth noting here that since the exchange fields of   unconventional magnets are momentum dependent, one needs to be careful when deriving the boundary conditions \cite{Sun23}: antisymmetrization of the unconventional magnet field is needed in order to maintain Hermicity of the Hamiltonian. In our case, this can be done  by replacing the momentum operator as $k_{x}\rightarrow\{k_{x},\theta(-x)\}/2$, where $\{\cdot\}$ is the anticommutator; see Refs.\,\cite{Sun23} and \cite{sukhachov2025} for AMs and  $p$-wave magnets, respectively.  In doing so, it is possible to find analytical solutions for the scattering processes, although the expressions are lengthy and, instead of writing them, we next discuss their behavior under the impact of unconventional magnetism.

\begin{figure*}
    \centering
    \includegraphics[width=0.98\linewidth]{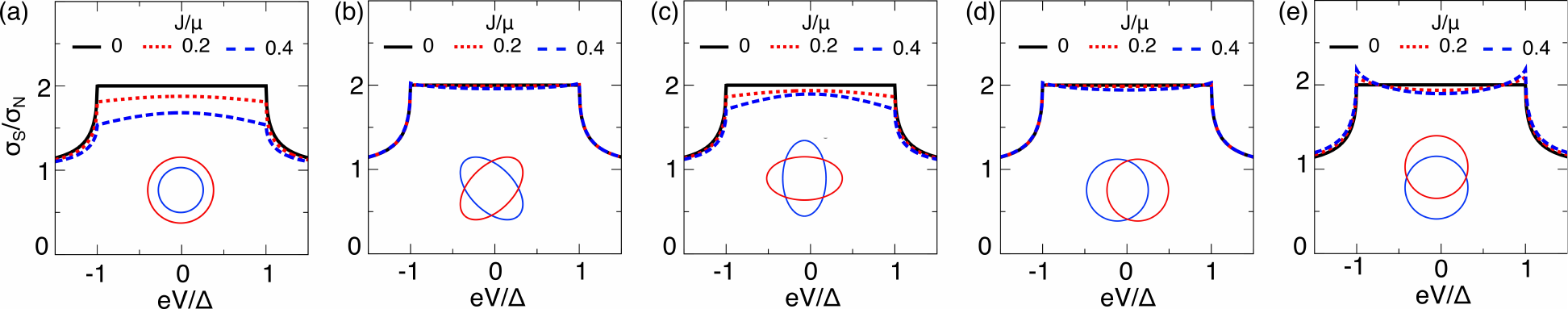}
    \caption{Tunneling conductance  integrated in $k_{y}$ as a function of energy $eV$ in normal-superconductor (NS) junctions formed by semi-infinite  unconventional magnets and a semi-infinite conventional superconductor, where N is a ferromagnet (a), $d_{xy}$-wave  altermagnet (AM) (b), a $d_{x^{2}-y^{2}}$-wave AM (c), a  $p_{x}$-wave  magnet (d), and a $p_{y}$-wave  magnet (e).   The conductance is normalized with the normal transmissivity $\sigma_{\rm N}$ obtained by setting $\Delta=0$ in Eq.\,(\ref{conductanceSN}), with $Z$ characterizing the strength of the barrier here taken $Z=0$. Distinct curves represent distinct values of the strength of the exchange field in the respective magnet, while the insets depict the Fermi surfaces in each case. Parameters: $\mu=1$, $\Delta=0.001\mu$,   $J=0.2\mu$.}
    \label{conductance_NS_UMs}
\end{figure*}

Thus, using the coefficients of the scattering states discussed above [Eq.\,(\ref{scatte1})], we are able to further inspect the conductance $\sigma_\mathrm{S}=\sigma_{\uparrow}+\sigma_{\downarrow}$, which  is obtained  as
\begin{equation}
\label{conductanceSN}
\sigma _{\uparrow \left( \downarrow \right) }=
\sigma_\mathrm{N}\sum_{k_{y}}
\left[ 1+\frac{v_{h\uparrow \left( \downarrow
\right) }}{v_{e\uparrow \left( \downarrow \right) }}\left\vert a_{\uparrow
\left( \downarrow \right) }\right\vert ^{2}-\left\vert b_{\uparrow \left(
\downarrow \right) }\right\vert ^{2}\right] ,
\end{equation}
where $\sigma_\mathrm{N}$ is the normal conductance~\cite{maeda2024}, $a_{\sigma}$ are the Andreev reflection amplitudes, $b_{\sigma}$ are the normal reflection amplitudes, and $v_{e\sigma}$, $v_{h\sigma}$ are the group velocities obtained by Eq.\ (\ref{groupvel}). The normal conductance  $\sigma_\mathrm{N}$  is obtained by using  Eq.\,(\ref{conductanceSN}) but with $\Delta=0$.  In Fig.\,\ref{conductance_NS_UMs}, we present the tunneling conductance $\Sigma_{S}$ obtained from Eq.\,(\ref{conductanceSN})  as a function of energy in   NS junctions without a potential barrier $Z=0$ and modelled as discussed in the previous paragraph: N and S  represent semi-infinite of a magnetic material with exchange field $J$ and a conventional spin-singlet $s$-wave superconductor. For completeness, in Fig.\,\ref{conductance_NS_UMs}(a) we consider N to be a ferromagnet, while in Fig.\,\ref{conductance_NS_UMs}(b,c) we take it to be a $d_{xy}$-wave and $d_{x^{2}-y^{2}}$-wave AMs, respectively. In Fig.\,\ref{conductance_NS_UMs}(d,e) we consider N to describe $p_{x(y)}$-wave  magnets. Addressing multiple types of magnetic order is useful in order to identify the role of unconventional magnetism on conductance. In NS junctions with N being a ferromagnet, the finite values of the exchange field $J$ reduce the constant subgap conductance while leaving a larger value at zero energy. The reduction of the subgap conductance can be understood to originate from the split Fermi surfaces due to $J$ (see inset), a situation that then suppress   Andreev reflection processes \cite{PhysRevLett.74.1657,Kashiwaya1999}. In the case of N being a $d$-wave AM, the situation has some similarities but also differences with respect to the ferromagnet case which depends on the orientation of the altermagnetic field, compare  Fig.\,\ref{conductance_NS_UMs}(b,c) and Fig.\,\ref{conductance_NS_UMs}(a). In fact, for NS junctions with $d_{xy}$-wave AMs [Fig.\,\ref{conductance_NS_UMs}(b)], the conductance within the superconducting gap only slightly feels the variations of the exchange field $J$, remaining roughly constant at values equal to $\sigma_{\rm S}\approx 2\sigma_{\rm N}$ \cite{Papaj23,Sun23}. 

The insensitivity of the conductance in $d_{xy}$-wave AMs to variations of the exchange field  originates because, when integrating over all incident angles towards the interface, the majority and minority spin bands contribute equally, which then gives rise to a vanishing total spin polarization of incident states (see also inset); that is why the conductance    for a $d_{xy}$-wave AM behaves as for a normal metal without spin polarization.    Such a behavior can be understood by using  symmetry arguments: Under a mirror reflection with respect to the $xz$ plane, the orbital part of the $d_{xy}$-wave AM changes its sign so that $\sigma _{\uparrow }$ and $\sigma
_{\downarrow }$ are connected via mirror reflection. After the sum of all transverse channels with positive and negative $k_y$, $\sigma _{\uparrow }$ become the same with $\sigma
_{\downarrow }$. On the other hand, the orbital part of the $d_{x^{2}-y^{2}}$-wave AM has no sign change under the mirror operation with respect to the $xz$ plane, leading to
a finite spin polarization $\sigma _{\uparrow }\neq \sigma _{\downarrow }$. This conductance behavior is thus different from what occurs when N is a ferromagnet but it is similar when N is a  normal metal [Fig.\,\ref{NSjunction_1}(c)], pointing out an intriguing difference of AMs with respect to ferromagnets despite both breaking time-reversal symmetry. For NS junctions where N region is a $d_{x^{2}-y^{2}}$ in Fig.\,\ref{conductance_NS_UMs}(c), the subgap conductance exhibits an overall reduction as the exchange field increases \cite{Sun23}, a behavior similar to the one when N is a ferromagnet, see Fig.\,\ref{conductance_NS_UMs}(a). \textcolor{red}{It thus means that the orientation of the altermagnetic order strongly affects the signature of Andreev reflection~\cite{Papaj23,Sun23}.} This happens because   the  trajectories with a dominant contribution to the charge transport are those exhibiting a normal incidence to the interface, a regime where the Fermi surface of the AM is similar to a ferromagnet because one spin species is larger than the other, compare insets in Fig.\,\ref{conductance_NS_UMs}(a,c). We have verified that when the barrier strength takes finite values, characterized by $Z\neq0$, the subgap conductance reduces.

 In relation to NS junctions where N is a semi-infinite $p$-wave magnet, we find slightly different behavior from the one discussed for $d$-wave AMs and ferromagnets, depending on the orientation of the exchange field of the unconventional magnet. When N is a $p_{x}$-wave  magnet, the conductance is largely insensitive to variations of the exchange field $J$ [Fig.\,\ref{conductance_NS_UMs}(d)], akin to what occurs in $d_{xy}$-wave AMs and indicating that Andreev reflections are dominant.  The almost constant value of the subgap conductance when N is a $p_{x}$-wave  magnet originates from the shift of the Fermi surfaces of the $p_x$-wave  magnet along the $k_x$-direction. The slight variation of the subgap conductance indicates that the mismatch of the Fermi surface exists in the used model given by Eq.(\ref{eq:AM_H}). In Ref.\ \cite{maeda2024}, however, the shape of spin-resolved Fermi surface does not change with $J$ and the subgap conductance is constant within the superconducting gap. Interestingly, when N is a $p_{y}$-wave  magnet, the increase of $J$ modifying the conductance in such a way that reduces it within the superconducting gap but enhances it at the gap edges $\pm\Delta$ and also outside the gap, as shown in Fig.\,\ref{conductance_NS_UMs}(e). This implies that Andreev scattering tends to be suppressed in NS junctions with N  being a $p_{y}$-wave  magnet.  The reason for this behavior is    that there are no available Fermi surfaces corresponding  to ARs for quasiparticles with large transverse momentum. The discussed effect was first predicted in Ref.\,\cite{maeda2024} and recently also confirmed in Ref.\cite{sukhachov2025}. Ref.\ \cite{maeda2024} also addressed conductance in NS junctions with N being a $p$-wave magnet and S being a   $p$-wave superconductor, a system shown to produce zero-bias conductance peaks which acquires a nontrivial dependence on the magnetic order. Thus,  unconventional magnets affect the charge conductance in NS junctions.
 
\begin{figure}
    \centering
    \includegraphics[width=0.98\linewidth]{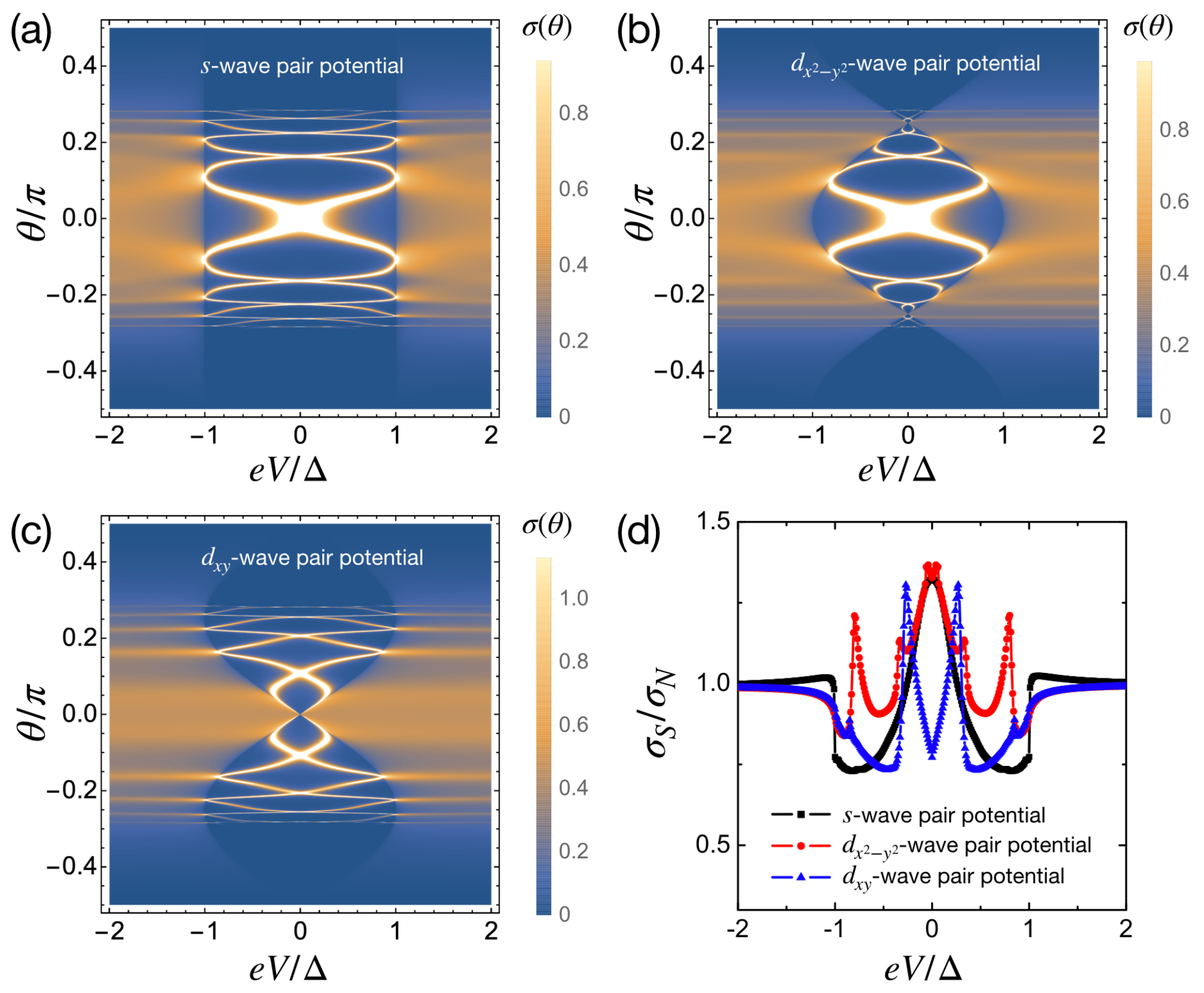}
    \caption{Tunneling conductance as a function of energy $eV$ and the injection angle to the interface $\theta$ in  N$_{1}$N$_{2}$S junctions, where N$_{1}$ is a semi-infinite normal metal,   N$_{2}$ is a $d$-wave  altermagnet of length $L$, and $S$ a semi-infinite  superconductor that can have  spin-singlet $s$-wave (a) or $d$-wave pair potentials (b,c).   A barrier potential characterized by $Z$ is assumed at the interface between N$_{1}$ and N$_{2}$; also, $\theta$  denotes the injection angle   defined as $k_y=k_\mathrm{F}\sin\theta$, with $k_{\rm F}$ being the Fermi wavevector.    (d) The total tunneling conductance integrated in $\theta$ as a function of eV corresponding to (a-c).  Parameters: $k_\mathrm{F}L=10$ and $Z=2$, $J=0.3\mu$, $\Delta=0.001\mu$. Adapted from Ref.\,\cite{Bo2025}.}
    \label{conductance_NS_AMs2}
\end{figure}

So far we have assumed that the  unconventional magnet is semi-infinite and it is thus natural to wonder what happens to the conductance when the  unconventional magnet has a finite length. These ideas were explored in Refs.\,\cite{Papaj23,Sun23,IkegayaAltermagnet,Niu2024,Bo2025} which involved   AMs but not  $p$-wave magnets. To understand the conductance features in superconducting junctions with   AMs of finite length, it is important to understand the underlying properties of  N$_{1}$N$_{2}$S junctions, where N$_{1}$ and S are semi-infinite  normal  and   superconducting regions, while N$_{2}$ is a finite normal region. In this case, an incident electron from N$_{1}$ gets transmitted to N$_{2}$, which then experiences confinement effects due to the finite length of N$_{2}$ and also Andreev reflections at the  N$_{2}$S interface \cite{PhysRevB.104.134507}. The interplay of confinement and  ARs give rise to the formation of discrete states within the superconducting gap having electron and hole characters, also known as  Andreev bound states (ABSs). This phenomenon already appears when  N$_{2}$ is just a normal metal, where ABSs have been revealed to produce conductance  features that help identify the type of superconductor \cite{PhysRevB.104.134507}. In the case of N$_{1}$N$_{2}$S junctions where N$_{1}$ and S are a semi-infinite normal metal and a spin-singlet $s$-wave superconductor, Ref.\,\cite{Papaj23} predicted that  ABSs appear and can even produce zero-bias peaks for $d_{x^{2}-y^{2}}$-wave AMs [Fig.\,\ref{conductance_NS_AMs2}]; this suggests a strong dependence of conductance   on the altermagnetic orientation. Moreover, in a similar type of N$_{1}$N$_{2}$S junction,  Ref.\,\cite{IkegayaAltermagnet} found that a special type of AR occurs, known as specular AR when the reflection angle is inverted with respect to the angle of incidence \cite{PhysRevLett.97.067007}, which induces a positive nonlocal conductance and can be useful for realizing Cooper pair splitters.

Later,  Ref.\,\cite{Bo2025} explored conductance and ABSs in N$_{1}$N$_{2}$S junctions with N$_{2}$ being a $d$-wave AM and S being a $d$-wave superconductor, a system  motivated by the need to understand the impact of the $d$-wave nature in both materials; the ABSs in Ref.\,\cite{Bo2025} are referred to as de Gennes and Saint-James states. Interestingly, while the $d$-wave nature of AMs induces ABSs already when S is a conventional spin-singlet $s$-wave superconductor \cite{Papaj23}, having a $d$-wave superconductor affects the levels around zero energy which then produce distinct features in the tunneling conductance, see Fig.\,\ref{conductance_NS_AMs2}(a-d). For a  N$_{2}$ finite region being a $d_{x^{2}-y^{2}}$-wave AM, the tunneling conductance develops a zero-bias peak due to the formation of zero-energy ABSs; this effect occurs because the spin-split bands in $d_{x^{2}-y^{2}}$-wave AMs enhance the mismatch between electron and hole wave vectors, resulting in a sizable phase accumulation in the process of sequential NRs and ARs inside the   altermagnetic region N$_{2}$; these zero-bias peaks originate due to confinement effects in the finite  region N$_2$ \cite{PhysRevB.104.134507}. In contrast, for a  N$_{2}$ finite region being a $d_{xy}$-wave AM, ABSs with a different profile appear and lead to a  V-shape profile of the tunneling conductance around zero energy, but with a finite value at zero energy, see Fig.\,\ref{conductance_NS_AMs2}(c,d). It is important to notice that other more recent works have also explored Andreev transport in  modified   junctions based on AMs \cite{PhysRevB.109.245424,Ping_Niu_2024}  and $p$-wave magnets $p$-wave \cite{soori2025}, which have enabled the inspection of nonlocal Andreev reflections also known as crossed Andreev reflections \cite{deutscher2002crossed,asano2021andreev}. While these studies indeed reveal the intriguing impact of unconventional magnetism in superconducting systems, there are still phenomena that need to be explored such as crossed ARs \cite{PhysRevB.109.245424} and also Andreev physics in  $p$-wave magnets  \cite{soori2025} as well as in  superconducting junctions with higher momentum  unconventional magnets. The so far reported strong dependence of the ABSs and their conductance signatures in junctions with $d$-wave altermagnetism and    superconductivity unveils  novel superconducting phenomena that can be further exploited for the design of e.g., Andreev qubits \cite{PhysRevLett.90.226806,PhysRevA.109.032601,PhysRevB.103.045410}, see also Refs.\,\cite{aguado2020majorana,kjaergaard2020superconducting}.

\subsubsection{Proximity effect and inverse proximity effect}
In the previous part, we have seen that Andreev processes in superconducting junctions are able to carry information about the superconductor deep into a non-superconducting  unconventional magnet material when placed in very close proximity. We have shown that the carried information strongly depends on the properties of both the  unconventional magnet and superconductor, raising a natural question about the nature of superconductivity when combining  unconventional magnets and superconductors in the form of hybrid junctions. This question was partially addressed in Subsection \ref{subsection32}, where we showed that distinct types of superconducting correlations emerge when an entire two-dimensional unconventional magnet has   homogeneous superconductivity, provided spatial invariance is preserved. However, in junctions, such as those where  ARs have been studied and where spatial invariance is broken,  the nature of superconductivity still remains unknown and we plan to study it in this part.

 \begin{figure*}[!t]
\centering
\includegraphics[width=0.99\textwidth]{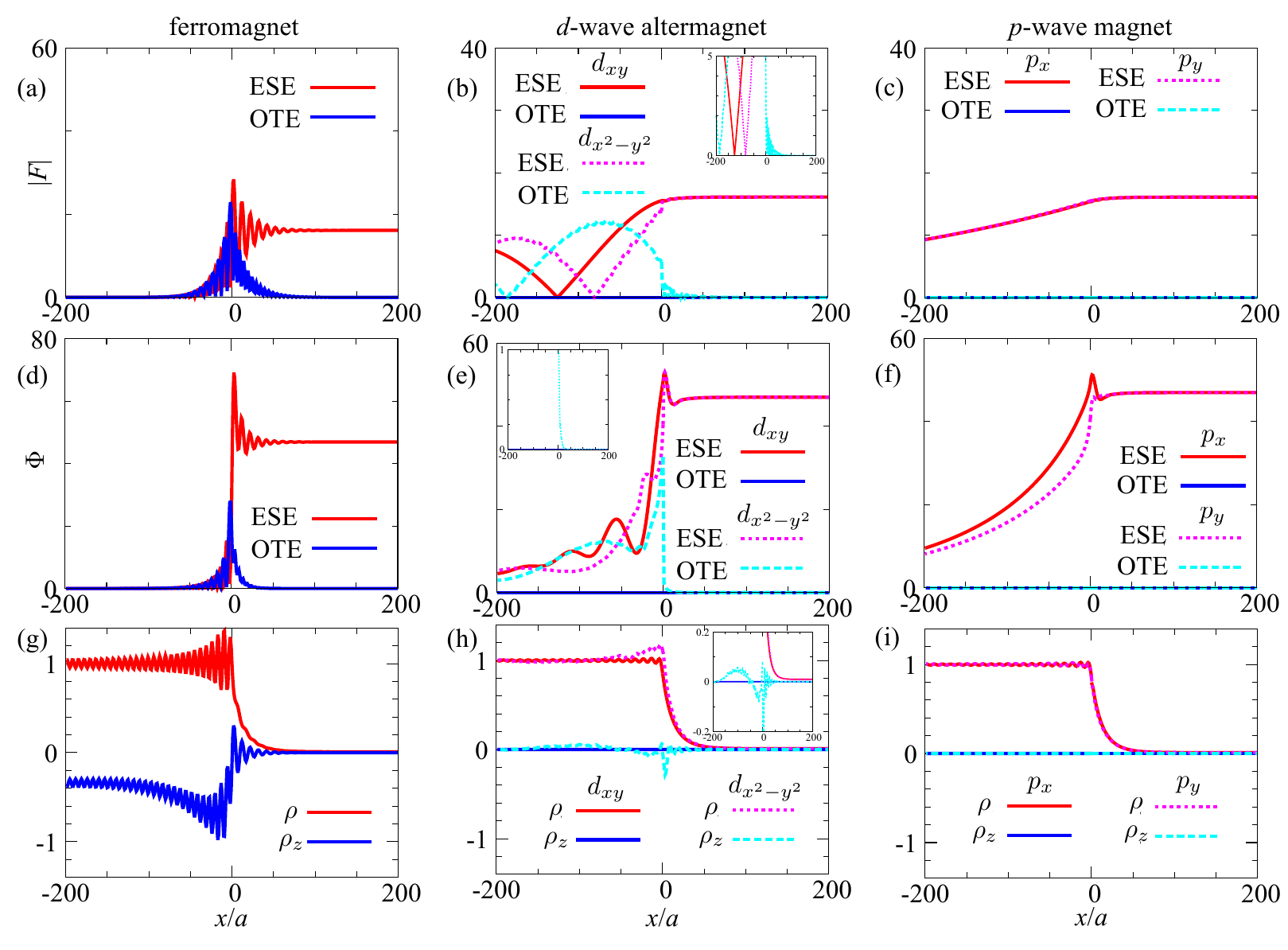}
\caption{(a-c) Absolute value of the  zero-frequency local pair amplitudes   as a function of space  at   $k_{y}=0.02\pi$ in  NS junctions without potential barrier, where the interface is located at $x=0$ and the S (N) regions correspond to a conventional spin-singlet $s$-wave superconductor (magnetic system) situated at $x>0 (x<0)$. Distinct panels correspond to distinct N regions, including   a ferromagnet (a), $d$-wave altermagnet (b), $p$-wave  magnet (c).  The pair amplitudes are obtained within a   recursive Green's function approach. Here, ESE (OSO) stands for even-frequency (odd-frequency) spin-singlet even-parity (odd-parity), while OTE (ETO) for odd-frequency (even-frequency) spin-triplet even-parity (odd-parity).  (d-f) Square absolute value of the pair amplitudes of (a-c) integrated in $k_{y}$.   (g-i) Space dependence of the electronic  local density of states (LDOS)  ($\rho$) and local spin-density along $z$ ($\rho_{z}$), both integrated in $k_{y}$ and normalized with respect to the   LDOS at $E=0$ and  $\Delta=0$ for the systems corresponding to (a-c). Tight-binding parameters: $a=1$, $\mu=0.1t$, $\Delta=0.02t$, $m_{z}= t_{x}=t_{y}=t_{xy}=t_{x^{2}-y^{2}}=0.1t$. Adapted from Ref.\,\cite{fukaya_proximity}.  
}
\label{NS_UM_F_local} 
\end{figure*}

For simplicity, let us first point out some basic features about superconducting correlations in  NS junctions located at  $x=0$ and with a pair potential $\Delta(x)=\theta(x)\Delta$; here N and S are semi-infinite regions of a normal metal and  a conventional superconductor, respectively. The superconducting correlations can then be obtained from the anomalous part of the Nambu Green's function $\hat{G}(x,x',k_{y},\omega_{n})$ associated with the BdG Hamiltonian of the system, as discussed in Subsection \ref{subsection32}; since momentum along $x$ is not a good quantum number, $\hat{G}(x,x',k_{y},\omega_{n})$ depends on two spatial coordinates \cite{tanaka2011odd_review,Cayao2020odd,triola2020role,tanaka2024review}. For NS junctions, it is useful to calculate the Nambu Green's function  by using scattering states \cite{TK96a,kashiwaya2000,lu2018study,PhysRevB.100.115433,Cayao2020odd}, so that the semi-infinite nature of N and S are taken into account. Alternatively, one can use a recursive Green's function approach within a tight-binding description, also proven useful for calculating the Green's function of semi-infinite superconducting systems   \cite{fukaya2024x,fukaya_proximity}. Under general conditions, even though the pair potential in N is zero, it is possible to induce a nonzero pair amplitude in N, with a characteristic length $\xi_{\rm N}=\hbar v_{\rm F}/E$ \cite{PhysRevB.69.174504,PhysRevB.73.014503}, which at zero energy is kept all the way to infinity and signals the superconducting coherence; at finite temperatures, one recovers  $\xi_{\rm N}\approx1/T$, after doing an appropriate analytic continuation $E\rightarrow i\omega_{n}$. Analogously, an Andreev reflected hole has the opposite velocity at zero energy and traces the same phase in a phase coherence fashion. Both the induced pair amplitude in N and the Andreev reflections have been shown to be directly related to each other and to be at the core of what is known as the superconducting \emph{proximity effect} \cite{golubov1988theoretical,klapwijk2004proximity,PhysRevB.96.155426,PhysRevB.98.075425,Cayao2020odd}, where the part of N closest to S acquires superconducting correlations, see light green region in N of the bottom panel of Fig.\,\ref{NSjunction_1}(a). Furthermore, in the same way as the N region gets affected by the superconductor, the region of S closest to N over a range $\xi_{\rm S}=\hbar v_{\rm F}/\Delta$ also suffers changes intimately  connected to the nature of N. For instance, given that S can be a conventional spin-singlet s-wave superconductor, it is possible to induce superconducting correlations with a distinct symmetry, which  gives rise to an effect known as the \emph{inverse proximity effect}; this effect has been mostly studied when N is a magnetic material which then induces  a magnetic moment in the superconductor \cite{PhysRevB.38.8823,PhysRevB.69.174504,PhysRevB.71.024510,PhysRevB.72.064524} but recently also addressed under the presence of ABSs \cite{PhysRevResearch.3.043148}. The inverse proximity effect is depicted by a blue-green region in S of the bottom panel of Fig.\,\ref{NSjunction_1}(a). 
 
To understand the proximity effect and the inverse proximity effect in NS junctions  formed by  unconventional magnets, in what follows we discuss the emergent pair amplitudes from Ref.\,\cite{fukaya_proximity} obtained for semi-infinite unconventional magnets and  semi-infinite spin-singlet $s$-wave superconductors. We consider a NS junction along $x$    located at $x=0$, such that the pair potential is finite only in S, $\Delta(x)=\theta(x)\Delta$; to model N and S, a tight-binding (TB) description of the unconventional magnets and superconductors is taken with an appropriate discretization along $x$ but keeping $k_{y}$ a good quantum number, as discussed in Ref.\,\cite{fukaya_proximity}, see also Section \ref{section2}.  Here, within a TB representation, we remind that the strength of the exchange fields for a ferromagnet, $d_{xy}$-wave AM, $d_{x^{2}-y^{2}}$-wave AM, $p_{x}$-wave magnet, and $p_{y}$-wave magnet are denoted by $m_{z}$, $t_{xy}$, $t_{x^2-y^2}$, $t_{x}$ and $t_{y}$, respectively; see  Subsection \ref{section2a}. Then,  the pair amplitudes are found within a recursive Green's function approach as discussed in Ref.\,\cite{fukaya_proximity}. Under general circumstances,  NS junctions formed by unconventional magnets host four pair symmetries, which include the ESE, OSO, ETO, and  OTE classes discussed in Subsection \ref{subsection32}, where the first, second, and third capital letters represent the even (odd) symmetry of frequency, spin, and momentum parity, respectively. Moreover, the main role of the interface in NS junctions is that it breaks the spatial invariance, a mechanism that leads to the coexistence of even- and odd-parity states even when the bulk of S is a conventional spin-singlet $s$-wave parity \cite{Cayao2020odd}. Thus,  here we also expect four pair  symmetry classes due to the interplay of unconventional magnetism and breaking of spatial invariance at the NS interface.  In Fig.\,\ref{NS_UM_F_local}, we show the   absolute value of the local pair amplitudes as a function of space $x=x'$ at   lowest frequency and fixed $k_{y}$ for NS junctions with N being either a ferromagnet, a $d$-wave AM, or a $p$-wave magnet; here $x<0$ and $x>0$ correspond to N and S, respectively.  In Fig.\,\ref{NS_UM_F_local}(d-f), we also plot the square absolute value of the pair amplitudes of panels Fig.\,\ref{NS_UM_F_local}(a-c) integrated in the transverse momentum $k_{y}$. 
For completeness,  in Fig.\,\ref{NS_UM_F_local}(g-i) we also present the zero-energy electronic LDOS and local spin density along $z$ integrated in $k_{y}$. For each energy $E$, we define the LDOS $\rho(E)$ and spin density along $z$ $\rho_z(E)$  as
\begin{align}
    \rho(E)&=\rho_\uparrow(E)+\rho_\downarrow(E),\\
    \rho_z(E)&=\rho_\uparrow(E)-\rho_\downarrow(E),
\end{align}%
with the LDOS with up-spin $\rho_\uparrow(E)$ and down-spin $\rho_\downarrow(E)$, respectively. The first feature to notice here is that, by symmetry, the odd-parity pair amplitudes OSO and ETO vanish locally at $x=x'$, leaving only   ESE and OTE pair classes as allowed superconducting correlations. Depending on the nature of magnetism in N, the ESE and OTE pair amplitudes, however, exhibit intriguing properties that reflect   the nature of both the proximity effect and the inverse proximity effect.  For NS junctions with N being a ferromagnet in Fig.\,\ref{NS_UM_F_local}(a,d), ESE and OTE pair amplitudes are induced locally in space with a leakage into N over  distances determined by $\xi_{\rm F}\approx1/\sqrt{J}$, which determines the proximity effect into the ferromagnet; note that penetration length $\xi_{\rm F}$ is much shorter than $\xi_{\rm N}$ in normal metals \cite{PhysRevB.69.174504,PhysRevB.73.014503}. There is also a finite value of OTE pairing in S, which decays in an oscillatory fashion  from the interface over $\xi_{\rm S}=100a$ for this case and is accompanied by a decaying LDOS as well as a finite local spin density along $z$ [Fig.\,\ref{NS_UM_F_local}(g)]; this altogether signals the inverse proximity effect. The finite value of the ESE pair amplitude in the S region of Fig.\,\ref{NS_UM_F_local}(a,d) simply reflects the spin-singlet $s$-wave nature of the considered parent superconductor.

\begin{figure}[!t]
\centering
\includegraphics[width=0.49\textwidth]{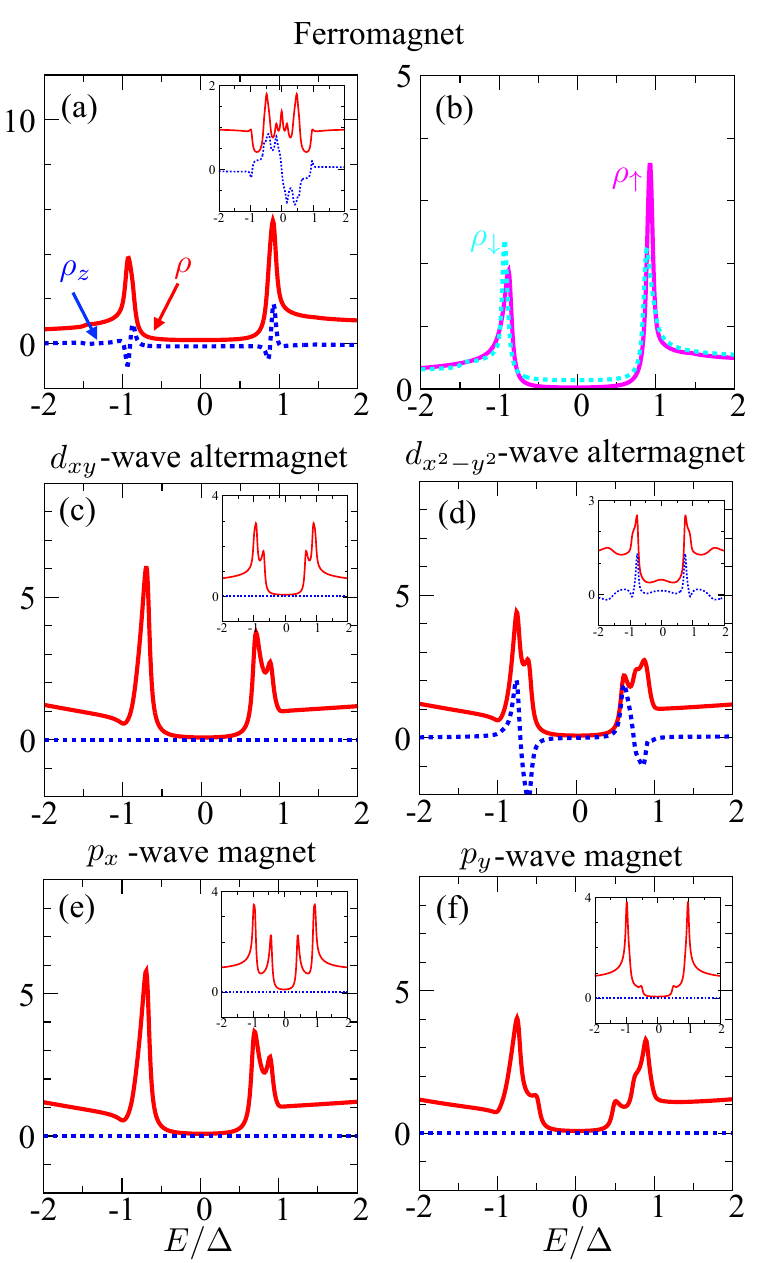}
\caption{Energy dependence of the local density of states (LDOS) ($\rho$) and spin density along $z$ ($\rho_{z}$)  in the S region close to the interface of a NS junction, where N is a magnetic material of finite length $L_{\rm N}$ and S a semi-infinite spin-singlet $s$-wave superconductor. The LDOS  and spin density   are integrated in $k_{y}$, and normalized with respect to its value at $E=0$ and $\Delta=0$. (a)   LDOS and spin density for a ferromagnet; (b) LDOS for up and down spins in a ferromagnet. (c-f) The same as in (a) but for $d$-wave altermagnets and $p$-wave  magnets. Insets show the LDOS and spin density at $\mu=4t$, $m_{z}=t_{x}=t_{y}=t_{xy}=t_{x^{2}-y^{2}}=0.4t$. Tight-binding parameters: $L_{\rm N}=10a$, $\mu=0.1t$, $\Delta=0.02t$,  $m_{z}=t_{x}=t_{y}=t_{xy}=t_{x^{2}-y^{2}}=0.1t$, $x=1a$. Adapted from Ref.\,\cite{fukaya_proximity}. 
}
\label{NS_UM_FiniteN_LDOS} 
\end{figure}

In the case of $d$-wave AMs in Fig.\,\ref{NS_UM_F_local}(b,e),  the ESE  pair amplitude induced in N develops an oscillatory profile with distinct periodicities  that depend on the interplay between the chemical potential, the transverse momentum $k_{y}$, and  the type of $d$-wave magnetic order; an apparent decay into N is also seen, which is, however, slower than in ferromagnets [Fig.\,\ref{NS_UM_F_local}(a,d)]. A similar oscillatory profile was reported before in NS junctions with Rashba SOC \cite{PhysRevB.98.075425}. Moreover,   an OTE pairing can be also induced in the $d_{x^{2}-y^{2}}$-wave AM, which exhibits  an oscillatory decaying profile from the interface towards the bulk of S  for distances within $\xi_{\rm S}$, see  Fig.\,\ref{NS_UM_F_local}(b,e). No OTE is induced  in NS junctions with $d_{xy}$-wave AMs because its even parity is formed by an odd dependence in space along $x$ and an odd dependence in $k_{y}$, which leads to a vanishing amplitude locally in $x$.   We note that, while a finite LDOS in both $d$-wave AMs is found,  a finite local spin density is only present  close to the interface of NS junctions with   $d_{x^{2}-y^{2}}$-wave AMs when OTE pairing is present, see Fig.\,\ref{NS_UM_F_local}(h). Thus, the absence of OTE in NS junctions with $d_{xy}$-wave AMs is accompanied by the absence of spin density along $z$, see Fig.\,\ref{NS_UM_F_local}(h). For NS junctions with $p$-wave magnets, an odd-parity symmetry in $k_{y}$ restricts the formation of OTE as for $d_{xy}$-wave AMs, while ESE is induced into N in an oscillatory decaying fashion that is distinct to what happens in $d$-wave magnets, see  Fig.\,\ref{NS_UM_F_local}(c,f). In this case, the LDOS in N develops weak oscillations with the same periodicity as the ESE pair amplitude and decays from the interface into S, while vanishing local spin density is obtained in NS junctions with both types of $p$-wave magnets, see Fig.\,\ref{NS_UM_F_local}(i). In all cases discussed here, the vanishing LDOS deep in S already indicates the conventional nature of the parent superconductor, which exhibits a U-shaped LDOS and is independent of the magnetic nature of N. We can therefore see that  the proximity effect in NS junctions based on unconventional magnets is governed by penetration lengths in the unconventional magnets that are entirely associated to their type of magnetic nature. Moreover, while a clear inverse proximity effect is predicted in $d_{x^{2}-y^{2}}$-wave AMs, the preliminary findings of  Ref.\,\cite{fukaya_proximity} suggests vanishing values for the local amplitudes in $d_{xy}$-wave AMs and  $p$-wave magnets.

Further insights on the inverse proximity effect can be obtained from Fig.\,\ref{NS_UM_FiniteN_LDOS}, where we show the LDOS ($\rho$) and local spin density ($\rho_{z}$) integrated over $k_{y}$  inside S but very close to the interface of a  NS junction. Since it is known that the inverse proximity effect is sensitive to the appearance of ABSs \cite{PhysRevResearch.3.043148}, we consider a finite length N region made of a magnetic material, which is expected to host ABSs due to confinement in N \cite{PhysRevB.91.024514,PhysRevB.104.L020501,PhysRevB.104.134507,baldo2023zero}. One of the main features is that   LDOS exhibits subgap peaks representing the emergence of ABSs, whose position however depends on the magnetic order and chemical potential, see Fig.\,\ref{NS_UM_FiniteN_LDOS} and also insets. Another interesting feature is that, while the   LDOS is always finite and develops a   U-shape profile, the local spin density develops finite values only for the $d_{x^{2}-y^{2}}$-wave AM and ferromagnet but vanishes otherwise.   The negative value of the spin density $\rho_z$ is a result of having that $\rho_\downarrow$ is larger than 
$\rho_\uparrow$, provided $\rho_{\uparrow\downarrow}>0$. This behavior holds even for  a stronger chemical potential and exchange field, supporting the absence of induced local pair amplitudes that contribute to the inverse proximity effect in NS junctions formed by $d_{xy}$-wave AMs and $p$-wave  magnets, see insets. These results, however, do not rule out  the formation of  nonlocal superconducting correlations as contributing sources to the inverse proximity effect in such systems \cite{PhysRevB.104.094503}, but their relevance still remains unexplored. Having discussed the proximity effect and the inverse proximity effect in terms of emergent superconducting correlations, it is important to remark some points needed for adding more fundamental understanding of superconductivity in unconventional magnets. Thus, while a direct relationship is known to exist between the discussed proximity effects and Andreev processes, the method used to obtain  Fig.\,\ref{NS_UM_F_local} does not allow to identify it; to discover such a relationship, the superconducting correlations need to be calculated by using a scattering Green's function approach, which involves scattering states defined in terms of Andreev and normal   processes  \cite{TK96a,kashiwaya2000,lu2018study,PhysRevB.100.115433,Cayao2020odd}. Furthermore, the understanding of the proximity effect as well as the inverse proximity effect would enormously benefit from the investigation of superconducting correlations  in hybrid systems with  unconventional magnets under realistic conditions, such as the unavoidable finite size \cite{PhysRevB.104.125405,ahmed2024oddfreABS} as well as the intrinsic multiple degrees of freedom \cite{PhysRevB.88.104514,PhysRevB.90.220501,PhysRevB.93.201402,PhysRevB.101.214507,triola2020role,PhysRevB.109.205406}. 
 
 \begin{figure*}[!t]
\centering
\includegraphics[width=0.99\textwidth]{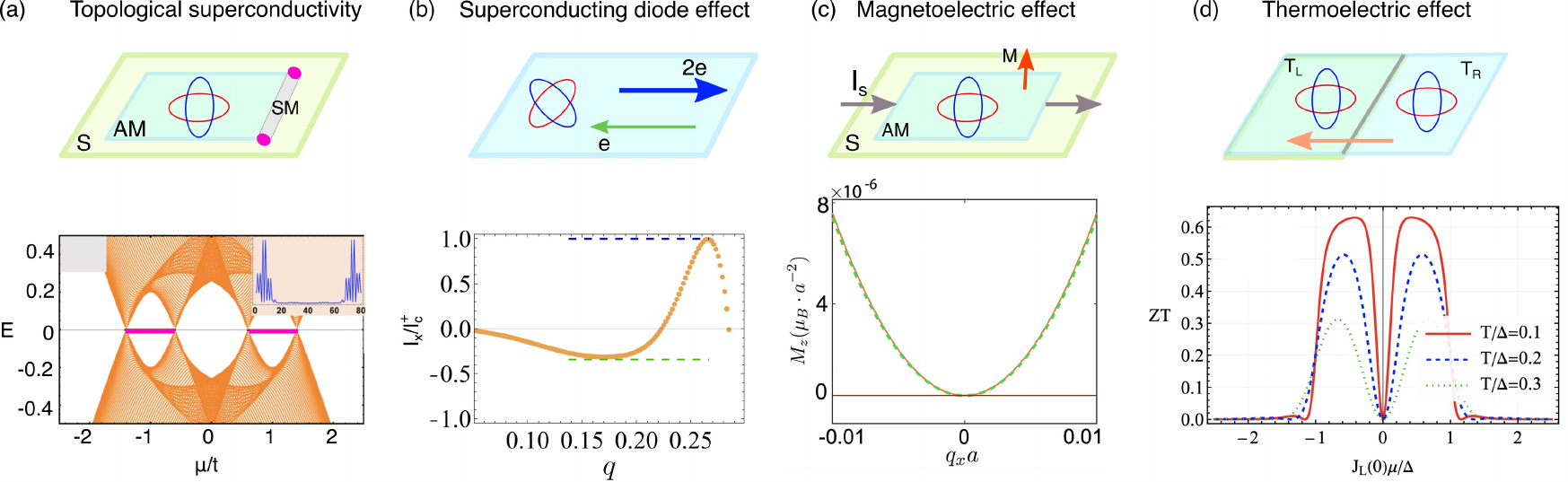}
\caption{Emergent  phenomena in superconducting systems with altermagnetism. (a) Top:  one-dimensional topological superconductivity in a hybrid system formed by a conventional superconductor S (light green), a $d_{x^{2}-y^{2}}$-wave  altermagnet (AM) (cyan), and a semiconductor  nanowire (gray); the semiconductor  (SM) becomes a one-dimensional topological superconductor with Majorana  modes at its ends (magenta filled circles).   Bottom: Low-energy spectrum as a function of the chemical potential, with Majorana  modes depicted in magenta; inset: wavefunctions of the Majorana  modes. Reprinted figure with permission from Sayed Ali Akbar Ghorashi, Taylor L. Hughes, and Jennifer Cano, Phys. Rev. Lett. 133, 106601  (2024) \cite{PhysRevLett.133.106601}; Copyright (2025) by the American Physical Society. (b) Top: Superconducting diode effect in a $d_{xy}$-wave AM with intrinsic conventional superconductivity, which originates due to a nonreciprocity in the  supercurrents (blue and green arrows). Bottom: Supercurrent along $x$ normalized by the positive maximum supercurrent $I_{c}^{+}$ as a function of the momentum $q$ for the Fulde–Ferrell state. The blue and green dashed lines mark $I_{c}^{+}$ and $I_{c}^{-}$, respectively.  Reproduced under the terms of the CC-BY 4.0 license and adapted with permission  from the authors of Ref.\,\cite{sim2024}. (c) Top: Magnetoelectric effect in a hybrid system formed by a $d_{x^{2}-y^{2}}$-wave AM and a conventional superconductor. By applying a supercurrent $I_{\rm S}$ induces an out of plane magnetization $M$ (red arrow) that characterizes the magnetoelectric effect.  Bottom: Magnetization along $z$ as a function of the momentum $q_{x}$ at a finite exchange field; the magnetization is represented by the red curve, while the green dashed curve shows the second-order spin susceptibility multiplied by $q_{x}^{2}$. Reprinted figures with permission from Jin-Xin Hu, Oles Matsyshyn, Justin C. W. Song, Phys. Rev. Lett. 134, 026001 (2025)   \cite{JXhu2024}; Copyright (2025) by the American Physical Society.  (d) Top: The thermoelectric effect in a hybrid system formed by two $d_{x^{2}-y^{2}}$-wave AMs  (cyan) at different temperatures $T_{\rm L,R}$ with the left AM coupled to a conventional superconductor (green). A finite temperature difference induces  an electric voltage which then allows the flow of an electric current (orange), signalling the  thermoelectric effect. Bottom: Figure of merit  $ZT$, which     characterizes the power conversion efficiency, as a function of the exchange field in the left AM for distinct temperature differences $T$. Reprinted figures with permission from Pavlo O. Sukhachov, Erik Wegner Hodt, and Jacob Linder, Phys. Rev. B 110, 094508  (2024) \cite{PhysRevB.110.094508}; Copyright (2025) by the American Physical Society.
}
\label{NS_UM_EmergentPhenomena} 
\end{figure*}
 
\subsection{Emergent superconducting phenomena in unconventional magnets}
\label{section23}
One particular aspect of combining unconventional magnetism with superconductivity is that it leads to novel phenomena even when considering conventional superconductors. One of the most fundamental consequences of such a combination was discussed in Subsection \ref{subsection32} and involves  the realization of  novel superconducting pair symmetries, where unconventional magnetism not only affects the spin symmetry but    also the parity in momentum, see  Table \ref{tab:pair_amp_symmetry}. In this regard, spin-triplet Cooper pairs can be realized in AMs and also in $p$-wave magnets \cite{maeda2025classifi}, which, under appropriate material design, can pave the way for superconducting spintronics \cite{eschrig2011spin,linder2015superconducting,Eschrig2015,yang2021boosting,mel2022superconducting,cai2023superconductor}  without net magnetization \cite{Bai_review24,ohldag2024hidden}. Besides spin-triplet Cooper pairs, the interest in unconventional magnets has already led to interesting studies that demonstrate the utility of AMs for even more intriguing superconducting phenomena. Among the most salient examples, of fundamental and applied relevance, we highlight the utilization of $d$-wave altermagnetism for realizing topological superconductivity \cite{PhysRevB.108.184505,CCLiu1,PhysRevB.109.224502,PhysRevLett.133.106601,chatterjee2025}, superconducting diodes \cite{chakrabortyd2024,sim2024,BanerjeePRB24}, magnetoelectric effects \cite{zyuzin2024,JXhu2024}, and thermoelectric effects \cite{PhysRevB.110.094508,chourasia2024}, see also Fig.\,\ref{NS_UM_EmergentPhenomena}. These studies demonstrate the impact of unconventional magnetism for advancing  the search of novel superconducting phenomena in distinct areas, an aspect that will very likely lead to   useful quantum applications in the near future. In this part, we   briefly review the four examples of novel superconducting effects highlighted above.
 
\subsubsection{Topological superconductivity}
Topological superconductivity is a topological phase of matter that is characterized by the formation of boundary states known as Majorana zero modes or simply as Majorana modes \cite{sato2016,sato2017topological,tanaka2024review}. The intrinsic spatial nonlocality and topological nature of Majorana states are useful for designing qubits that are robust against local sources of decoherence \cite{sarma2015majorana,beenakker2019search,aguado2020majorana,MarraJAP2022}, which highlights the enormous potential of topological superconductivity. In terms of ingredients for topological superconductivity, initial proposals showed that such a topological phase needs unconventional spin-triplet $p$-wave superconductivity \cite{sato2016,sato2017topological}, which is however scarce in nature \cite{RevModPhys.63.239}.  By now, it is well-understood that topological superconductivity can be engineered  by using common ingredients, such as spin-orbit coupling, conventional spin-singlet $s$-wave superconductivity, and magnetism \cite{tanaka2024review,Aguadoreview17}.  In this case, the topological superconducting phase hosting Majorana modes emerges after a topological phase transition as the external magnetic field exceeds a critical value \cite{tanaka2024review}. Following these ideas, several experiments  have reported measurements with signatures that seem  consistent  with   Majorana physics \cite{lutchyn2018majorana,zhang2019next,frolov2019quest}, although several studies have also challenged the interpretation of such results as they might come from trivial states and hence not related to Majorana modes \cite{PhysRevB.86.100503,PhysRevB.91.024514,prada2019andreev}.  Moreover, the topological phase transition, which is the key to identify the formation of the topological phase, has not been observed yet; in this regard, a recent study suggested that it can be measured by Raman spectroscopy \cite{mizushima2025Raman}, but still awaits experimental confirmation. Another concern in all these works is that the  magnetism needed to reach the topological superconducting phase,  coming from either a ferromagnet or external magnetic fields, is detrimental for conventional superconductivity e. g., due to Zeeman depairing effects. All these challenges suggest the need to explore other detection protocols and, most importantly, alternative materials with less detrimental effects for superconductivity.

With the advent of AMs, a new door opened in the field of topological superconductivity because such unconventional magnets exhibit a spin-splitting of energy bands like ferromagnets while maintaining zero net magnetization, thus being more friendly towards  superconductivity. In this regard, topological superconductivity was initially addressed as an intrinsic phenomenon   \cite{PhysRevB.108.184505} by taking into account $d_{x^{2}-y^{2}}$-wave altermagnetism and Rashba SOC with an extended attractive Hubbard interaction for spin-singlet and spin-triplet pairing channels. Ref.\,\cite{PhysRevB.108.184505} found that altermagnetism is favorable for spin-triplet $p$-wave superconductivity, which then can realize  topological superconductivity with distinct types of Majorana modes, including chiral and helical propagating Majorana modes as well as localized Majorana corner modes. Interestingly, the   combination of time-reversal symmetry and the four-fold rotation along the $z$-axis, inherent to AMs, enforces the emergence of Majorana corner modes, thus unveiling the potential of altermagnetism for topological superconductivity and Majorana physics. Beyond intrinsic  topological superconductivity, unconventional magnetism has shown to be promising for realizing Majorana physics in hybrid systems formed by unconventional magnets and superconductors. For instance, Majorana corner modes  were predicted in Ref.\,\cite{CCLiu1} by simply coupling a conventional superconductor, a two-dimensional topological insulator and a  AMs, where the Majorana states can be controlled by   the direction of the N\'{e}el vector; see also Ref.\,\cite{PhysRevB.109.224502} for a similar study by combining DIII topological superconductors and $d$-wave AMs. The combination of  AMs with conventional superconductors was more detailed addressed in Ref.\,\cite{PhysRevLett.133.106601}, which predicted the realization of topological superconductivity without net magnetization in one and two dimensions  with Majorana edge modes and Majorana chiral (and corner) modes, respectively. Of particular relevance is the realization of one-dimensional topological superconductivity  by coupling a one-dimensional semiconductor to a $d_{x^{2}-y^{2}}$-wave AM with conventional spin-singlet $s$-wave superconductivity Ref.\,\cite{PhysRevLett.133.106601}, see the top panel of Fig.\,\ref{NS_UM_EmergentPhenomena}(a). The one-dimensional hybrid model belongs to the BDI symmetry class, which has an integer topological invariant $\mathbb{Z}$ and realizes a topological phase when the exchange field when $\sqrt{\Delta^{2}+(t-\mu)^{2}}<J<\sqrt{\Delta^{2}+(t+\mu)^{2}}$, where $J$ is the exchange field,  $\mu$ the chemical potential in a tight-binding model, and $t$ the hopping strength, see Ref.\,\cite{PhysRevLett.133.106601}.  The emergence of Majorana modes is shown in the bottom panel of Fig.\,\ref{NS_UM_EmergentPhenomena}(a), where the energy spectrum of a finite size one-dimensional hybrid system is presented and the Majorana modes appear at zero energy and localized at the system ends. Ref.\,\cite{sun2025pseudoIsingSCpWaveUM} also showed that $p$-wave magnetism can be utilized for realizing topological superconductivity, which, together with Ref.\,\cite{PhysRevLett.133.106601}, clearly demonstrate the tremendous potential of unconventional magnets for realizing Majorana modes.   More recently, $p$-wave magnets   have   been shown to be a crucial ingredient for the emergence of Majorana flat bands when combining them in a hybrid system together with  a conventional superconductor \cite{nagae2025Majo}. The authors showed  that a zero-bias conductance peak signals the anomalous proximity effect \cite{Proximityp,proximityp2,proximityp3,odd1,Asano2013,Ikegaya2016,ahmed2025AnomalousMajoAndreev} and can be an indicator of Majorana flat band physics \cite{nagae2025Majo}, paving the route for another front to explore topological superconductivity. Interestingly, more recently, $p$-wave magnetism was shown to be useful for realizing Majorana  modes \cite{sun2025pseudoIsingSCpWaveUM}.
It is worth noting that the field requires further investigations to uncover the overall potential of unconventional magnetism for designing Majorana modes without net magnetization, which could involve  addressing higher angular momentum unconventional magnets.

\subsubsection{Superconducting diode effect}
\label{section34}
The superconducting diode effect (SDE) has recently attracted  great fundamental and application-oriented interests, largely due to its experimental demonstration in several materials \cite{FAndoNature2020,PhysRevLett.131.027001,liu2024superconducting,li2024field,qi2025high}; see also Ref.\,\cite{nadeem2023superconducting}. Broadly speaking, the diode effect is based on the conduction of current primarily along one direction and shown to be key building blocks for several electronic components \cite{coldren2012diode,mehdi2017thz,semple2017flexible,tokura2018nonreciprocal,kim2020analogue,loganathan2022rapid}. The SDE, on the other hand, is based on  the supercurrents of superconductors which become different  in opposite  directions, see top panel of Fig.\,\ref{NS_UM_EmergentPhenomena}(b). Thus,  superconducting diodes exhibit  functionalities that are superior than those of  diodes in the normal state which suffer from energy losses due to unavoidable finite resistance \cite{braun1875ueber,sze2008semiconductor}. The SDE can therefore enable   energy-saving applications using superconducting circuits in the future \cite{nadeem2023superconducting}. 
It is by now known that both the time-reversal and the inversion symmetry breaking are necessary conditions for the realization of the SDE although not sufficient \cite{tokura2018nonreciprocal,nadeem2023superconducting,nagaosa2024nonreciprocal,he2022phenomenological,DaidoPRL2022,Yuan22}, an idea that has already led to the prediction of the SDE in bulk superconductors \cite{FAndoNature2020,DaidoPRL2022,PhysRevLett.131.027001,PhysRevB.106.104501,he2022phenomenological,lin2022zero,scammell2022theory,banerjee2023enhanced}. For a time-reversal symmetric system, the opposite-going currents have the relation $I^{+}_{\sigma}=- I^{-}_{\bar{\sigma}}$, where the index $\sigma$ and $\bar{\sigma}$ are connected by time-reversal operation. If the system has inversion symmetry, $I^{+}_{\sigma}=- I^{-}_{\sigma}$ is expected. Thus, it is necessary to break both time- and inversion-symmetry to generate superconducting diode effect; however, it is not sufficient to obtain the superconducting diode effect by breaking these two symmetries. Together with breaking time-reversal and inversion symmetries, Ref.\,\cite{he2022phenomenological} showed that breaking $x$-inverting symmetries  are also needed to realize the superconducting diode effect.  The symmetry constraints for the SDE have shown to also result in the emergence of  finite momentum Cooper pairs, like  those in Fulde-Ferrell-Larkin-Ovchinnikov states, as well as different depairing effects between supercurrents with opposite directions \cite{DaidoPRL2022,davydova2022universal}.  To quantify the efficiency of the SDE, and hence measure the nonreciprocity of the system, it is often used the   quality factor defined as,
\begin{equation} 
\label{qualityfactor}
\eta =\frac{I_{c}^{+}-|I_{c}^{-}|}{I_{c}^{+}+|I_{c}^{-}|}\,, 
\end{equation} 
where $I_{c}^{+}$ and $I_{c}^{-}$ are the maximum and minimum supercurrents (or critical currents) in opposite directions. A nonzero value of the quality factor, $\eta\neq0$, indicates the emergence of the SDE; a perfect diode efficiency is reached when $\eta=1$, which would imply having a supercurrent only along one direction.

Most of the recent activity on the SDE with relatively high quality factors has involved systems with Rashba SOC, conventional superconductivity,  and applied magnetic fields  \cite{DaidoPRL2022,he2022phenomenological,PhysRevB.106.104501,nadeem2023superconducting}. While these are indeed relevant, the applied magnetic fields are often detrimental to superconductivity. Motivated by this fact, the SDE has very recently been explored in AMs \cite{sim2024,chakrabortyd2024,BanerjeePRB24}, which, as discussed in Section \ref{section2}, break time-reversal symmetry without an external magnetic field and exhibit zero net magnetization \cite{Bai_review24}, conditions that are friendly with superconductivity. AMs can thus be useful for realizing field-free superconducting diodes. In   Ref.\,,\cite{BanerjeePRB24}, the authors     studied the intrinsic SDE and also explored it in hybrid systems formed by AMs and superconductors, where  an electric field can induce and control the SDE. In the absence of electric fields, Ref.\,\cite{BanerjeePRB24} showed that the SDE can also appear for some  non-centrosymmetric point groups.   Ref.\,\cite{sim2024} addressed the intrinsic SDE in a $d_{xy}$-wave AM with superconductivity due to  attractive interactions as in  the BCS theory. Within the framework of Ginzburg-Landau theory, Ref.\,\cite{sim2024} finds the stabilization of pair density wave states, such as the Fulde-Ferrell and Fulde-Ferrell$^{*}$ states,  which intrinsically break inversion symmetry and carry a finite momentum. In Ref.\,\cite{sim2024}, a Fulde-Ferrell state is characterized by a single superconducting phase, while the Fulde-Ferrell$^{*}$  state by two   phases. Notably,  when exploring the supercurrents, the authors of Ref.\,\cite{sim2024} found a nonreciprocal behavior that  not only distinguishes between Fulde-Ferrell and Fulde-Ferrell$^{*}$ states but also signals the emergence of the SDE in the absence of any external magnetic field. The bottom panel of Fig.\,\ref{NS_UM_EmergentPhenomena}(b) shows the supercurrent for the Fulde-Ferrell  states as a function of momentum, revealing that $I_{c}^{+}$ and $I_{c}^{-}$  are distinct and enable a field-free SDE with efficiency of $\eta=0.5$ \cite{sim2024}. A similar idea was a bit later addressed in Ref.\,\cite{chakrabortyd2024} but with a $d_{x^{2}-y^{2}}$-wave AM and spin-singlet pairing on nearest-neighbor bonds, which, treated within a self-consistent approach, allowed the authors to find a  SDE having perfect efficiencies of $\eta=1$ under an external magnetic field.  The results of the perfect SDE are ascribed to the native finite-momentum superconductivity in AMs \cite{chakrabortyd2024}, which the authors also show to be connected to a topological phase transition in AMs under external magnetism \cite{PhysRevB.109.024404}.  We can therefore conclude that the unique spin properties of AMs are promising for realizing superconducting diodes with a broad range of efficiencies at zero and finite magnetic fields.  Nevertheless, we note that most of the theoretical works about the discussed SDE consider,  in addition to altermagnetism,  spin-orbit couplings which breaks the inversion symmetry.  Even though the search for superconducting diodes in  unconventional magnets is still in its infancy, it is very likely that other types of  unconventional magnets will enrich the device functionality without relying on complicated mechanisms.

\subsubsection{Magnetoelectric effect}
The intrinsic altermagnetic properties have also been shown to be the key for inducing a magnetoelectric effect \cite{zyuzin2024,JXhu2024}, a key mechanism for superconducting spintronics \cite{eschrig2011spin,linder2015superconducting,Eschrig2015,yang2021boosting,mel2022superconducting,cai2023superconductor}. In fact, the magnetoelectric effect  in superconductors occurs when an applied supercurrent   generates a finite spin polarization \cite{PhysRevLett.75.2004,PhysRevB.72.024515,PhysRevB.72.172501}, thus offering a powerful way to induce a net magnetization and a good control of spins without dissipation; see also \cite{he2019spin, PhysRevLett.118.016802,PhysRevResearch.2.012073,PhysRevB.102.214510,PhysRevResearch.3.L032012,PhysRevLett.128.217703} for recent works on the magnetoelectric effect in superconductors. One of the key ingredients for the magnetoelectric effect is the breaking of inversion symmetry, e.g., due to   SOC \cite{PhysRevLett.75.2004,PhysRevB.72.024515,PhysRevB.72.172501}, where spins couple to the motion of Cooper pairs and lead to a net spin polarization tied to the applied supercurrent. These conditions are often satisfied in time-reversal invariant noncentrosymmetric superconductors \cite{yip2014noncentrosymmetric}, where     an applied supercurrent $\bm{I}_s$ generates a magnetization $\bm{M}$ with a linear relationship between them, $\bm{M}\propto \bm{c}\times \bm{I}_s$, where $\bm{c}$ indicates the polar axis of the superconductor with Rashba SOC. 

With the discovery of AMs, the exploration of the superconducting magnetoelectric effect has experienced another impetus since the  altermagnetism offers unique spin-dependent properties without relying on SOC and at zero net magnetization.
This allowed the study of  the superconducting magnetoelectric effect   in systems formed by centrosymmetric superconductors and   AMs~\cite{zyuzin2024,JXhu2024}; these studies  predict a nonlinear superconducting magnetoelectric effect where an out of plane magnetization  induced due to an applied supercurrent in an AM-superconductor hybrid system, see top panel of Fig.\,\ref{NS_UM_EmergentPhenomena}(c). To be more precise, Ref.\,\cite{JXhu2024} found that the emergent net magnetization is a second-order nonlinear effect given by $\delta M^{(2)}_{c}=\chi^{c}_{ab}q_{a}q_{b}$, where $a,b=x,y$ is the supercurrent direction, $c=x,y,z$ the magnetization direction, and $q_{a,b}$ the Cooper pair momentum due to the applied supercurrent $\bm{I}_{s}\propto \bm{q}$. Moreover,  the coefficient $\chi^{c}_{ab}$ is the second-order spin susceptibility and is nonzero only when time-reversal symmetry is broken due to altermagnetism \cite{JXhu2024}. To visualize the nonlinear behavior of the magnetization, the bottom panel of Fig.\,\ref{NS_UM_EmergentPhenomena}(c) shows the results of Ref.\,\cite{JXhu2024} for an applied supercurrent along $x$ as a function of momentum. The authors of Ref.\,\cite{JXhu2024}  also argue that such a second-order nonlinear superconducting magnetoelectric effect can be the dominant contribution in the out-of-plane magnetization due to the large spin splitting of AMs.
Thus, unlike noncentrosymmetric superconductors with SOC, AMs allow for a nonlinear magnetoelectric effect in centrosymmetric   superconductors that can be useful for a nondissipative control of  magnetization. We expect that the predictions of Refs.\,\cite{zyuzin2024,JXhu2024} will motivate further studies to fully exploit the distinct parities of  unconventional magnets towards an efficient approach for superconducting spintronics.

\subsubsection{Thermoelectric effect}
Another line of research that has recently benefited from the interplay between altermagnetism and superconductivity is thermoelectricity, explored by combining a hybrid system formed by $d$-wave AMs and a conventional superconductor in Ref.\, \cite{PhysRevB.110.094508}, see top panel of Fig.\,\ref{NS_UM_EmergentPhenomena}(d). Under generic circumstances, thermoelectricity is based on   the direct conversion of temperature gradients into   electrical quantities, a phenomenon  known as the thermoelectric effect \cite{RevModPhys.26.237,galperin1974thermoelectric,ginzburg1978thermoelectric,macdonald2006thermoelectricity}; see also Refs.\,
\cite{falco1981thermoelectric,van1982thermoelectric,PhysRevB.53.6605,PhysRevB.65.064531,majumdar2004thermoelectricity,sothmann2014thermoelectric,PhysRevB.98.161408,PhysRevB.99.045428,PhysRevResearch.2.022019,yang2023role,artini2023roadmap,liu2023interplay}. One of the key requirements for inducing a thermoelectric effect   is the breakdown of   particle-hole symmetry, which, although intrinsic in superconductors, can be broken   by magnetic fields. A large body of works   considered hybrids systems between superconductors and ferromagnets, which managed to produce high efficiencies with  figures of merit reaching even $ZT\approx40$ in hybrid systems based on two spin-split superconductors \cite{PhysRevB.93.224509}, see also Ref.\,\cite{PhysRevLett.112.057001}. These findings put  ferromagnet-superconductor hybrid systems as promising for energy harvesting and cooling of electronics  at cryogenic temperatures. Despite the notable properties, ferromagnets introduce undesirable  stray magnetic fields  that are detrimental when combining multiple structures \cite{PhysRevLett.81.2344,yang2004domain,PhysRevLett.96.247003,PhysRevLett.99.227001,rouco2017competition,liu2019semiconductor,paschoa2020role}, thus limiting the scalability of thermoelectric devices based on superconductors. The drawbacks of ferromagnets clearly point towards the need of finding other materials without stray fields for the thermoelectric effect in superconductors.

Ref.\,\cite{PhysRevB.110.094508} showed that the above issue can be resolved in superconducting junctions with $d_{x^{2}-y^{2}}$-wave AMs, as schematically shown in Fig.\,\ref{NS_UM_EmergentPhenomena}(d). Unlike ferromagnets, AMs exhibit a momentum dependent spin splitting of energy bands without any net magnetization,  which was shown to be useful for stray-field-free devices with high storage densities. The momentum dependent spin splitting of AMs enables the breaking of particle-hole symmetry for each spin, while coupling to another AM renders the interface spin active, mechanisms  that permit to realize a thermoelectric response without any magnetic field. One of the main findings of Ref.\,\cite{PhysRevB.110.094508} is that the AM-superconductor hybrid junction can reach a figure of merit with reasonable efficiencies, as seen in the bottom panel of Fig.\,\ref{NS_UM_EmergentPhenomena}(d). The authors further showed that the behavior of the figure of merit, as well as the Seebeck coefficient, is similar to that a ferromagnet-superconductor system when temperature is varied \cite{PhysRevB.110.094508}. This study thus opens new possibilities for exploring thermoelectricity in superconductors with unconventional magnetism, which could even  further offer new ways for caloritronics  in superconductors \cite{Mart_nez_P_rez_2014,PhysRevB.93.224509,PhysRevB.93.224509,fornieri2017towards,hwang2020phase}.

\section{Josephson junctions with unconventional magnets}
\label{section3}
In the previous section, we have delved into the understanding of the interplay between superconductivity with unconventional magnetism and also discussed the emergent phenomena when  unconventional magnets are placed in contact with a superconductor. In this part, we focus on  superconducting junctions  formed by two superconductors   linked by an  unconventional magnet, also known as Josephson junctions (JJs). In these junctions, Andreev reflections at both superconducting interfaces lead to the formation of Andreev bound states (ABSs) when there is a finite phase difference between the pair potentials of the superconductors \cite{sauls2018andreev,asano2021andreev}. The phase-dependent ABSs due to Andreev processes are responsible for the transfer of Cooper pairs between superconductors without any applied voltage \cite{kulik1969macroscopic,ishii1970josephson,ishii1972thermodynamical,bardeen1972josephson,svidzinsky1973concerning,kulik1975,RevModPhys.51.101,furusaki1991dc,furusaki1990unified,FurusakiPRB1991,PhysRevB.45.10563,PhysRevLett.66.3056,Beenakker:92,Furusaki_1999}, which  signals the flow of a dissipationless supercurrent  also known as the Josephson effect \cite{Josephson}. Due to the superconducting nature, the ABSs and Josephson effect have been shown to be extremely useful for identifying the type of emergent superconductivity \cite{kashiwaya2000,PhysRevB.64.224515,PhysRevLett.96.097007,RevModPhys.76.411,sauls2018andreev,mizushima2018,seoane2024subgap} and also relevant for designing superconducting devices \cite{devoret2005implementing,acin2018quantum,krantz2019quantum,aguado2020perspective,benito2020hybrid,aguado2020majorana,kjaergaard2020superconducting,PRXQuantum.2.040204,siddiqi2021engineering}. In this section, we address the emergence of ABSs in JJs formed by superconductors and  unconventional magnets and how the distinct types of magnetic order produce measurable signatures in the phase-biased Josephson effect. 
 
 \begin{figure}[!t]
\centering
\includegraphics[width=0.45\textwidth]{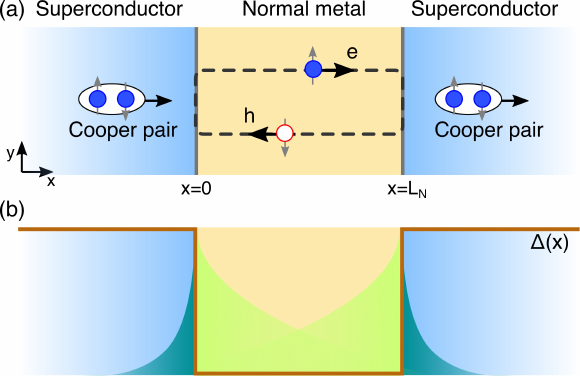}
\caption{Sketch of a Josephson junction along $x$ formed by two superconductors connected by a normal metal (N)  of length $L_{\rm N}$, with interfaces located at $x=0$ and $x=L_{\rm N}$. (a)  Within the superconducting gap, a right moving electron in N is Andreev reflected   as a hole at the right interface picking up the phase of the right superconductor and transferring a Cooper pair as well; the  reflected hole   travels across N and is then Andreev reflected as an electron at the left interface which picks up the phase and a Cooper pair from the left superconductor. These processes continue and lead  to the formation of an Andreev bound state  within the superconducting gap in N, as well as permitting the transfer of Cooper pairs between superconductors. (b) Due to the proximity effect, the normal region acquires superconducting correlations (light green), while the superconducting correlations in the superconductors acquire properties of the normal region (blue-green). Here, the amplitude of the pair potential  $\Delta(x)$ in the junction is indicated in brown color.}
\label{ABS_sketch} 
\end{figure}

\subsection{Modelling Josephson junctions}
To capture the fundamental aspects of phase-biased JJs, such as ABSs and the Josephson effect, there are two complementary approaches that are useful for pedagogical purposes. For this, it is important to specify that the JJs we consider here are formed along $x$ by two conventional spin-singlet $s$-wave superconductors coupled via an unconventional magnet, with a pair potential profile given by
\begin{equation}
\label{SNSDelta}
\hat{\Delta}(x)=
\begin{cases}
\Delta\,{\rm e}^{i\varphi_{\rm L}}\,,\quad x\in {\rm S_L}\\
0\,,\quad \quad\quad\, x\in {\rm UM}\,,\\
\Delta\,{\rm e}^{i\varphi_{\rm R}}\,,\quad x\in {\rm S_R}\,,\\
\end{cases}
\end{equation} 
where $\Delta$ is the  pair potential amplitude assumed to be the finite and the same in both left (L) and right (R) superconductors, while no superconductivity is taken in the  unconventional magnet (UM).  Moreover, $\varphi_{\rm L,R}$ is the  phase in the pair potentials of the  left/right superconductor, whose phase difference $\varphi=\varphi_{\rm R}-\varphi_{\rm L}$ plays an important role in JJs as we will see and consider in the following subsections of this part.  For now, we can specify the length of the  unconventional magnet to be $L_{\rm N}$, a quantity that is often used to distinguish between short ($L_{\rm N}\ll\xi_{\rm S}$) and long ($L_{\rm N}\gg\xi_{\rm S}$) JJs, with $\xi_{\rm S}$ being the superconducting coherence length \cite{Beenakker:92}, see also Refs.\,\cite{altshuler1987mesoscopic,PhysRevLett.67.132}.  We can also model the interfaces   by delta-like potentials $U(x)=V_{\rm L}\delta(x)+V_{\rm R}\delta(x-L_{\rm N})$. Note that, while the pair potential needs to be in principle obtained   self-consistently \cite{schmidt2013physics,zagoskin,tinkham2004introduction}, assuming the profile given by Eq.\,(\ref{SNSDelta}) is reasonable and  proven to help understand the formation of ABSs and the Josephson effect \cite{RevModPhys.76.411,sauls2018andreev,TextTanaka2021}. In a very similar way one can model the pair potential of unconventional superconductors \cite{tanaka2024review,TextTanaka2021}. In relation to the description of the superconductors and  unconventional magnets, we can model them by using either  continuum models or their tight-binding representations, as we explain next.

\subsubsection{Continuum model}
In this situation, the superconductors and the  unconventional magnets are described by the continuum models given by Eqs.\,(\ref{eq:mp}) and (\ref{eq:am}). Then, the properties of JJs can be explored via the scattering approach in the same way as we discussed in Subsection \ref{section23}. For simplicity, we   assume that the interfaces of the JJ along $x$ are located at $x=0$ and $x=L_{\rm N}$, with $L_{\rm N}$ being the length of the  unconventional magnet, while momentum $k_{y}$ is still a good quantum number along the transverse direction. Taking a BdG Hamiltonian for each section of the JJ, with the pair potential given by Eq.\,(\ref{SNSDelta}), one constructs scattering states incident on the interfaces, which are then appropriately matched at such interfaces and scattering coefficients are found. These scattering coefficients can  then be used to calculate the ABSs, Green's functions, and supercurrents; for more details see e. g.,\,Refs.\,\cite{furusaki1991dc,FurusakiPRB1991,Beenakker:92,PhysRevB.45.10563,Furusaki_1999,kashiwaya2000,lu2018study,PhysRevB.100.115433,Cayao2020odd,tanaka2024review}. See also Refs.\,\cite{McMillan1968}. Within the scattering formalism, the supercurrent can be obtained as \cite{furusaki1991dc,Furusaki_1999}
\begin{equation}
\label{EqIphi}
\begin{split}
I(\varphi)&=\int_{-k_{\rm F}}^{k_{\rm F}} \bar{I}(\varphi,k_{y})dk_{y}\,,\\
\bar{I}(\varphi,k_{y})&=\frac{e\Delta}{2\hbar\beta}\sum_{\omega_{n},\sigma}\bigg[\frac{q^{n}_{e}+q^{n}_{h}}{\Omega_{n}}\bigg]\bigg[\frac{a_{e\sigma}(\varphi)}{q^{n}_{e}}-\frac{a_{h\sigma}(\varphi)}{q^{n}_{h}}\bigg]
\end{split}
\end{equation}
where $\beta=1/(\kappa_{\rm B}T)$, with $\kappa_{\rm B}$ being the Boltzmann constant and $T$ the temperature, $\Omega_{n}=\sqrt{\omega_{n}^{2}+\Delta^{2}}$, $\omega_{n}=\pi\kappa_{\rm B}T$, $q_{e(h)}^{n}=k_{\rm F}\sqrt{1\pm i(\Omega_{n}/\mu)-(k_{y}/k_{\rm F})^{2}}$, while $a_{e\sigma}$ and  $a_{h\sigma}$ represent  four types of Andreev reflection coefficients from by injections of electron-like and hole-like quasiparticles with $\sigma=\uparrow,\downarrow$   from the left superconductor. As explained above and in Subsection \ref{section23}, the Andreev reflection coefficients are obtained by matching the scattering states at the interfaces, see Ref.\,\cite{PhysRevLett.133.226002} for the Andreev coefficients in JJs based on AMs. Eq.\,(\ref{EqIphi}) is known as the Furusaki-Tsukada formula \cite{furusaki1991dc,Furusaki_1999} and nicely reflects the direct relationship between the amplitude of Andreev processes with the Josephson current $I(\varphi)$. To evaluate $I(\varphi)$ in Eq.\,(\ref{EqIphi}), we can then  parametrized the transverse component of the wave vector $k_{y}=k_{\rm F}{\rm sin}(\theta)$ by the angle of incidence $\theta\in[-\pi/2,\pi/2]$ on the leftmost  region, such that the integration is carried out over the angle $\theta$. We would also like to point out  that the scattering coefficients form the so-called S-matrix, and that is why their poles define the formation of ABSs \cite{Beenakker:92,Furusaki_1999,RevModPhys.69.731}. Thus, by finding the Andreev coefficient   $a_{e\sigma}$, we can also calculate the ABSs within this scattering approach. We lastly note that it is even possible to take into account the finite size of the superconductors under appropriate construction of the required scattering states accompanied by their boundary conditions, see Refs.\,\cite{PhysRevB.96.155426,PhysRevB.106.L100502,PhysRevResearch.6.L012062,tanaka2024review}.

\subsubsection{Tight-binding model}
When multiple degrees of freedom are present, the continuum approach to treat JJs becomes challenging. An alternative way is to treat the system in real space within 
a tight-binding approach, where the JJ Hamiltonian is discretized along $x$ and written as
\begin{equation}
\label{EqJJ}
H_{\rm JJ}(k_{y})=H_{\rm UM}+H_{\rm S_L}+H_{\rm S_R}+H_{\rm T}\,,
\end{equation}
where each element on the right-hand side is a matrix in real space and their dependence on the transverse momentum $k_{y}$ is omitted for simplicity. In Eq.\,(\ref{EqJJ}), $H_{\rm UM}$ describes the  unconventional magnet, while $H_{\rm L(R)}$ describes the left (right) superconductor and $H_{\rm T}$ the coupling between  the unconventional magnet (UM) and superconductors. We consider that the superconductors are of length $L_{\rm S}=Ma$, with $M$ the number of sites in each superconductor, while $L_{\rm N}=Na$ the length of the unconventional magnet with $N$ being the number of lattice sites; $a$ is the lattice spacing, taken to be the same in each region. The elements of $H_{\rm JJ}(k_{y})$ are given by
\begin{equation}
\begin{split}
H_{\rm UM}(k_{y})&=\sum_{j}c^{\dagger}_{j}\hat{u}_{\rm N}(k_{y})c_{j}+c^{\dagger}_{j+1}\hat{t}_{\rm N}(k_{y})c_{j}+{\rm h. c.}\,,\\
H_{\rm S_{\alpha}}(k_{y})&=\sum_{j}c^{\dagger}_{\alpha_j}\hat{u}_{\rm S}(k_{y})c_{\alpha_j}+c^{\dagger}_{\alpha_{j+1}}\hat{t}_{\rm S}(k_{y})c_{\alpha_j}+{\rm h. c.}\,,\\
&+\sum_{j}\Delta{\rm e}^{i\varphi_{\rm \alpha}}[c_{\alpha_{j\uparrow}}c_{\alpha_{j\downarrow}}-c_{\alpha_{j\downarrow}}c_{\alpha_{j\uparrow}}]+{\rm h. c.}\,,\\
H_{\rm T}(k_{y})&=\hat{V}_{\rm L}c^{\dagger}_{1}c_{\rm L_{M}}+
\hat{V}_{\rm R}c^{\dagger}_{\rm R_1}c_{\rm N}+{\rm h. c.}
\end{split}
\end{equation}
where $c_{j}=(c_{j\uparrow},c_{j\downarrow})^{\rm T}$, $c_{\alpha_j}=(c_{\alpha_{j\uparrow}},c_{\alpha_{j\downarrow}})^{\rm T}$, while $c_{j\uparrow}$   annihilates an electronic states at site $j$ and spin $\sigma$ in the  unconventional magnet; similarly, $c_{\alpha_{j\uparrow}}$   annihilates an electronic states at site $j$ and spin $\sigma$ in the superconductor $\alpha={\rm L/R}$, and $\Delta$ is the amplitude of the spin-singlet $s$-wave pair potential. 
\begin{equation}
\label{HNMElemen}
\begin{split}
    \hat{u}_\mathrm{N}&=[-\mu+4t_1-2t_1\cos{k_y}]{\sigma}_{0}\\
    &+[-t_{x^2-y^2}\cos{k_y}+t_{y}\sin{k_y}]\sigma_z\,,,\\
    \hat{t}_\mathrm{N}&=-t_1{\sigma}_{0}+\Big[-it_{xy}\sin{k_y}+\frac{t_{x^2-y^2}}{2}-i\frac{t_{x}}{2}\Big]{\sigma}_{z}\,,\\
\hat{u}_\mathrm{S}&=[-\mu+4t_1-2t_1\cos{k_y}]{\sigma}_{0},\\
    \hat{t}_\mathrm{S}&=-t_1{\sigma}_{0}\,,\\
    \hat{V}_\mathrm{L,R}&=-t_\mathrm{int}t_{1}\sigma_{0}\,,
    \end{split}
\end{equation}%
where $\mu$ and $t_1$ are the chemical potential and the hopping integral, $\sigma_{0,x,y,z}$ is the Pauli matrices in spin space and $t_\mathrm{int}\in[0,1]$ is the parameter that controls the tunneling amplitude at the interface. Having the tight-binding Hamiltonian given by Eq.\,(\ref{EqJJ}), we can calculate the spectrum $E_{n}$ by diagonalizing $H_{\rm JJ}(k_{y})$. Moreover, this information can be used to find the free energy $F$, which then allows us to calculate the  zero temperature Josephson current  as
\begin{equation}
\label{IphiTB}
I(\varphi)=\frac{e}{\hbar}\int dk_{y}  \frac{dF(\varphi,k_{y})}{d\varphi}
\end{equation}
where $F(\varphi,k_{y})=\sum_{n<0}E_{n}(\varphi,k_{y})$.

While Eq.\,(\ref{IphiTB}) is useful for finite length systems, it gets computationally expensive when the systems are large. In that case, we can opt e. g., for incorporating the superconductors via self-energies within a recursive Green's function approach, see Refs.\,\cite{PhysRevB.105.094502,fukaya2024x}.  Within this approach, the Green's function in the middle two sites of the  unconventional magnet can be obtained, which then allows us to calculate the LDOS and Josephson current as \cite{KawaiPRB2017,FukayaPRB2020,Fukayanpj2022}
\begin{equation}
\label{IphirhoSNS}
 \begin{split}
 \rho(\varphi)&=-\frac{1}{\pi}\int{\rm Im}{\rm Tr'}\hat{G}_{00}(\varphi,k_{y},E)dk_{y}\,,\\
 I(\varphi)&=\frac{ie}{\hbar\beta}\int {\rm Tr'}\sum_{\omega_n}[\tilde{t}^{\dagger}_{\rm N}\hat{G}_{01}(\varphi,k_{y},i\omega_{n})\\
 &-\tilde{t}_{\rm N}\hat{G}_{10}(\varphi,k_{y},i\omega_{n})]dk_{y}\,,
 \end{split}
\end{equation}
where $\beta=1/(\kappa_{\rm B}T)$,  $\tilde{t}_{\rm N}$ is the hopping matrix in the UM in Nambu space obtained by using $\hat{t}_{\rm N}$ in Eq.\,(\ref{HNMElemen}), and $\hat{G}_{ij}$ is  the Green’s function in Nambu space in the two middle sites ($i,j=0,1$) of the  unconventional magnet, {\rm Tr'} means that the trace is taken only over the electron subspace, and the integration is carried out within $[-\pi,\pi]$. Moreover,  in the first equation we have carried out the analytic continuation to real energies $i\omega_{n}\rightarrow E+i\delta$. We also point out that having the   Green's function in the middle sites of the  unconventional magnet is also useful for exploring  the ABSs, which are obtained from the poles of the Green's function by solving ${\rm det}[\hat{G}^{-1}_{00}(k_{y},\omega)]=0$ for $\omega$, see also Ref.\,\cite{zagoskin}. In what follows, we discuss the formation of ABSs, the phase-dependent supercurrents, and the emergent phenomena due to the interplay between unconventional magnetism and the Josephson effect that was reported so far.
 
\begin{figure*}[!t]
\centering
\includegraphics[width=0.99\textwidth]{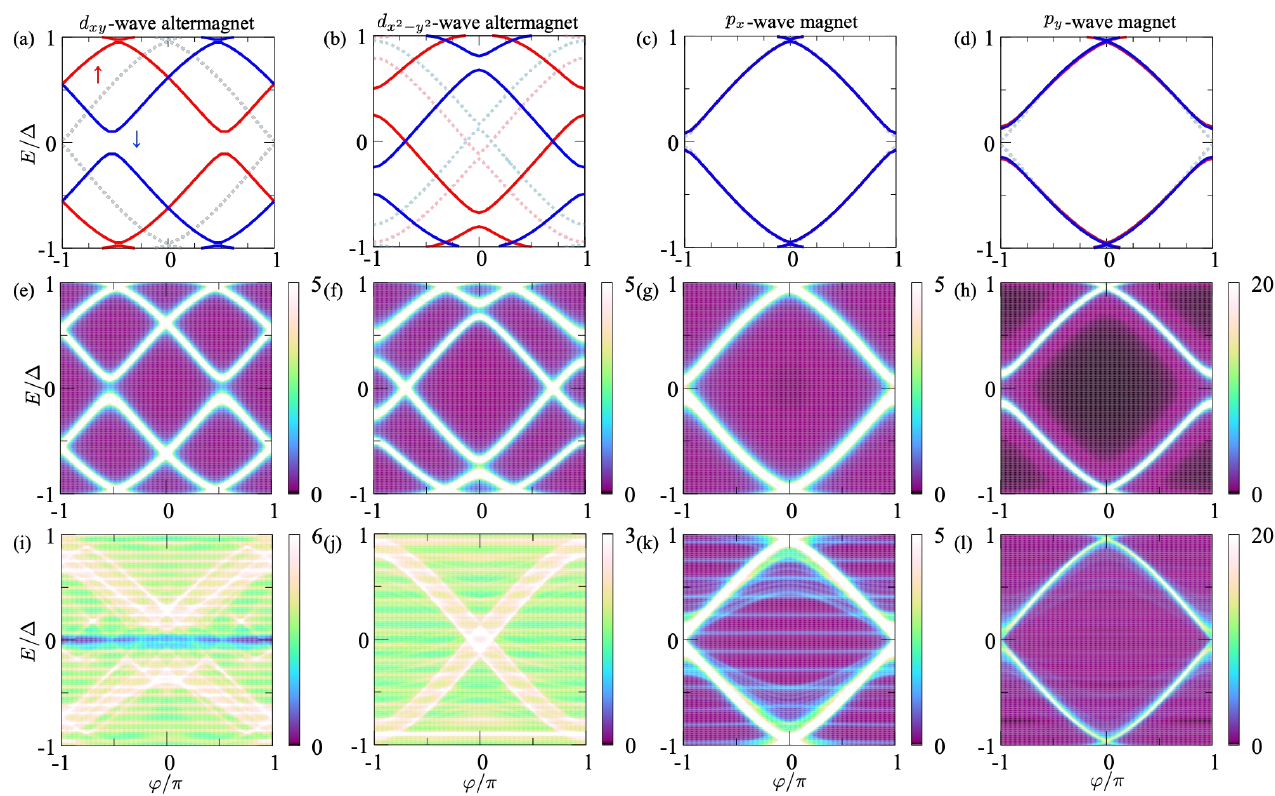}
\caption{Andreev bound states (ABSs) and their spectral signatures via  local density of states (LDOS) in Josephson junctions based on  unconventional magnets obtained from a tight-binding model. (a-d) ABSs as a function of  the superconducting phase difference $\varphi$, obtained from the poles of the Green's function in the middle of the  unconventional magnet at  $k_{y}=0.2\pi$. Light red and light blue curves represent the ABSs at $k_{y}=0$. (e-h) Electronic LDOS obtained from the Green's function used in (a-d) as a function of energy and phase difference at $k_{y}=0.2\pi$ in the middle of the  unconventional magnet ($x=14a$).  (i-l) Electronic LDOS for (e-h) integrated over the transverse momentum $k_{y}$ in the middle of the unconventional magnet at $x=14a$  as a function of energy and phase difference.  The LDOS is normalized by its value in the normal state at $E=0$.  Tight-binding parameters: $a=1$, $L_{\rm N}=30a$, $t_{x(y)}=0.4t$, $t_{xy}=0.42t$, $t_{x^{2}-y^{2}}=0.4t$, $\Delta=0.02t$, $t_{\rm int}=0.95$, $\mu=1.5t$, $\delta=0.01\Delta$. Here, $\tau$ controls the transparency across the  unconventional magnet. Adapted from Ref.\,\cite{fukaya2024x}. }
\label{ABS_JJs} 
\end{figure*}

\subsection{Formation of Andreev bound states and their spectral signatures}
\label{subsection4b}
To understand the impact of unconventional magnetism on the ABSs, we first remind their formation in JJs formed by semi-infinite conventional spin-singlet $s$-wave superconductors and a normal metal \cite{kulik1969macroscopic}, see Fig.\,\ref{ABS_sketch}(a). For a short normal metal, there is a pair of spin-degenerate ABSs that emerge within the pair potential $\Delta$  and given by $E_{\pm}(\varphi)=\pm\Delta\sqrt{1-\sigma_{\rm N}{\rm sin}^{2}(\varphi/2)}$, see Ref.\,\cite{kulik1969macroscopic}; here,  $\varphi$ is the superconducting phase difference and $\sigma_{\rm N}$ the normal transparency. The first feature to stress that the ABSs are $2\pi$-periodic, with $E_{\pm}(\varphi=0)=\pm\Delta$ and $E_{\pm}(\varphi=\pi)=\pm\Delta\sqrt{1-\sigma_{\rm N}}$.  Moreover,  in the full transparent regime $\sigma_{\rm N}=1$, the ABSs are $E_{\pm}=\pm\Delta{\rm cos}(\varphi/2)$, implying that the subgap spectrum is gapless at $\varphi=\pi$ since the ABSs reach zero energy. In the tunneling regime $\sigma_{\rm N}=0$,  the ABSs are independent of the phase  and are located near the gap edges $E_{\pm}=\pm\Delta$. For a nonzero normal transmission below full transparencies, the subgap spectrum is always gapped; for instance, at $\varphi=\pi$, the ABSs are given by $E_{\pm}(\varphi=\pi)=\pm\Delta\sqrt{1-\sigma_{\rm N}}$, hence the subgap spectrum exhibits an energy gap. When the normal metal becomes longer, more energy levels appear at subgap energies and the cosine-like dispersion with the phase changes to a linear dispersion in the long junction regime \cite{kulik1969macroscopic,cayao2018andreev,zagoskin}.

The features discussed in the previous paragraph are well-known and will help understand how  the behaviour of the ABSs is affected when considering  unconventional magnets instead of a normal metal connecting semi-infinite spin-singlet $s$-wave superconductors in a JJ, a  question   addressed in Ref.\,\cite{fukaya2024x}. The ABSs of a short JJ as a function of the superconducting phase difference are shown in  Fig.\,\ref{ABS_JJs}(a-d) at fixed transverse momentum $k_{y}$, obtained from the poles of $\hat{G}_{00}$ in Eq.\,(\ref{IphirhoSNS}). The first property to highlight is that, under general conditions, the JJ hosts four ABSs that are spin split and strongly dependent on the transverse momentum $k_{y}$ but with a distinct phase dependence   depending on the type of unconventional magnet. For a JJ with a $d_{xy}$-wave AM, the ABSs are spin split only when $k_{y}\neq0$, with avoided crossings (gap) around zero energy between positive and negative ABSs of the same spin, very likely due to a normal transmission effect across the  unconventional magnet, see  Fig.\,\ref{ABS_JJs}(a). The authors of Ref.\,\cite{fukaya2024x} also showed that the ABSs profile changes when $k_{y}$ takes other values, where the energy gaps due to the avoided crossings reduce and even reach zero energy at distinct $\varphi$, including $\varphi=0$; this signals   a phase shift of the ABSs due to   $d_{xy}$-wave altermagnetism. For JJs with a $d_{x^{2}-y^{2}}$-wave AM, the ABSs also exhibit a spin splitting as in the $d_{xy}$-wave AM but their phase-dependence  is slightly different, see Fig.\,\ref{ABS_JJs}(b). 
 For instance, even at $k_{y}=0$, the ABSs are  split in spin since there already is a finite strength of the $d_{x^{2}-y^{2}}$-wave   field at this transverse momentum, see  Eqs.\,(\ref{MTB}) and Ref.\,\cite{fukaya2024x}. In this case, ABSs with distinct spins develop zero-energy crossings at distinct values of $\varphi$, an effect that   depends on $k_{y}$ and also occurs at $\varphi=0$, unveiling also a phase shift due to $d_{x^{2}-y^{2}}$-wave altermagnetism in JJs.  It is interesting to note that the gapless ABSs in $d_{x^{2}-y^{2}}$-wave AM based JJs are protected by a combined mirror and time-reversal symmetry. This is because     the JJ Hamiltonian can be decomposed into two by exploiting  the combination of mirror and time-reversal symmetries. Because the coupling between two decomposed Hamiltonian is absent, the Andreev bound states do not develop any energy gap at zero energy and they are hence gapless in Fig.~\ref{ABS_JJs} (b). These features of the ABSs were also reported in Refs.\,\cite{Beenakker23,alipourzadeh2025ABSJE}.  Thus,   ABSs in JJs with $d$-wave AMs exhibit  a distinct phase dependence with respect to JJs with superconductors   linked by a normal metal.
 
 When it comes to JJs with  $p$-wave magnets, the ABSs behave slightly similar to the ABSs in JJs with a normal metal but also exhibit some differences, see Fig.\,\ref{ABS_JJs}(c,d). In both types of  $p$-wave magnets the ABSs develop a cosine-like profile with a finite energy gap at $n\pi$, with $n\in\mathbb{Z}$, but the spin splitting is distinct; only the ABSs in JJs with a $p_{y}$-wave  unconventional magnet seem to form a visible   spin splitting since $k_{y}$ here enters as a magnetic field, see Fig.\,\ref{ABS_JJs}(c,d). Ref.\,\cite{fukaya2024x} shows that other values than the chosen in Fig.\,\ref{ABS_JJs}(d) give a larger spin splitting. With this, we can say that the ABSs in JJs formed by conventional superconductors and  $p$-wave magnets are different from what we discussed for JJs with AMs. At the same time, the ABSs in JJs with AMs  and  $p$-wave  magnets exhibit  features that are distinct from JJs with normal metals, hence uncovering the impact of unconventional magnetism on the formation of ABSs.
 
  \begin{figure*}[!t]
\centering
\includegraphics[width=0.95\textwidth]{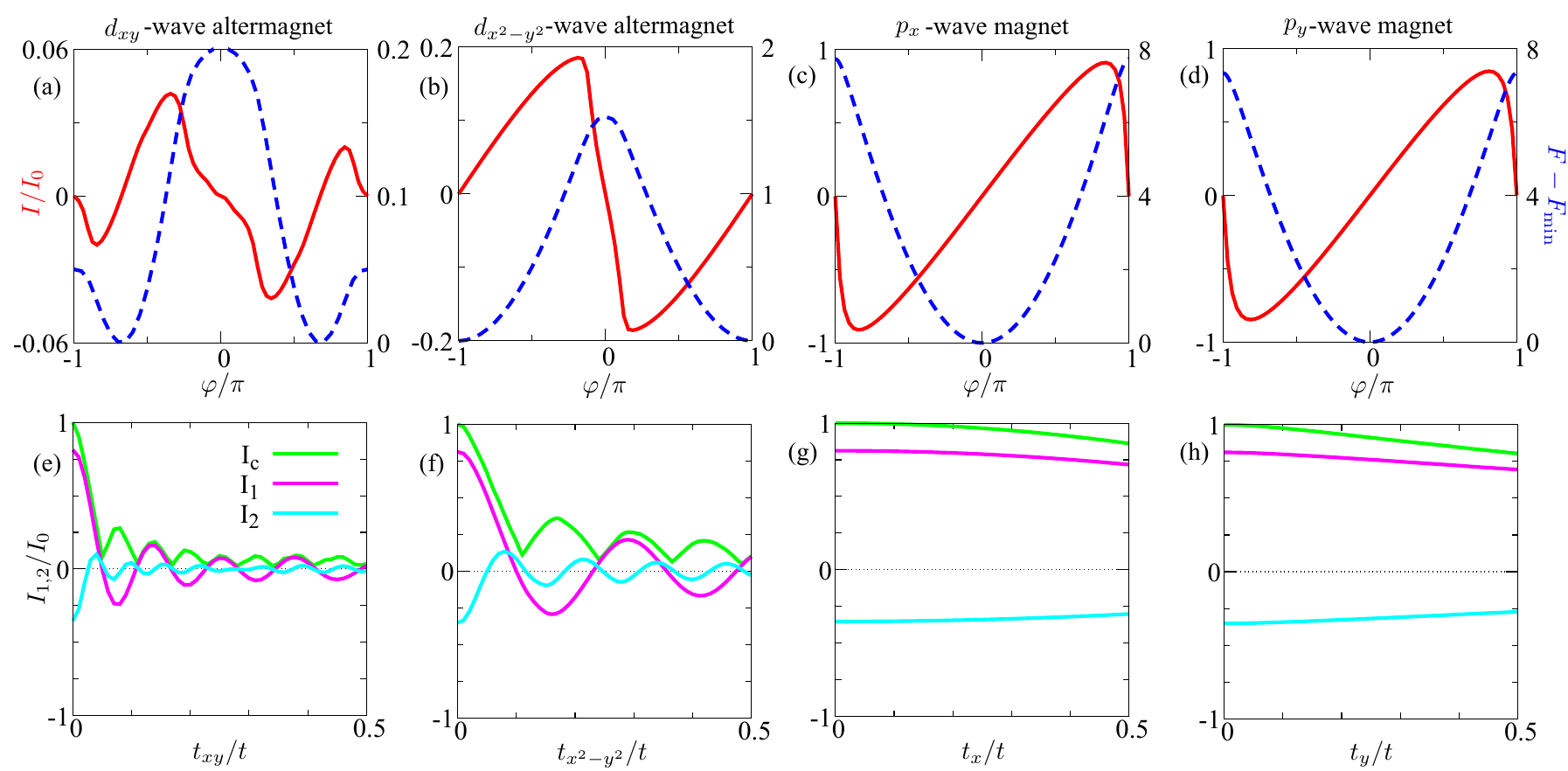}\\
\caption{Josephson effect in  Josephson junctions (JJs) formed by conventional spin-singlet $s$-wave superconductors and  unconventional magnets. (a-d) Current-phase curves  indicated by red curves for JJs with $d_{xy}$- and $d_{x^{2}-y^{2}}$-wave altermagnets as well as for JJs with  a $p_{x}$-wave  magnet, with a $p_{y}$-wave  magnet.  The blue curves depict the respective free energy. The strength of the exchange field in (a-d) is the same as in  Fig.\,\ref{ABS_JJs}(a-d), and $I_{0}$ is the critical current without magnetic order.
(e-f) Critical current $I_{\rm c}$ and  first and second harmonics $I_{1,2}$ of the Josephson current $I(\varphi)$ as a function of the exchange field in the respective unconventional magnet. Here, $I_{c}$, $I_{1}$, and $I_{2}$ are shown by green, magenta, and cyan curves, respectively.  Parameters: $T=0.025T_{c}$, $T_{c}=0.01t$; rest of parameters are same as in Fig.\,\ref{ABS_JJs}.  Adapted from \cite{fukaya2024x}. 
}
\label{IphiJE} 
\end{figure*}

In terms of detection schemes of ABSs, one way is to explore their spectral signatures e. g., in the LDOS of the  unconventional magnet, which can be accessed     via conductance by tunnel coupling a normal probe to the neighbourhood of the junction \cite{datta1997electronic}. Following this idea, Ref.\,\cite{fukaya2024x} obtained the LDOS by using the first expression of Eqs.\,(\ref{IphirhoSNS}), which we present in Fig.\,\ref{ABS_JJs}(e-h) for the LDOS at  fixed $k_{y}$ and in Fig.\,\ref{ABS_JJs}(i-l) for the LDOS integrated in $k_{y}$. In general, the high intensity regions at fixed transverse momentum in the LDOS of Fig.\,\ref{ABS_JJs}(e-h) clearly correspond to the formation of ABSs in Fig.\,\ref{ABS_JJs}(a-d) for $d$-  and $p$-wave  magnet, where the visibility of the     gaps around zero energy depends on the broadening which is inevitable under realistic conditions. At fixed $ky$, the ABSs reaching zero energy are very likely to produce zero-energy LDOS peaks, see e. .g., Fig.\,\ref{ABS_JJs}(f,g). The  signatures of the ABSs become more intriguing when looking at the integrated LDOS over $k_{y}$, where, due to the strong dependence of the ABSs on $k_{y}$, the LDOS features are not simple to analyze, see Fig.\,\ref{ABS_JJs}(i-l). Nevertheless,  it is still possible to identify a rather soft gap around zero energy in JJs with a $d_{xy}$-wave AM [Fig.\,\ref{ABS_JJs}(i)], which Ref.\,\cite{fukaya2024x} showed it  to have a V-shape. In contrast, JJs with $d_{x^{2}-y^{2}}$-wave AMs exhibit    subgap structure that develops large values near to $\varphi=0$, see Fig.\,\ref{ABS_JJs}(j). In the case of JJs with $p$-wave  magnet in Fig.\,\ref{ABS_JJs}(k,l), the ABSs are less challenging to identify, albeit the gap around zero energy does not survive the transverse momentum integration.  It is also worth noting that the horizontal branches in Fig.\,\ref{ABS_JJs}(k) correspond to dispersionless traces of ABSs appearing at distinct $k_{y}$ but do not affect the overall visibility of the phase-dependent branches. The discussed spectral features in AMs and  $p$-wave magnets correspond to a fixed exchange field; when such a field increases,   Ref.\,\cite{fukaya2024x} found that the subgap spectrum oscillates for JJs with AMs, with the periodicity determined by the exchange field of the respective AM. In JJs with $p$-wave  magnets, increasing the exchange field remains roughly constant the LDOS, with rather small variations in the case of $p_{y}$-wave   magnet at large exchange fields. We thus close this part by stressing that the formation of ABSs in JJs with   unconventional magnets strongly depends on the type of unconventional magnetism and on the transverse direction, signatures that can be with caution identified via LDOS. It remains an open problem the behavior of ABSs in JJs with unconventional magnets having higher angular momentum dependence which will definitely shed more light on the importance of the interplay between superconductivity and unconventional magnetism.

\subsection{The Josephson effect and current-phase characteristics}
\label{section3AA}
Having identified   the effect of unconventional magnetism on the formation of ABSs, we now discuss
 the phase-dependent supercurrents $I(\phi)$ characterizing Cooper pair transport across JJs with  unconventional magnets, studied in Ref.\,\cite{fukaya2024x} for AMs and  $p$-wave magnets.   The  Josephson effect in JJs with AMs was also studied in Refs.\ 
\cite{Ouassou23,Beenakker23,PhysRevLett.133.226002,zhang2024,Cheng24,WeiEUPhys24,sun2024}. Before addressing the Josephson effect in  unconventional magnets, it is worth reminding the Josephson effect when conventional superconductors are linked by a short normal metal or simply when  two conventional superconductors are directly coupled. In this case,  the supercurrent is given by  \cite{PhysRevLett.66.3056}
\begin{equation}
\label{IphisSCs}
I(\varphi)=\frac{I_{c}\sigma_{\rm N}\,
{\rm sin}(\varphi)}{\sqrt{1-\sigma_{\rm N}{\rm sin}^{2}(\varphi/2)}}
{\rm tanh}\left[\frac{E(\varphi)}{2\kappa_{\rm B}T}\right]\,,
\end{equation}
where $E(\varphi)=\Delta \sqrt{1-\sigma_{\rm N}{\rm sin}^{2}({\varphi}/{2})}$,  $I_{c}=e\Delta/2\hbar$, $\sigma_{\rm N}$ the normal transparency introduced in Eq.\,(\ref{subsection4b}) and $T$ the temperature. Thus, in the tunneling regime ($\sigma_{\rm N}\ll1$), the supercurrent in conventional JJs  is given by $I(\varphi)=I_{\rm c}{\rm sin}(\varphi)$, first derived by \cite{PhysRevLett.10.486}, which has a maximum at   $\varphi=\pi/2$;   in the full transparent regime ($\sigma_{\rm N}=1$), the supercurrent is given by $I(\varphi)=I_{\rm c}{\rm sgn}({\rm cos}(\varphi/2)){\rm sin}(\varphi/2)$, first derived Kulik and Omelyanchuk \cite{kulik1978josephson}, with a maximum at $\varphi=\pi$. In both cases, the supercurrent is an odd function of the phase, $I(\varphi)=-I(\varphi)$, it is $2\pi$-periodic with the phase, it vanishes at integer values of the phase $I(\varphi)=0$ for $\varphi=n\pi$, with $n\in\mathbb{Z}$, and the positive and negative maxima are equal $I_{c}=I_{c}^{\pm}$, with $I_{c}^{\pm}={\rm max}_{\varphi}[\pm I(\varphi)]$. Below we will see that these seemingly standard properties of conventional JJs are strongly affected by unconventional magnetism.

For JJs with unconventional magnets, the supercurrent  in Ref.\,\cite{fukaya2024x} is obtained using the second expression of Eqs.\,(\ref{IphirhoSNS}) for the system associated to Fig.\,\ref{ABS_JJs}, where two semi-infinite spin-singlet $s$-wave superconductors are linked via an unconventional magnet. The respective  supercurrents as a function of the superconducting phase difference $\varphi$ are presented in Fig.\,\ref{IphiJE}(a-d) for JJs with $d$-wave AMs and $p$-wave magnets. Therein, we also show the corresponding free energy $F(\varphi)$ as a function of $\varphi$, whose minimum value $F_{\rm min}={\rm min}_{\varphi}[F(\varphi)]$ is  subtracted for visualization purposes, see blue curves in Fig.\,\ref{IphiJE}(a-d).  The first observation in  Fig.\,\ref{IphiJE}(a-d) is that the current-phase curves and the free energies share some similarities  and also strong differences depending on the type of  unconventional magnet.  Among the similarities is that the supercurrent $I(\varphi)$ and the free energy   $F(\varphi)$ exhibit a $2\pi$-periodicity with  $\varphi$, while $I(\varphi)$ in all cases exhibits zero value  at $n\pi$ with $n\in\mathbb{Z}$, and the positive/negative maxima are equal. However, $I(\varphi)$ and   $F(\varphi)$  acquire a distinct dependence with respect to the superconducting phase difference that unveils the type of JJ. In the case of JJs with a $d_{xy}$-wave AM, the current-phase curve develops minima and maxima around   $\varphi=\pm\pi$ and $\varphi=0$, respectively, see Fig.\,\ref{IphiJE}(a); this supercurrent behaviour is accompanied by a free energy with minima at $\varphi\neq0,\pi$, and hence signals the emergence of a $\varphi$-JJ as a result of $d$-wave altermagnetism, see also Ref.\,\cite{PhysRevLett.133.226002}. We note that $\varphi$-JJs were initially predicted in Ref.\,\cite{Buzdin03}.  In JJs with $d_{x^{2}-y^{2}}$-wave AMs, the current-phase curve has a sine-like behaviour but is shifted by $\pi$, which is also accompanied by  minima of the free energy at $\varphi=\pm\pi$, see Fig.\,\ref{IphiJE}(b) and Ref.\,\cite{fukaya2024x}; this behaviour corresponds to a $\pi$-JJ, also reported in Refs.~\cite{Ouassou23,Beenakker23,PhysRevLett.133.226002,zhang2024,Cheng24,WeiEUPhys24,alipourzadeh2025ABSJE}. One of the  interesting predictions in Ref.\,\cite{Ouassou23} is that the maximum Josephson current as a function of   temperature exhibits a non-monotonic due to the $d_{x^2-y^2}$-wave altermagnetic order, thus opening an intriguing effect for identifying altermagnetism. Furthermore, Refs.\ \cite{Beenakker23,PhysRevLett.133.226002,zhang2024}  showed that   0-$\pi$ phase transitions can  occurr by changing the altermagnetic order and the length of the junction. The angle of the Fermi surface in AMs and the direction of the N\'{e}el vector were also shown to  affect the behavior of the Josephson current~\cite{Cheng24,WeiEUPhys24,alipourzadeh2025ABSJE}. In the case of JJs with $p$-wave  magnets, the phase-dependent supercurrents $I(\varphi)$ have a regular sine-like behavior with a sharp transition across $\varphi=\pm\pi$ due to a presumably high transmission as seen in the Andreev spectrum of Fig.\,\ref{ABS_JJs}(k,l). The corresponding free energy for JJs with $p$-wave  magnets has a single minimum at $\varphi=0$ and hence represents a regular $0$-JJ. 
 
We can gain further understanding of the supercurrent  by inspecting the maximum current $I_{\rm c}={\rm max}_{\varphi}[I(\varphi)]$, known as critical current, presented in Fig.\,\ref{IphiJE}(e-h) as a function of the exchange field for each UM \cite{fukaya2024x}. Moreover, since the Josephson current can be decomposed in terms of even and odd harmonics of $\varphi$ as $ I(\varphi)=\sum_{m}[I_{m}\sin{n\varphi}+J_{m}\cos{n\varphi}]$, in Fig.\,\ref{IphiJE}(e-h) we also show the nonzero first and second harmonics  $I_{1,2}$. We note that Refs.\,\cite{PhysRevLett.133.226002,fukaya2024x} showed that  $J_{n}=0$ in JJs with AMs and  $p$-wave magnets due to their intrinsic symmetries, see also Ref.\,\cite{tanaka971}. The first observation is that the critical currents for $d$-wave AMs exhibit an oscillatory decay as the exchange fields increase, with a distinct periodicity tied to the type of altermagnetism, see Fig.\,\ref{IphiJE}(e,f)  and Ref.\,\cite{fukaya2024x}. The oscillatory profile can be understood in terms of the oscillatory profile of their first and second harmonics as depicted by cyan and magenta curves  in Fig.\,\ref{IphiJE}(e,f). In the case of $p$-wave magnets, the critical currents do not oscillate but develop a slow decay as the exchange fields increase, a phenomenon explained in Ref.\,\cite{fukaya2024x} to be the result of the non-detrimental effect of preserving time-reversal symmetry in  $p$-wave magnets and the  spin-singlet Cooper pairs in the parent conventional superconductors.   The oscillations in  critical currents are common in magnetic systems with a net magnetization \cite{RevModPhys.77.935}  and also in topological junctions with Majorana states \cite{PhysRevB.96.205425,cayao2018andreev,cayao2018finite,PhysRevLett.123.117001,PhysRevB.104.L020501,PhysRevB.105.054504,baldo2023zero,PhysRevB.109.L081405}, which both phenomena are absent here in  the discussed JJs based on  unconventional magnets and conventional superconductors.   Taking the critical current properties and the discussion made in the previous paragraph, we can   say that  unconventional magnets represent a rich playground for anomalous current-phase characteristics and also for designing distinct types of JJs by simply using conventional superconductors.

 \begin{figure}[t!]
    \centering
    \includegraphics[width=8.5cm]{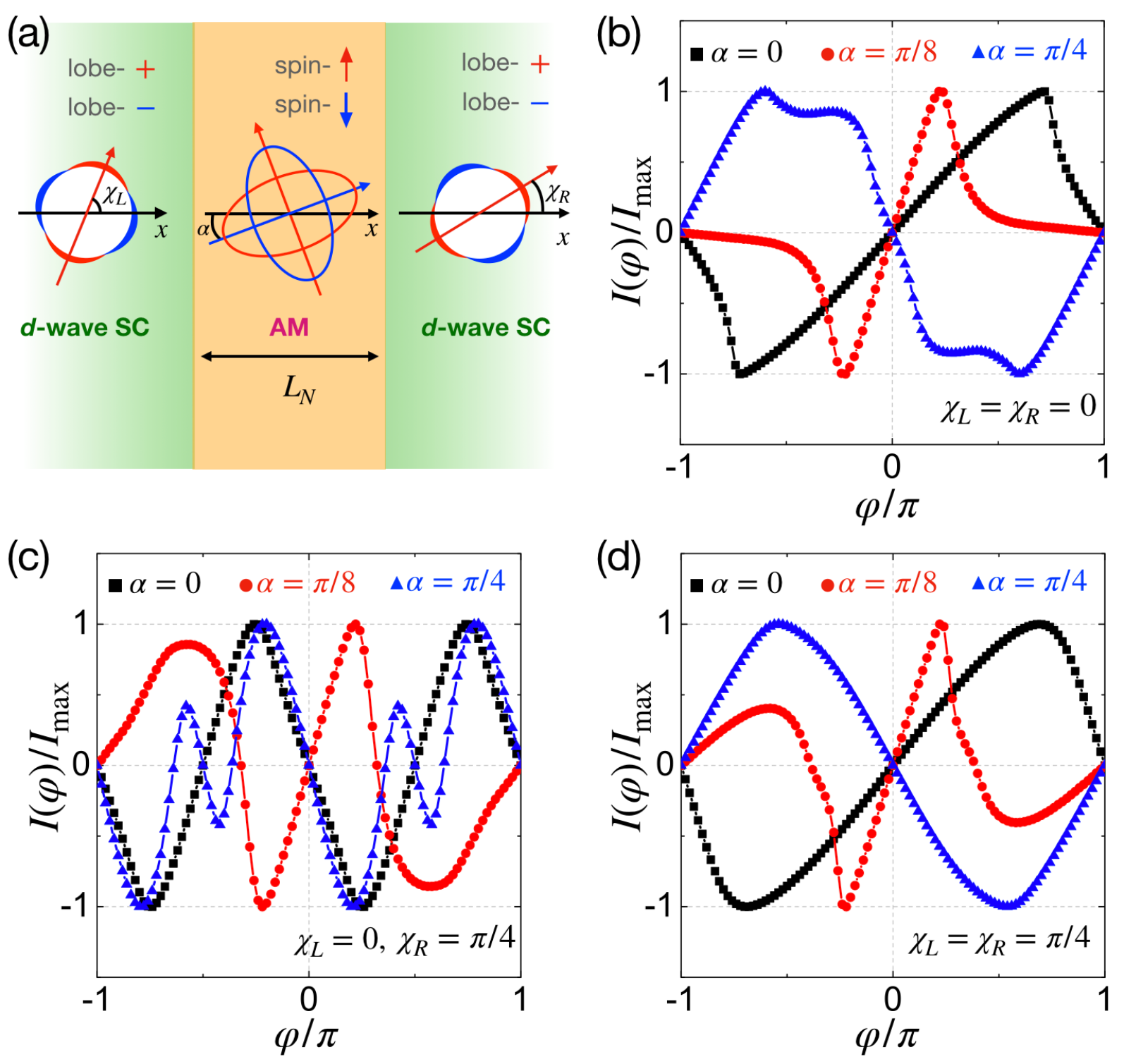}
    \caption{Current-phase characteristics $I(\varphi)$ in  Josephson junctions (JJs) formed by    semi-infinite $d$-wave superconductors  (SCs) and a  $d$-wave  altermagnet (AM) of finite length $L_{\rm N}$. (a) Sketch of the JJ without interface barriers, where the angles $\chi_{L/R}$ indicate the orientation of the pair potential lobe with respect to the $x$-axis   in the left/right superconductor. Also, $\alpha$ denotes the orientation of the  exchange field lobe with respect to the $x$-axis. (b-d) $I(\varphi)$ for $\chi_{L/R}=0$ (b), $\chi_{L}=0$ and $\chi_{R}=\pi/4$ (c), and $\chi_{L/R}=\pi/4$ (d). Parameters:   $k_\mathrm{F} L_{\rm N}=30$, $J=0.2\mu$, 
    $\kappa_{\rm B}T=0.0108\Delta$.  Adapted from Ref.\,\cite{Bo2025}.}
    \label{dJJ}
\end{figure}%

Besides JJs formed by conventional superconductors and  unconventional magnets,  Ref.\ \cite{Bo2025} very recently studied in JJs formed by $d$-wave superconductors and a $d$-wave AM [Fig.~ \ref{dJJ} (a)] and the main findings we show in Fig.~ \ref{dJJ} (c-d). In this case, the sign of the $d$-wave pair potential changes on the Fermi surface, in contrast to what occurs in conventional spin-singlet $s$-superconductors. This property  can enhance the $\sin(2\varphi)$ component of the Josephson current $I(\varphi) $ and realize the $\varphi $-junction \cite{tanaka961,tanaka971} in asymmetric junctions \cite{Buzdin03, Pugach10}.
 In the JJs based on $d$-wave superconductors and a $d$-wave AM, the current-phase curves are generally expressed as $\sum\nolimits_{n}I_{n}\sin \left( n\varphi \right) $. Since the lobe angles with respect to the interface change, the current-phase curves exhibit  $0$-$\pi $ transitions in the symmetric junctions [Fig.~\ref{dJJ} (b)(d)]. Particularly, for asymmetric $0^{\circ }$-$ \pi/4$ junction, the first-order harmonic of Josephson current $I_{1}$ reappears when the AM has neither $d_{x^{2}-y^{2}}$ nor $d_{xy}$ symmetry, as shown in Fig.~\ref{dJJ} (c)
where the Josephson current is finite at $\varphi=\pm \pi/2$ for $\alpha=\pi/8$. Such behavior is in sharp contrast to that in a $0^{\circ }$-$\pi/4$ junction with a ferromagnet \cite{TanakaJosephson2000}, where the first-order sinusoidal component never appears in the current-phase curve. It is shown in Ref.\,\cite{Bo2025} that the intriguing behavior of the Josephson current in  JJs formed by $d$-wave superconductors and a $d$-wave AM is strongly tied to
 the underlying symmetries in AMs,  which involve the combination of   time-reversal symmetry $\mathcal{T}$ as the time-reversal operator, $\mathcal{M}_{xz}$ as the mirror reflection operator with respect to $xz$-plane, and $\mathcal{C}_{4}$ the fourfold rotation symmetry with respect to $z$ axis. More precisely, Ref.\,\cite{Bo2025} showed that the symmetries $\mathcal{M}_{1}=\mathcal{T}\mathcal{M}_{xz}$ and $\mathcal{M}_{2}=\mathcal{T}\mathcal{M}_{xz}\mathcal{C}_{4}$ give rise to the protected nodal points at $\varphi =\pm \pi /2$ for $d_{x^{2}-y^{2}}$- and $d_{xy}$-AM, respectively.
As a result, the interplay between $d$-wave superconductivity and $d$-wave altermagnetism offers a good controllability of distinct types of JJs, which might be of utility for designing Josephson devices \cite{krantz2019quantum,kjaergaard2020superconducting,siddiqi2021engineering}.

\subsection{Emergent  phenomena due the interplay between unconventional magnetism and the Josephson effect}
Apart from the already discussed unusual current-phase characteristics, the combination of unconventional magnetism and the Josephson effect also leads to 
novel superconducting correlations and can be used to engineer nonreciprocal Josephson transport. In particular, we stress the prediction of odd-frequency spin-triplet pairing \cite{fukaya2024x} and the realization of the Josephson diode effect \cite{Qiang24}, emergent phenomena that hold fundamental relevance for understanding superconductivity in JJs with  unconventional magnets and also applied importance for superconducting electronics. In the following, we discuss these two discoveries.

\begin{figure}[!t]
\centering
\includegraphics[width=0.49\textwidth]{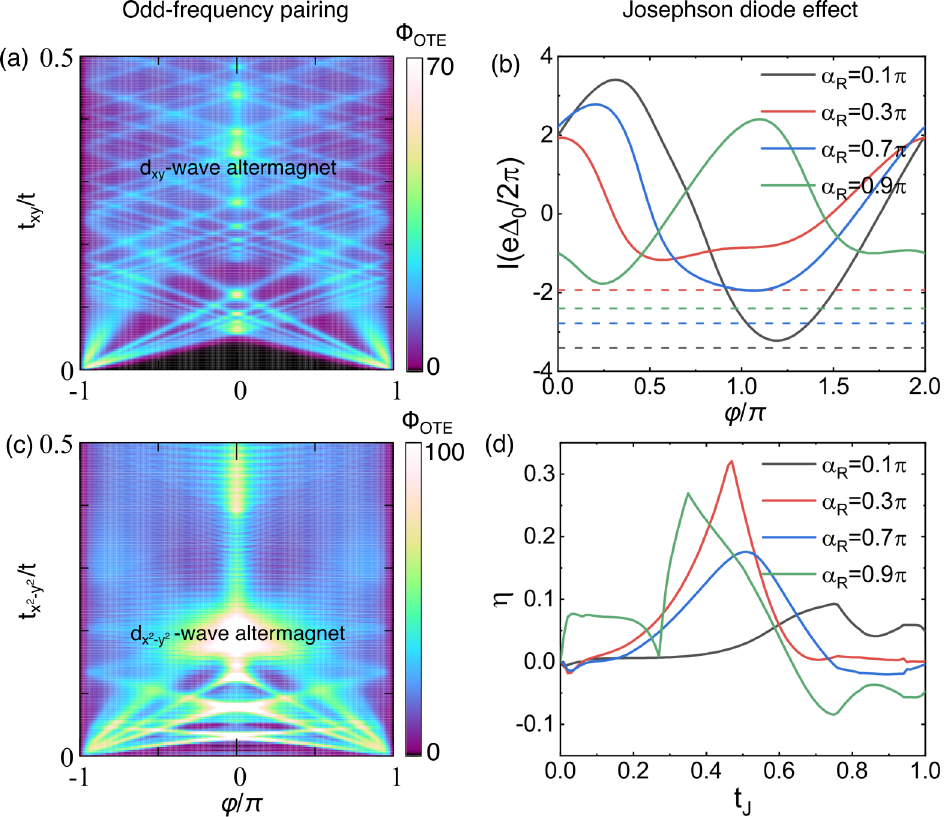}
\caption{Emergent phenomena in  Josephson junctions (JJs) based on  altermagnets (AMs). (a,c) Odd-frequency spin-triplet even-parity  (OTE) pairing: Square  absolute value of the  OTE pair amplitude integrated over $k_{y}$ as a function of the strength of the altermagnetic field and the superconducting phase difference $\varphi$. The OTE pair amplitude is obtained in the middle of a $d$-wave AM connecting two semi-infinite conventional superconductors at lowest Matsubara frequency within a recursive Green's function approach. Tight-binding parameters: $a=1$, $L_{\rm N}=30a$, $\Delta=0.02t$, $t_\mathrm{int}=0.95$, $\mu=1.5t$, $T_\mathrm{c}=0.01t$, and $T=0.025T_\mathrm{c}$. Adapted from Ref.\,\cite{fukaya2024x}. (b,d) Josephson diode effect: (a): Current-phase curves for distinct values of the orientation angle $\alpha_{\rm R}$ of the AM lobes with respect to the $x$-axis. (d): Josephson diode efficiency $\eta$ as a function of the strength of the altermagnetic field for distinct values of $\alpha_{\rm R}$. Reprinted figures with permission from Qiang Cheng, Yue Mao, and Qing-Feng Sun, Phys. Rev. B. 110, 014518 (2024) \cite{Qiang24}; Copyright (2025) by the American Physical Society.
}
\label{IphiJE_phenomena} 
\end{figure}

\subsubsection{Odd-frequency spin-triplet pairing}
As discussed in Subsection \ref{subsection32},  the interplay between superconductivity and unconventional magnetism leads to novel superconducting correlations. Motivated by these results, Ref.\,\cite{fukaya2024x} predicted the formation   of several superconducting pair symmetries by exploiting the Josephson effect and unconventional magnetism. The superconducting correlations were obtained from the anomalous electron-hole component of the Nambu Masubara Green's functions  as discussed in Section \ref{subsection32} for a JJ formed along $x$ by semi-infinite spin-singlet $s$-wave superconductors linked by  unconventional magnets; periodic boundary conditions along the transverse direction is taken, such that  $k_{y}$ is a good quantum number.  Among    unconventional magnets, Ref.\,\cite{fukaya2024x}  considered $d$-wave AMs and $p$-wave  magnets, which, when coupled to the superconductors, were shown to host superconducting correlations having symmetries similar but also distinct to that of the parent superconductor. Notably, for $p$-wave  magnets and $d$-wave AMs, Ref.\,\cite{fukaya2024x} finds spin-triplet superconducting correlations belong to the ETO and OTE symmetry classes discussed in Subsection \ref{subsection32}, respectively, whose parity symmetry is determined by the symmetry of unconventional magnetism. For instance, for JJs with $d_{xy}$-wave AMs, the OTE pair amplitude has an even parity that is  determined by the combination of the odd dependence under the exchange of spatial coordinates and  odd in $k_{y}$; the oddness in $k_{y}$ entirely arises due to the magnetic order, given that the altermagnetic field of a $d_{xy}$-wave AM is odd under $k_{y}$. Similarly, the parity symmetry of the OTE pairing in JJs with $d_{x^{2}-y^{2}}$-wave AMs is determined by the evenness in space and in momentum along the transverse direction. The role of AMs for affecting the parity of emergent superconducting correlations was already discussed in Subsection \ref{subsection32}. However, the most relevant finding of Ref.\,\cite{fukaya2024x} is that the formation of OTE pairing can be controlled by the interplay between altermagnetism and the Josephson effect. To visualize this fact, we show in Fig.\,\ref{IphiJE_phenomena}(a,b) the square absolute value of the OTE pairing in the middle of a $d$-wave AM integrated over $k_{y}$ as a function of the altermagnetic field and the superconducting phase difference $\varphi$, see also Ref.\,\cite{fukaya2024x}. As observed, a finite OTE pairing is induced only when both the altermagnetic field and $\varphi$, which not only unveils the formation of novel superconducting correlations but also suggest that they are important  in the proximity effect of JJs formed by AMs. An interesting open problem is perhaps to explore how long in space survive these pair correlations, which could take us further to understand and  realize a long-range proximity effect in JJs with AMs and their extension to  $p$-wave magnets as well. Studies of this kind are necessary for envisaging   Josephson devices hosting spin-polarized supercurrents \cite{eschrig2011spin} useful for superconducting spintronics \cite{eschrig2011spin,linder2015superconducting,Eschrig2015,yang2021boosting,mel2022superconducting,cai2023superconductor}, which can be achieved using  unconventional magnets at zero net magnetization.

\subsubsection{The Josephson diode effect}
In the Josephson diode effect,  nonreciprocity occurs in the supercurrents $I(\varphi)$ driven by the Josephson effect \cite{Jiangping07,MisakiPRB2021,davydova2022universal,JiangPhysRevX22,Tanakadiode1,Tanakadiode2,PhysRevLett.129.267702,PhysRevB.108.054522,PhysRevLett.130.266003,PhysRevB.107.245415,PhysRevB.109.L081405,10.1063/5.0210660,fu2023fieldeffect}, which by now has already motivated an intense experimental activity
\cite{wu2022field,baumgartner2022supercurrent,pal2022josephson,mazur2022gatetunable,Turini2022,PhysRevLett.131.027001,nadeem2023superconducting,Yuying23,valentini2023parityconserving}. More specifically, in a similar way as for the superconducting diode effect in bulk superconductors, the Josephson diode effect involves the nonreciprocity of maximum and minimum critical currents 
$I_{c}^{+}\neq I_{c}^{-}$, where  $I_{c}^{\pm}={\rm max}_{\varphi}[\pm I(\varphi)]$. Thus, the Josephson diode efficiency is quantified by quality factor $\eta=(I_{c}^{+}-|I_{c}^{-}|)/(I_{c}^{+}+|I_{c}^{-}|)$ given by Eq.\,(\ref{qualityfactor}) but involving the maximum and minimum Josephson currents. We can also see that the antisymmetric dependence of the supercurrent with the phase $\varphi$ is broken, namely, $I(\varphi) \neq-I(-\varphi)$; hence, the Josephson diode effect requires current-phase curves that are distinct to what is obtained in conventional JJs.  As for the superconducting diode discussed in  Subsection \ref{section34}, the Josephson diode effect also requires breaking time-reversal and inversion symmetries as necessary but not sufficient conditions.  Ref.\,\cite{he2022phenomenological} showed that, breaking time-reversal and inversion symmetries, needs to be combined with breaking $x$-inverting symmetries   in order to induce nonreciprocal Josephson transport. These conditions can be easily achieved in JJs formed by semiconductors which exhibit intrinsic SOC under the presence of external magnetic fields breaking time-reversal and $x$-inverting symmetries. In terms of supercurrent harmonics, when the Josephson current is decomposed as $I(\varphi)=\sum_{m}[I_m\sin(m\varphi)+J_m\cos(m\varphi)]$, the discussed symmetry Josephson diode requirements were shown in Ref.\,\cite{Tanakadiode1} to be directly connected to the Josephson current exhibiting the simultaneous coexistence of the harmonics $\sin(\varphi)$, $\sin(2\varphi)$, and $\cos(\varphi)$ \cite{Tanakadiode1}. Thus, a Josephson diode effect is possible when $I_{1}$, $I_{2}$ and $J_{1}$ coexist, maximizing the  efficiency $\eta$ when such harmonics are comparable in magnitude \cite{he2022phenomenological}.

Given the requirements for Josephson diodes, JJs formed by AMs  represent an interesting and promising ground. This is because AMs exhibit broken time-reversal symmetry and have a momentum dependent magnetic field akin to SOC but of not relativistic origin, already shown to be important for   superconducting diodes in bulk superconductors and whose extension to JJs is hence natural, see Subsection \ref{section34}. Exploring AMs for Josephson diodes is further supported by the findings of Refs.\,\cite{fukaya2024x,Bo2025} discussed in Subsection \ref{section3AA}, where the current-phase characteristics of JJs with AMs were shown to exhibit multiple harmonics ($I_{1,2}$), thus being promising for the Josephson diode effect according to Ref.\,\cite{Tanakadiode1} if an additional mechanism is able to allow for $J_{1}$.  In this regard, $J_{1}$ can be induced   e. g., by an external magnetic field or perhaps other spin-dependent fields, which then implies that $I(\varphi) \neq-I(-\varphi)$ and the Josephson diode effect can be realized. So far, there exist limited studies  reporting the Josephson diode effect in a JJ formed by  conventional spin-singlet $s$-wave superconductors with $d$-wave altermagnetism and also Rashba SOC \cite{Qiang24,jiang2025josephson}; see Refs.\,\cite{sharma2025tunableJDE,boruah2025fieldfreeJD} for AM-based Josephson diodes using unconventional superconductors. In the top panel of Fig.\,\ref{IphiJE_phenomena}(b), the current-phase curves of Ref.\,\cite{Qiang24} are shown for distinct values of the orientation angle that controls the $d$-wave altermagnetism and the Rashba SOC. The distinct values of maximum and minimum values of $I(\varphi)$ and the finite quality factors in the bottom panel of Fig.\,\ref{IphiJE_phenomena}(b) clearly show the emergence of the Josephson diode effect using $d$-wave AMs.  While these findings are indeed encouraging, Ref.\,\cite{Qiang24} showed that the Josephson diode needs the simultaneous presence of altermagnetism and Rashba SOC, vanishing if SOC is zero.  More recently, it has been predicted the possibility to realize a field-free  Josephson diode effect  by combining AMs and Ising superconductor in a JJ~\cite{boruah2025fieldfreeJD}, while Ref.\,\cite{sharma2025tunableJDE} combined AMs with spin-singlet and spin-triplet superconductors in a JJ.   This then makes us wonder if it is possible at all to realize Josephson diodes by simply using AMs or they always need to be combined with other materials. One way could be to combine them with   $p$-wave magnets  or other higher angular momentum unconventional magnets since, under certain circumstances, they also offer a spin splitting of energy bands. Pursuing these ideas have an immediate impact in superconducting electronics where Josephson diodes can be utilized.

\begin{figure*}[!t]
\centering
\includegraphics[width=0.9\textwidth]{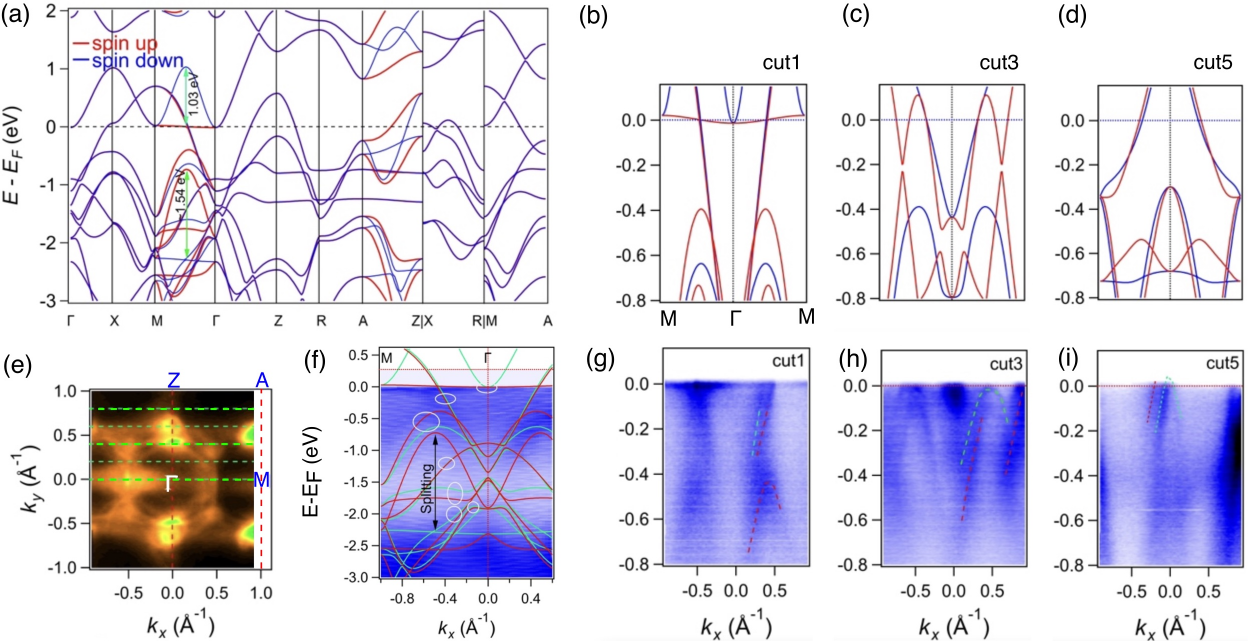}\\
\caption{Measurement of the band splitting in the $\Gamma{\rm MAZ}$ plane of RuO$_{2}$. (a) Electronic band structure along high symmetry lines for the altermagnetic phase of RuO$_{2}$, shifted upward by $0.36$\,eV; spin up and down bands are indicated by red and blue, respectively. (b-d) Calculated band structure along cuts 1, 3, 5, indicated by the brightest green dashed lines in the Fermi surface in the $\Gamma{\rm MAZ}$ plane (e).  (f)  Angle-resolved photoemission spectroscopy (ARPES) spectra with a wide energy range along $\Gamma{\rm M}$ direction, with the calculated band structure overlaid; the blue color is changed to green to avoid confusion with the blue background. The band splitting is indicated by the double black arrow. (g-i) ARPES spectra of cuts parallel to the $\Gamma{\rm M}$ direction with an equal $k_{z}$ offset of $0.2(\Gamma{\rm Z}$) between each other. Reproduced under the terms of the CC-BY 4.0 license and adapted with permission  from the authors of Ref.\,\cite{lin2024observation}.
}
\label{FigBandExp} 
\end{figure*}

\section{Experimental advances}
\label{section4}
Having discussed the theoretical efforts due to unconventional magnetism in superconductors, this part addresses  the experimental advances. Since the field is still at its infancy, almost all the reported experiments involve normal state  AMs \cite{PhysRevLett.118.077201,ghimire2018large,PhysRevLett.122.017202,NakaNatCommun2019,Ahn2019,LiborSAv,PhysRevB.104.024401,MaNatcommun2021,reichlova2021macro,shao2021spin,Rafael21,PhysRevLett.128.197202,PhysRevB.106.064435,Bose_2022,PhysRevLett.129.137201,PhysRevLett.130.216701,Feng_2022,PhysRevB.108.024410,Tschirner_2023,liu2023inverse,wang2023emergent,PhysRevLett.132.056701,Lee24,Osumi2024,PhysRevB.109.094413,PhysRevLett.132.176701,PhysRevB.109.224430,fedchenko2024observation,reichlova2024,KrempaskyNature2024,reimers2024,han2024SciAdv,jiang2024,lin2024observation,zhu2024observation,lu2024,PhysRevLett.133.056701,fzhang2024,hu2024,gray2024timeres,KIMEL2024172039,li2024spinsplittingaltermruo2,pan2024unvei}; see also Refs.\,\cite{Bai_review24,tamang2024} for recent reviews on AMs. These experiments measured the band splitting  in AMs \cite{liu2023inverse,KrempaskyNature2024,lin2024observation,reimers2024,Lee24,Osumi2024,fedchenko2024observation,PhysRevLett.133.056701} as well as anomalous and spin transport effects \cite{ghimire2018large,reichlova2021macro,Feng_2022,LiborSAv,Tschirner_2023,wang2023emergent,PhysRevLett.132.056701,Lee24,PhysRevB.109.224430},   spin currents \cite{NakaNatCommun2019,shao2021spin,Rafael21,PhysRevLett.128.197202,Bose_2022,PhysRevLett.129.137201,PhysRevLett.130.216701,PhysRevB.108.024410,li2024spinsplittingaltermruo2,pan2024unvei}, magnetoptical effects \cite{PhysRevB.104.024401,PhysRevB.106.064435,gray2024timeres,PhysRevLett.132.176701,PhysRevB.109.094413,KIMEL2024172039}. So far,  unconventional magnets with an odd-parity magnetic order  have not been experimentally studied, although there exist interesting predictions regarding Mn$_{3}$GaN and CeNiAsO for $p$-wave  magnets \cite{hellenes2024P}.  In the case of superconductivity and unconventional magnetism, as we have already mentioned in the introduction, superconductivity has been observed in thin films of strained RuO$_2$  \cite{PhysRevLett.125.147001,ruf2021strain,PhysRevMaterials.6.084802} and indications of superconductivity also exists in monolayer FeSe \cite{mazin2023inducedmoFeSe}. Since these materials are also promising for altermagnetism \cite{Bai_review24,tamang2024}, it seems that intrinsic superconductivity in altermagnets, or the coexistence of superconductivity and altermagnetism, might be indeed possible. In superconducting junctions, only a single study has very recently been reported \cite{Kazmin_2025}, which addresses superconducting transport via Andreev reflections in junctions formed by MnTe, expected to be altermagnetic, and In superconductor. 

In what follows, we briefly discuss the normal state experiments on AMs reporting their spin splitting and $d$-wave nature as well as the generation of spin currents, of particular importance for novel superconducting phenomena as we argue below. Moreover, we also delve into the experiments in the superconducting state, on strained superconductivity in RuO$_{2}$ and the Andreev reflection in MnTe/In junctions.

\subsection{Spin splitting and $d$-wave altermagnetic nature in the normal state}
One of the first properties that attracted experimental interest in AMs  has been the spin splitting of energy bands \cite{liu2023inverse,KrempaskyNature2024,lin2024observation,reimers2024,Lee24,Osumi2024,fedchenko2024observation,PhysRevLett.133.056701}. As we have discussed in the introduction and also in Section \ref{section1},  the band splitting has a  nonrelativistic origin  and is, therefore,   distinct to what happens in spin-orbit coupled materials. This band splitting is a result of opposite-spin sublattices linked by the crystal-rotation symmetries \cite{landscape22}. Moreover, since there is zero net magnetization in AMs, the band splitting is also different from what occurs in ferromagnets. As a result, the band splitting in AMs is a unique altermagnetic property. While several experiments have reported consistent evidence of band splitting in AMs, there have also been studies that challenge the altermagnetic nature of ${\mathrm{RuO}}_{2}$. In this regard, a recent study \cite{PhysRevLett.133.176401} reported the absence of altermagnetic spin splitting   in  ${\mathrm{RuO}}_{2}$, while Refs.\,\cite{PhysRevLett.132.166702,kessler2024absence} indicate that the ground state of  ${\mathrm{RuO}}_{2}$ is nonmagnetic. Ref.\,\cite{kessler2024absence} also   suggested   multiple scattering as the source of the magnetic signals reported earlier  in ${\mathrm{RuO}}_{2}$ \cite{PhysRevLett.118.077201} and argued that the altermagnetic signals seen in such a material likely have extrinsic origin. It is worth noting that a recent theoretical study suggested that  the adequate characterization of the amount of Ru vacancies can actually promote the formation of a magnetic state in RuO$_2$ \cite{PhysRevB.109.134424} at a lower and more realistic Hubbard $U$ than that used in Refs.\,\cite{LiborSAv,Ahn2019}; see also Refs.\,\cite{fedchenko2024observation,jeong2024AMpolarMetphase,PhysRevLett.132.086701,weber2024OPExSpPol,chen2024altermSpinSPLITmagne} for recent supporting evidence of altermagnetism in RuO$_2$. These studies thus reflect the intense activities  in the field of altermagnetism, highlighting that    more experiments are needed to expand the understanding of RuO$_2$ and perhaps extend this motivation to measure the spin splitting in other altermagnetic materials as   already reported for MnTe \cite{KrempaskyNature2024}.

\begin{figure*}[!t]
\centering
\includegraphics[width=0.8\textwidth]{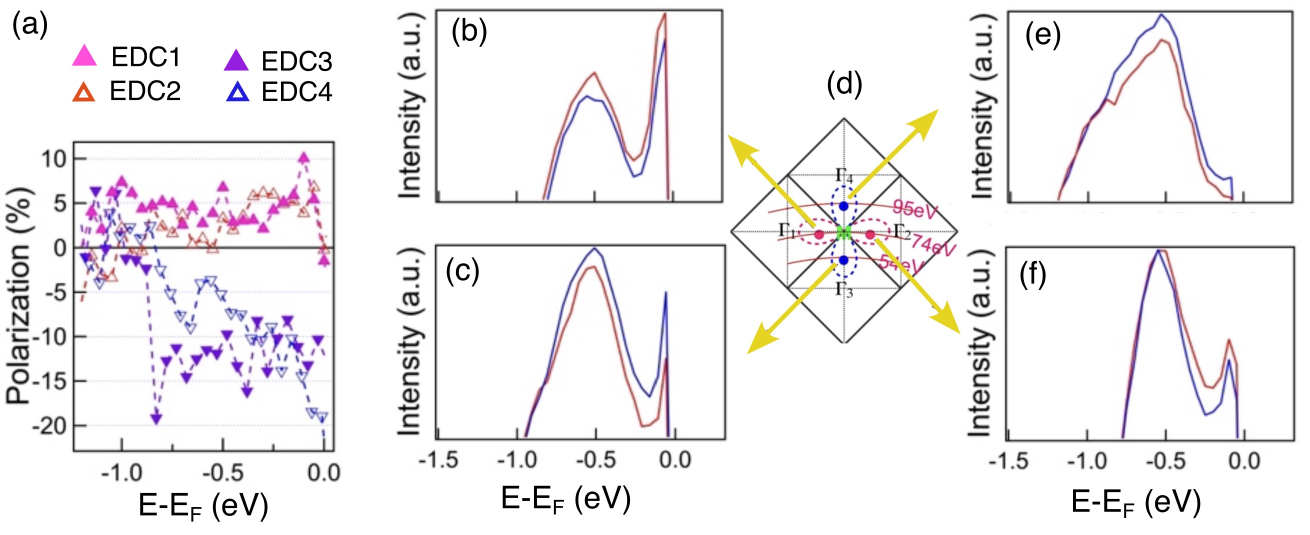}\\
\caption{Experimental identification of the $d$-wave altermagnetic order in  RuO$_{2}$. (a) Spin polarization along EDC 1-4 symmetrically around the M point, at momenta on the left/right and above/below M, after removing the linear background. EDC stands for energy distribution curve.   (b) Zoomed-in view of the intensity of spin-polarized EDCs along EDC1 at the left point of M. The red and blue curves indicate  up and down spins, respectively. (c,e,f) The same as in (b) but for momenta at the right, above, and below  the M point. (d) shows the   cross-section of the Brillouin Zone  and the high symmetry points, where the four points where the spin EDCs are obtained. The left/right and above/below   momenta points  are depicted by red and blue filled circles in (d), respectively, and are located along $\Gamma_{1}{\rm M}\Gamma_{2}$ and $\Gamma_{3}{\rm M}\Gamma_{4}$. (d) also indicates the photon energies  at which spin-ARPES spectra was acquired.   Reproduced under the terms of the CC-BY 4.0 license and adapted with permission  from the authors of Ref.\,\cite{lin2024observation}.
}
\label{FigdwaveExp} 
\end{figure*}

In spite of the controversy exposed in the previous paragraph, the confirmation of altermagnet-induced band splitting would be a great breakthrough in the field. In light of this, now we would like to describe   the band splitting and identification of the $d$-wave parity  in  the altermagnet ${\mathrm{RuO}}_{2}$ reported in Ref.\,\cite{lin2024observation}; the great interest in this material justifies its consideration.    Single crystals of RuO$_{2}$ were grown by chemical vapor transport method and conventional synchrotron-based  angle photoemission spectroscopy (ARPES) was used to measure the spin band splitting  \cite{lin2024observation}, see Fig.\,\ref{FigBandExp}. In  Fig.\,\ref{FigBandExp}(a), the bandstructure   along high symmetry lines in the altermagnetic phase shows a distinct spin splitting along $\Gamma{\rm M}$ and AZ directions; along    $\Gamma{\rm M}$ direction, an energy splitting of $1.54$eV is predicted, see also Ref.\,\cite{landscape22}. The spin splitting is perhaps more clearly seen in Fig.\,\ref{FigBandExp}(b-d), which shows the bandstructure  along three cuts parallel to   $\Gamma{\rm M}$ and along $\Gamma{\rm Z}$ in the Fermi surface shown in Fig.\,\ref{FigBandExp}(e). The band splitting seen in Fig.\,\ref{FigBandExp}(a-d) is confirmed in Fig.\,\ref{FigBandExp}(f-i), where the ARPES spectra   are presented and measured with energy and angular resolutions set to $\sim20$meV and $0.1\degree$, respectively at $22\degree$K. By overlaying the theoretical bandstructure, the authors of Ref.\,\cite{lin2024observation} estimate the largest band splitting of $1.54$eV to be within the energy range of $-0.7$eV and $-2.5$eV, see Fig.\,\ref{FigBandExp}(f). The band dispersion from ARPES measurements can be seen to adjust well to the bandstructure calculations, albeit experimental data seem to have large broadening effects and low energy resolution, compare Fig.\,\ref{FigBandExp}(b-d) and Fig.\,\ref{FigBandExp}(g-i). These findings thus support the induced band splitting in altermagnetic ${\mathrm{RuO}}_{2}$, even though more studies are necessary to settle the controversy about the altermagnetic nature of this effect. 

 \begin{figure*}[!t]
\centering
\includegraphics[width=0.8\textwidth]{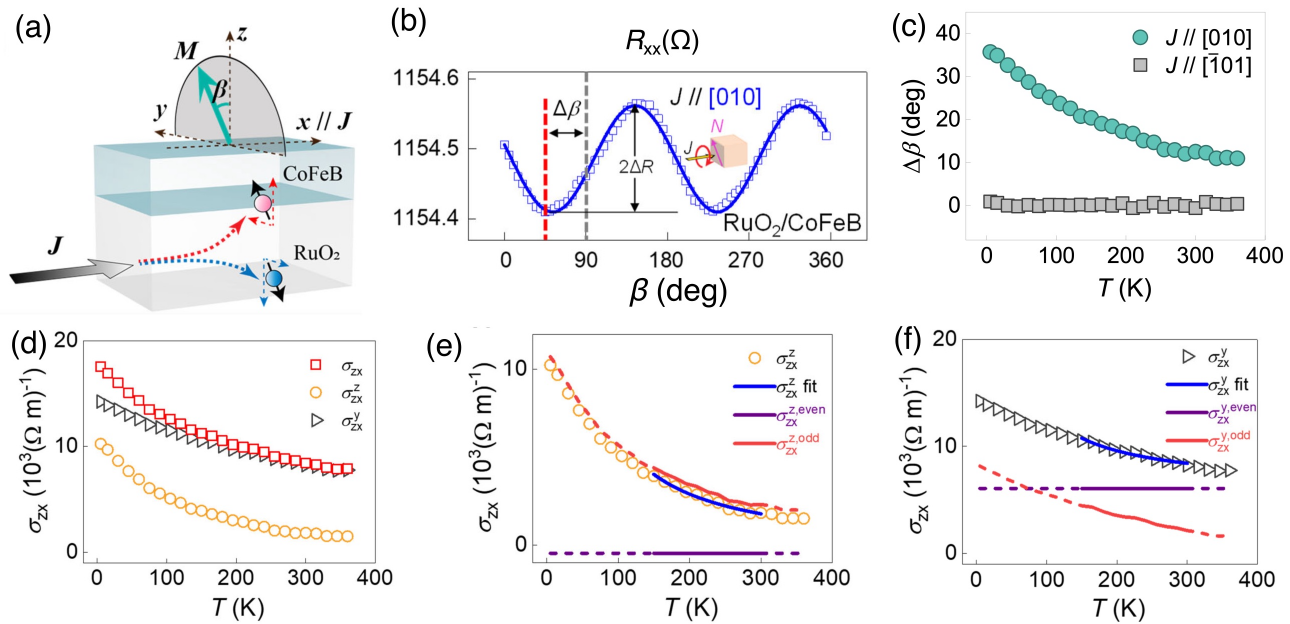}\\
\caption{Detection of spin currents in altermagnetic RuO$_{2}$.  (a) Scheme of the spin Hall magnetoresistance measurement in a RuO$_{2}$/CoFeB heterostructure, while scanning a $4$T magnetic field $M$ in a plane perpendicular to the electric field; here, $\beta$ is the angle between the magnetic field and the normal direction $z$, $\Delta\beta$ is a phase shift, and  $x$ is the  direction of the dc current $J$ parallel to the [010] crystallographic direction. The current $J$ favors the separation of spins, indicated by dotted red and cyan arrows. (b) Spin Hall magnetoresistance as a function of $\beta$ at $5\degree$K and with $J$ parallel to [010]. The magenta arrow indicates the direction of the N\'{e}el vector.  (c) Temperature dependence of the phase shift $\Delta\beta$ obtained from the spin Hall magnetoresistance measurements. (d-e) Spin Hall conductivities as a function of temperature, showing the total and the $y$- and $z$ polarized components  $\sigma_{zx}$, $\sigma_{zx}^{y}$, and $\sigma_{zx}^{z}$. Reproduced  with permission  from the authors of Ref.\,\cite{pan2024unvei}.
}
\label{FigspincurrentExp} 
\end{figure*}

A key   property of  AMs, which distinguishes them from conventional magnets, is that their magnetic order has an even-parity symmetry. One of the simplest AMs have $d$-wave symmetry, which was predicted for ${\mathrm{RuO}}_{2}$ \cite{landscape22}. This property was also addressed in the experimental study of Ref.\,\cite{lin2024observation} by means of spin-ARPES with an energy resolution set to $\sim100$meV at $11\degree$K.  The key findings of  Ref.\,\cite{lin2024observation} that reveal the $d$-wave nature of  the altermagnet ${\mathrm{RuO}}_{2}$ are shown in Fig.\,\ref{FigdwaveExp}. As we discussed in Section \ref{section1}, the $d$-wave parity of AMs is reflected in the four lobe structure with an alternating spin whose direction flips after a $\pi/2$ rotation, see Fig.\,\ref{FigdwaveExp}(d). To measure the $d$-wave altermagnetic texture, the authors of Ref.\,\cite{lin2024observation} chose four points, one per  lobe (filled blue/red  circles), and measured the spin polarization by means of spin-ARPES at distinct photon energies. The measured polarization is largely positive and reaches a maximum of $10\%$ for the  the points along  $\Gamma_{1}{\rm M}\Gamma_{2}$, while it is negative and with a maximum of up to $20\%$ for the points along $\Gamma_{3}{\rm M}\Gamma_{4}$. These features already reveal an opposite behavior for the spin polarization, which supports  the $d$-wave nature of ${\mathrm{RuO}}_{2}$. More insights are obtained from Fig.\,\ref{FigdwaveExp}(b,c) and Fig.\,\ref{FigdwaveExp}(e,f), which show the intensity of the spin-polarized energy distribution curves for each spin. In this case, the contribution from spin up at the left point of M along $\Gamma_{1}{\rm M}$ exhibits a larger value than its spin-down counterpart,   which indicates a spin-up polarization, see Fig.\,\ref{FigdwaveExp}(b). For the bottom point of M along $\Gamma_{3}{\rm M}$, the spin-down part is larger than the spin-up contribution, thus favouring a spin-down  polarization that is of opposite sign as the one obtained for the point along $\Gamma_{1}{\rm M}$, see Fig.\,\ref{FigdwaveExp}(c). Following a similar analysis, at the right and top points along the  ${\rm M}\Gamma_{2}$ and ${\rm M}\Gamma_{4}$, respectively,  the authors of Ref.\,\cite{lin2024observation} measure a spin-up and spin-down  polarizations. Thus, the measurements discussed here support the altermagnet nature of ${\mathrm{RuO}}_{2}$. Despite the ongoing controversy,  it is important to highlight these experiments as they can inspire other studies that  can then help understand the   symmetries of the unconventional magnetic order.  It is also worth noting that the spin splitting of AMs and their altermagnetic nature have also been  recently measured in other materials, such as   MnTe \cite{Osumi2024,KrempaskyNature2024,Lee24} and CrSb \cite{reimers2024,lu2024,li2024crsb,yang2025,Zeng_2024,PhysRevLett.133.206401}, which are $g$-wave AMs. These studies highlight the universality of the spin-splitting in AMs. Establishing the band splitting and detecting the parity symmetries of  AMs would be very useful when combining them with superconductors, as these properties are crucial for engineering novel superconducting phenomena such as spin-triplet   higher momentum Cooper pairs \cite{maeda2025classifi,fukaya2024x} and controllable $0-\pi$ transitions in Josephson junctions \cite{PhysRevLett.133.226002,fukaya2024x,Bo2025}. 
 
\subsection{Spin currents in the normal state}
A spin current is characterized by the flow of the spins of electrons in the absence of a charge current and represents a unique way to carry information \cite{Sarma_RevModPhys_2004,HIROHATA2020166711,Newhorizons_spintronics}.  The ability to generate spin currents with spin torques in magnets offers  a robust way for controlling the magnetization, a crucial task towards efficient spintronics \cite{Newhorizons_spintronics,kim2024spin}. In AMs, the spin splitting of energy bands, and their associated anisotropic spin distribution of Fermi surface, has been shown to be an interesting way  for inducing transverse spin currents \cite{NakaNatCommun2019,shao2021spin,Rafael21,PhysRevLett.128.197202,PhysRevLett.129.137201,Bose_2022,PhysRevLett.130.216701,PhysRevB.108.024410,li2024spinsplittingaltermruo2,pan2024unvei,chen2024altermSpinSPLITmagne,yang2025MagnetoResistanceAM,fu2025SpintronicsAM}, thus reflecting the importance of altermagnetism; see also Refs.\,\cite{PhysRevB.108.L140408,PhysRevB.109.174438}. This spin current has a nonrelativistic origin, is  odd under time-reversal symmetry, can be reversed by switching the N\'{e}el vector, and exhibits an unusual spin polarization that makes AMs promising for spin-torque devices. We remind that  spin currents induced by  relativistic spin-orbit coupling, as in the spin Hall effect,  are    even under time-reversal symmetry \cite{RevModPhys.87.1213}. 

As pointed out above, the nonrelativistic spin currents have attracted the interest of several groups and, very recently, their magnetic origin in altermagnetic RuO$_{2}$ as well as their direct connection to the N\'{e}el vector have been experimentally addressed in Ref.\,\cite{pan2024unvei}. The key results of  Ref.\,\cite{pan2024unvei} are shown in Fig.\,\ref{FigspincurrentExp}, where spin Hall magnetoresistance measurements were carried out to identify the spin current polarization in a RuO$_{2}$/CoFeB heterostructure [Fig.\,\ref{FigspincurrentExp}(a)]. The magnetoresistance $R$ was measured using the four-point method under an applied dc current of $100\mu$A along $x$, while an external magnetic field $M$ was rotated in the $y-z$ plane having an angle $\beta$ with $z$, see Fig.\,\ref{FigspincurrentExp}(a). The magnetoresistance at $5\degree$K  as a function of $\beta$  is   presented in Fig.\,\ref{FigspincurrentExp}(b), which develops a angular dependence profile given by ${\rm cos}(\beta-\Delta\beta)$; here, $\Delta\beta$ is  a phase shift of the angle $\beta$, which is absent in conventional magnetoresistance measurements \cite{PhysRevLett.110.206601}. For the applied dc current along $x$ in RuO$_{2}(101)$,  the spin current develops terms along $y$ and $z$ directions, which leads to a magnetoresistance minimum  when the magnetization is parallel to the spin current polarization and away from $\beta=90\degree$. When measuring   the temperature dependence of the phase shift $\Delta\beta$ in Fig.\,\ref{FigspincurrentExp}(c), the authors find that it acquires a vanishing value when the applied current $J$ is parallel to [$\bar{1}01$], while $\Delta\beta$ develops large values when the current is along [010] and decrease as temperature increases. These findings   suggest that the temperature dependence has an impact  on the corresponding spin Hall conductivities, whose temperature dependence is presented in    Fig.\,\ref{FigspincurrentExp}(d-f). For clarity, Fig.\,\ref{FigspincurrentExp}(d) shows the total, the $z$ polarized, and the $y$ polarized spin Hall conductivities as a function of temperature, while Fig.\,\ref{FigspincurrentExp}(e,f) shows the respective conductivity components along $z$ and $y$ as well as their time-reversal odd and even contributions.  
Since the even spin conductivity is independent of scattering due to its intrinsic  bandstructure origin, its temperature dependence is expected to be weak \cite{Bose_2022}. Thus, the strong temperature dependence  of the total spin conductivities can be only attributed to the odd contribution, which, as we discussed before is tied to the altermagnetic band splitting in  RuO$_{2}$. By taking the ratio between odd spin conductivity contributions along $y$ and $z$,  the authors of Ref.\,\cite{pan2024unvei} have also estimated the  angle between the direction of the  time-reversal odd spin current polarization  and $z$; they found that time-reversal odd spin current polarization is roughly parallel to the N\'{e}el vector for a range of temperatures within $5\degree$K and $360\degree$K. The results of Ref.\,\cite{pan2024unvei} support the magnetic origin of the nonrelativistic spin currents in  RuO$_{2}$ and its polarization along the N\'{e}el vector of such a material.  

Moreover, there have   been other experimental studies reporting spin currents in   RuO$_{2}$ \cite{PhysRevLett.129.137201,PhysRevLett.128.197202,Feng_2022,pan2024unvei} and recently also in other altermagnetic materials \cite{Bai_review24}. 
In the case of $d$-wave AMs, spin currents have been measured in  Mn$_{5}$Si$_{3}$
\cite{PhysRevB.109.224430,reichlova2024}, V$_{2}$Te$_{2}$O \cite{PhysRevB.108.024410}, 
KV$_{2}$Se$_{2}$O \cite{jiang2024},  Rb$_{1-\delta}$V$_2$Te$_2$O \cite{fzhang2024}. In the case of AMs with $g$-wave symmetry, spin currents have already  been experimentally found    in   MnTe \cite{PhysRevLett.130.036702,Lee24} and CrSb \cite{PhysRevMaterials.8.084412}, further extending the family of materials with key universal altermagnetic properties that can be used for  spintronics \cite{wolf2001spintronics,Baltz2024,10.1063/5.0023614,ohldag2024hidden,10.1063/5.0009482,jungwirth2018multiple}. Nevertheless, we stress that more studies are needed  to further support the altermagnetic origin of spin currents in the predicted AMs, which will hopefully  trigger the interest to assess also materials for AMs with higher angular momentum and perhaps also odd-parity unconventional magnets. Experiments of this type will be extremely useful  when contacting AMs  (or their odd-parity counterparts) with superconductors towards a new era of superconducting spintronics, where spin currents in the superconducting state are  fundamental \cite{linder2015superconducting,Eschrig2015}.

\subsection{Intrinsic superconductivity in strained RuO$_2$ films}
The possibility of intrinsic superconductivity in   unconventional magnets has not been directly addressed yet experimentally, but there exist theoretical proposals suggesting altermagnetic spin fluctuations as a mechanism to induce superconductivity \cite{mazin2022notesaSC}  as well as the possibility of finite momentum superconductivity without any net magnetization  due to the coexistence of superconductivity and altermagnetism \cite{PhysRevResearch.5.043171,chakraborty2024_1,bose2024altermagn,sim2024,hong2024,mukasa2024}. Despite the lack of experimental evidence of intrinsic superconductivity, there exist measurements of superconductivity in materials which are believed to be altermagnets. This is the case of RuO$_2$, where strain-induced superconductivity was reported in Ref.\,\cite{PhysRevLett.125.147001,ruf2021strain,PhysRevMaterials.6.084802}. Moreover, a recent theoretical work \cite{mazin2023inducedmoFeSe} has suggested that monolayer FeSe, known to host superconductivity \cite{zaki2021time,PhysRevLett.130.046702,qiu2023concFeSe}, can manifest altermagnetism due to their intrinsic time-reversal symmetry breaking. Altermagnetism was also predicted to appear in  the parent cuprate La$_2$CuO$_4$ of a high-temperature superconductor \cite{LiborPRX22}, while very recently also the ground state of the strongly correlated CuAg(SO$_4$)$_2$ \cite{domanski2023unique} has been reported to be altermagnetic \cite{jeschke2024highly}  as well as in Sr$_{2}${RuO}$_{4}$ \cite{autieri2025}. It is worth noting that there exist other high-temperature cuprate superconductors  possessing signatures of time-reversal symmetry breaking \cite{weber1990evidence,zhao2023time,andersen2024spontaneous}, suggesting them as  promising  superconducting materials  for hosting altermagnetism.  

\begin{figure}[!t]
\centering
\includegraphics[width=0.49\textwidth]{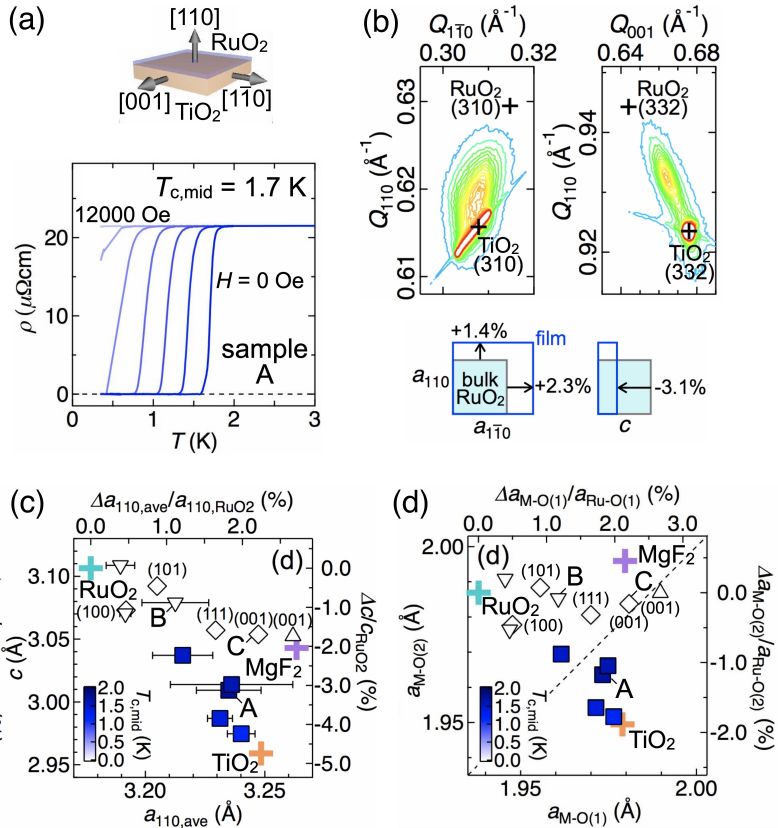}\\
\caption{Measurement of superconductivity induced by strain in RuO$_{2}$ films. (a) The top panel shows a sketch of a RuO$_{2}$ film (light blue) grown on TiO$_{2}$ (light orange) along  the [110] direction, referred to as sample A. The bottom panel depicts the in-plane resistivity $\rho$ as a function of temperature for distinct values of  an applied magnetic field parallel to the out-of-plane direction in steps of $2000$Oe. A superconducting transition at $T_{\rm c, zero}=1.6\degree$ K, with a midpoint temperature of $T_{\rm c, mid}=1.7\degree$ K.  (b)  X-ray diffraction maps in reciprocal space taken for asymmetric reflections along [1$\bar{1}$0] and [001] in-plane directions. The cross symbol  ``+'' indicates the peak positions obtained from bulk lattice parameters for RuO$_{2}$ and TiO$_{2}$. Bottom panels depict the variations in the lattice parameters ($a_{110}$ and $a_{1\bar{1}0}$) with respect to  their bulk values in RuO$_{2}$. (c)  Mapping of the average   lattice parameter along [110]  and [1$\bar{1}$0]    versus   the lattice parameter $c$ along [001], with the color indicating $T_{\rm c, mid}$, while the ends of the horizontal bars mark $a_{110}$ and $a_{1\bar{1}0}$. (d). Bond lengths between metal (M) atoms and oxygen (O) for RuO$_{2}$ films, including the bulk values of RuO$_{2}$, TiO$_{2}$, and MgF$_{2}$, while the symbols color coded by $T_{\rm c, mid}$.  Reprinted figures with permission from Masaki Uchida, Takuya Nomoto, Maki Musashi, Ryotaro Arita, and Masashi Kawasaki,  Phys. Rev. Lett. 125, 147001 (2020) \cite{PhysRevLett.125.147001}; Copyright (2025) by the American Physical Society.
}
\label{FigstrainSCExp} 
\end{figure}

Given the interest in RuO$_2$, we now discuss the experiment performed in Ref.\,\cite{PhysRevLett.125.147001} reporting intrinsic strain-induced superconductivity. Thin films of RuO$_{2}$ were epitaxially grown   on single-crystal TiO$_{2}$ along the [110] direction, see top panel of  Fig.\,\ref{FigstrainSCExp}(a). The film thickness was adjusted in the range of $26-32$nm in order to be able to apply large values of epitaxial strain. To identify the superconducting state, the authors measure the resistivity $\rho$ as a function of temperature by a   four-probe method in a cryostat having a $9$T superconducting magnet and a $^{3}H_{e}$ refrigerator. They found that $\rho$ drops to zero at the critical temperature $T_{\rm c,zero}=1.6\degree$K and with a midpoint temperature of $T_{\rm c,mid}=1.7\degree$K; by applying an external magnetic field, the transition temperature moves towards lower values. The authors also considered  RuO$_{2}$ films on other substrates but did not obtain signs of superconductivity, which points towards the importance of strain for inducing superconductivity.  From reciprocal space mappings, the authors show how epitaxial strain modifies the lattice parameters ($a_{110}$, $a_{1\bar{1}0}$, and $c$) of the RuO$_{2}$ films along three directions orthogonal to each other, achieving an anisotropic extension of $a_{110}$ and $a_{1\bar{1}0}$ and a reduction of $c$, see  Fig.\,\ref{FigstrainSCExp}(b). To identify the link between epitaxial strain and superconductivity, Fig.\,\ref{FigstrainSCExp}(c) presents the mapping of the average lattice parameters $a_{110}$, $a_{1\bar{1}0}$ versus $c$, with the main signature being the appearance of superconductivity only in RuO$_{2}$ films grown along [110] on TiO$_{2}$ when $c$ is reduced by $+2\%$ or more. This behavior was also identified by looking at the bond lengths between e. g., Ru and O atoms, denoted by $a_{\rm M-O(1)}$ and $a_{\rm M-O(2)}$, shown in Fig.\,\ref{FigstrainSCExp}(d); the most important feature  in this case is the decrease of $a_{\rm M-O(2)}$ as the bulk value of  TiO$_{2}$ is reached, while $a_{\rm M-O(1)}$ undergoes an increase. The authors further explore that soft phonon modes might be playing an important role for the emergence of superconductivity in strained RuO$_{2}$ films. 

The interesting results  of Ref.\,\cite{PhysRevLett.125.147001} motivate us to raise a natural question: is altermagnetism, predicted in  RuO$_{2}$ \cite{landscape22} and with already experimental evidence \cite{lin2024observation}, present in the strain-induced superconducting state reported in Ref.\,\cite{PhysRevLett.125.147001}?. Moreover, it is worth noting  that the critical temperature below which RuO$_{2}$ becomes superconducting is much lower than those predicted to realize altermagnetism \cite{landscape22}, hence raising questions about the coexistence of altermagnetism and superconductivity in  RuO$_{2}$. If superconductivity and altermagnetism coexist, would it be possible to raise the superconducting critical temperature   e. g., by strain engineering \cite{ruf2021strain}, such that the emergent effects are of utility  for realistic quantum applications?. For now, we believe that realizing strain induced superconductivity in other altermagnetic candidates is a promising ground.  To close this part, we stress again that similar questions as for  the RuO$_{2}$   discussed here naturally arise in other superconducting systems  with hints of altermagnetism, such as in La$_2$CuO$_4$ \cite{LiborPRX22}, FeSe \cite{mazin2023inducedmoFeSe}, CuAg(SO$_4$)$_2$ \cite{domanski2023unique,jeschke2024highly}, and Sr$_{2}${RuO}$_{4}$ \cite{autieri2025}.  Yet more pressing is the realization of intrinsic superconductivity in $p$-wave magnets, which, to date, has not been reported. We hope that all these questions  will motivate the community to establish  intrinsic    superconductivity in unconventional magnets.

\subsection{Signature of Andreev transport in altermagnet-superconductor junctions}
The interplay between unconventional magnetism and superconductivity is a promising ground to explore   emergent  superconducting phenomena, as suggested by many theoretical works during the past two years, see e. g., Refs.\, \cite{Sun23,Papaj23,maeda2024,PhysRevLett.133.226002,IkegayaAltermagnet,fukaya2024x,maeda2025classifi,Bo2025}. Perhaps the simplest devices where this interplay can be addressed involves superconducting junctions formed by known superconductors and  unconventional magnets. As discussed in Section \ref{section2}, when placing  an  unconventional magnet in contact with a superconductor, superconducting correlations are induced in the unconventional magnet  by means of the proximity effect, which is directly linked to   Andreev reflections \cite{klapwijk2004proximity,Cayao2020odd}. We remind that, in the Andreev reflection process,  an incoming electron  is reflected back as a hole  at the normal-superconductor interface  and thereby enhancing the differential conductance $dI/dV$ \cite{klapwijk2004proximity,PhysRevB.25.4515}. Thus,   differential conductance is able to reveal the emergence of superconductivity and also the symmetry of its underlying order parameter \cite{kashiwaya2000}. In the case of AMs and  $p$-wave magnets, theoretical studies showed a strong dependence of  Andreev transport on the orientation of the  unconventional magnets \cite{maeda2024,Bo2025}, which can be taken as a signature of emergent superconducting effects \cite{maeda2025classifi}. 

\begin{figure}[!t]
\centering
\includegraphics[width=0.49\textwidth]{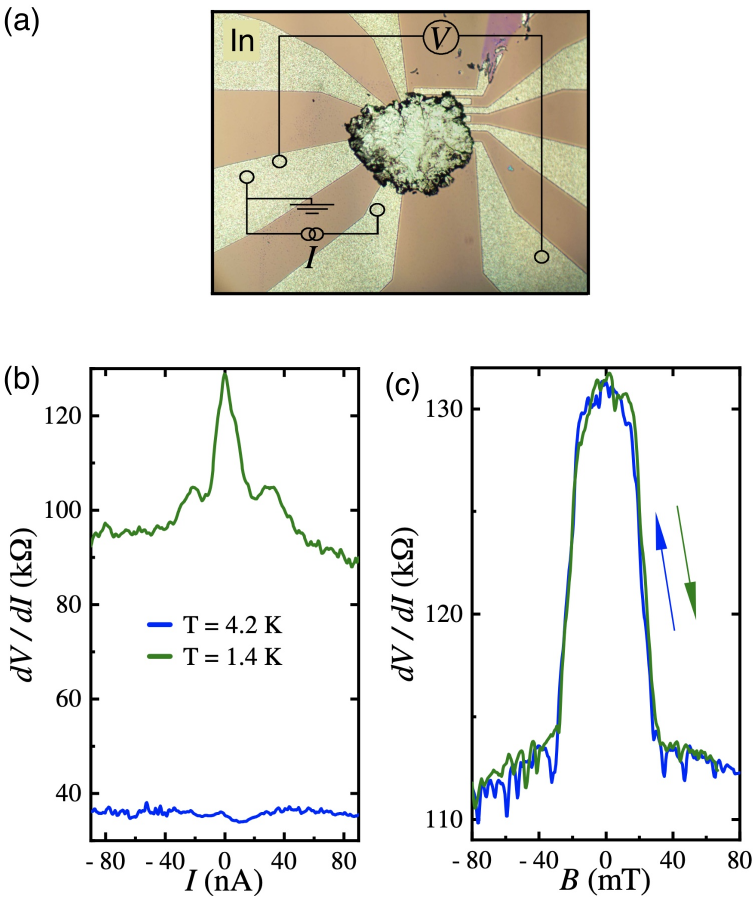}\\
\caption{Andreev reflection in a   NS junction  where N is a normal state material made of altermagnetic MnTe and  S is a superconductor made of In. (a) Image of the sample with electrical connections. A single crystal MnTe flake is transferred to the In leads pattern, so planar In-MnTe junctions are formed at the bottom surface of the flake. (b) Differential resistance of a single In-MnTe junction as a function of the dc current $I$ for two distinct temperatures as. (c) Zero-bias resistance peak as a function of the magnetic field  for two sweep directions. Reprinted figures from Physica B: Condensed Matter 696, 416602 (2025),  Andreev reflection for MnTe altermagnet candidate, D. Yu. Kazmin, V.D. Esin, Yu. S. Barash, A. V. Timonina, N. N. Kolesnikov, E. V. Deviatov \cite{Kazmin_2025}, Copyright (2025), with permission from Elsevier.}
\label{FigARExp} 
\end{figure}

The simplicity of the recipe for exploring Andreev transport has already motivated an experimental study in altermagnet-superconductor junctions, which was recently published in Ref.\,\cite{Kazmin_2025}. In particular, the authors of Ref.\,\cite{Kazmin_2025} considered a NS junction, where N is made of   MnTe in the normal state  and S is determined by Indium (In), see Fig.\,\ref{FigARExp}(a). To remind, MnTe is a $g$-wave  altermagnet \cite{landscape22} and its nonrelativistic band splitting has already been experimentally   reported  \cite{belashchenko2024,Lee24,KrempaskyNature2024}   
as well as its optical properties \cite{gray2024timeres} and a spontaneous Hall effect \cite{PhysRevLett.130.036702}; the quality of  MnTe  was also studied in magnetization measurements  \cite{Orlova_2024}. Moreover, indium  becomes superconducting below a critical temperature of $T_{\rm c}\approx3.4\degree$K and has a superconducting gap of $\Delta_{\rm In}=0.55$meV \cite{PhysRev.123.442}; indium is a type I spin-singlet $s$-wave  BCS superconductor   \cite{RevModPhys.26.277,RevModPhys.35.1}. The authors of Ref.\,\cite{Kazmin_2025} fabricated  MnTe-In planar junctions with transparent NS interfaces and measure electron transport across the junction by a common three-point technique, obtained in a dilution refrigerator equipped with a superconducting solenoid. Fig.\,\ref{FigARExp}(b) shows the measured differential resistance $dV/dI$ as a function of the applied current $I$  across a single MnTe-In junction at $4.2\degree$K and $1.4\degree$K: this reveals that the differential resistance (of the order of $40{\rm k}\Omega$) has a extremely weak dependence on the applied current but its high value reflects the nature of the MnTe-In interface \cite{Kazmin_2025}. At low temperatures $1.4\degree$K, where indium is a superconductor, the differential resistance is further increased following a nonlinear behaviour that reaches  values  even larger than $100{\rm k}\Omega$ [Fig.\,\ref{FigARExp}(b)]. Interestingly, for currents within $\pm40$A, the differential resistance develops a symmetric increased profile centred at $I=0$, with symmetric  small peaks at $\pm40$A that resembles the superconducting gap and a large zero-bias resistance peak resembling standard Andreev physics at disordered interfaces \cite{PhysRevB.25.4515,PhysRevB.93.041303,PhysRevB.96.245304}. The authors also explain that the unusual $dV/dI$ can be also seen to arise due to edge electrostatics which promote the depletion of carriers concentration at the NS interface, hence affecting the transport \cite{Batov2004}.  Moreover, when measuring the differential resistance as a function of an applied magnetic field,     the zero-bias resistance appears for magnetic fields within $\pm40$T, which is an relevant feature given that these values are close to the critical field of indium \cite{PhysRev.123.442}. The authors of Ref.\,\cite{Kazmin_2025}  also verified these findings in two distinct samples at $1.2\degree$K.

Since the behavior of $dV/dI$ reported in Ref.\,\cite{Kazmin_2025} only emerges when indium is superconducting, they are very likely reflecting Andreev transport but it is also fair to say that  there exist some intriguing questions. The most pressing concern is perhaps the size of the induced superconducting gap, which seems to be much larger than the bulk value in indium ($0.55$meV), see Fig.\,\ref{FigARExp}(b,c). The specific profile of the differential resistance, and the respective conductance, still deserves to be further explored since, to date, no theoretical study has addressed transport in NS junctions with a $g$-wave AM such as MnTe; a recent study, however, already anticipates higher angular momentum spin triplet  pairs when combining $g$-wave AMs with conventional superconductivity \cite{maeda2025classifi} but Andreev transport needs to be addressed in order to provide further understanding of the experiment carried out in Ref.\,\cite{Kazmin_2025}. Similarly, it would be important to clarify about the presence of altermagnetism at the conditions of the experiment of Ref.\,\cite{Kazmin_2025}, specially whether altermagnetism still survives at the low temperatures. Another open question is about the possible effects  due to the finite size of MnTe could be also playing a role in the subgap structure of differential resistance via e.g. confinement \cite{PhysRevB.91.024514,PhysRevB.104.134507}; it has recently been shown that confinement  favors the presence of  ABSs in AMs \cite{Bo2025}.   Novel superconducting phenomena  might  already be present at the interface of the MnTe-In junctions studied in Ref.\,\cite{Kazmin_2025}  and its confirmation   would require additional efforts on both the experimental and the theoretical fronts. Nevertheless, experiments like the one carried out in Ref.\,\cite{Kazmin_2025}  is indeed encouraging to further exploit the potential of unconventional magnetism for realizing novel superconducting effects. We hope that the physics discussed here will motivate to address the Josephson effect and perhaps the realization of superconducting devices by combining superconductors and unconventional magnets.


\section{Conclusions and outlook}
\label{section5}
In this review article, we have summarized the recent advances on emergent superconducting phenomena emerging due to the  interplay between unconventional magnetism and superconductivity. We have first provided an introduction to the field of unconventional magnetism in the normal state, making emphasis on their theoretical understanding and how to model magnetic orders with   even- and odd-parities. We have pointed out the fact that $d$-wave altermagnets and $p$-wave unconventional magnets represent the magnetic counterparts of unconventional $d$- and $p$-wave superconductors. We have then discussed how the interplay of unconventional magnetism  and superconductivity gives rise to the emergence of novel superconducting phenomena. In this part we have stressed that unconventional magnets   not only convert  spin-singlet into spin-triplet superconducting correlations but also transfer their parity, thus producing novel superconducting states with higher angular momentum that would not exist otherwise. Therein, we have also highlighted the importance of altermagnetism for realizing   finite momentum superconductivity, topological superconductivity, superconducting diodes, nontrivial light-matter coupling, magnetoelectric and thermoelectric effects. 

Later, we have presented a thorough discussion of emergent physics in hybrid systems formed by unconventional magnets and superconductors, which we have accompanied by details on how to model such hybrid systems. In particular, we have clarified the signatures of unconventional magnetism in Andreev transport, proximity and inverse proximity effects, and also in the Josephson effect, where altermagnetism is e. g., the key to induce $0-\pi$ and $\varphi$-Josephson junctions. Here, we have also stressed that   unconventional magnetism in Josephson junctions promotes the emergence of highly controllable odd-frequency pairing by means of the Josephson effect. Furthermore, we have summarized the recent experimental progress, highlighting the need of experimental signatures of the band splitting in altermagnets and the identification of their altermagnetic nature as well as the challenges  to fabricate systems where altermagnetism coexists with superconductivity and its impact when contacting with superconductors.

Even though the field that combines unconventional magnetism and superconductivity is still in its infancy, the great efforts of the community bring us closer to understand its potential for truly novel superconducting phenomena that can be useful for quantum applications. Of particular relevance is that the special spin-momentum locking in unconventional magnets already establishes new approaches for engineering spin-triplet Cooper pairs as well as their utilization for superconducting spintronics, topological superconductivity,  thermoelectricity  and multifunctional Josephson circuits, all at zero net magnetization and hence favorable for scalable superconducting devices. Despite the great interest, most of the research  has addressed $d$-wave altermagnets and $p$-wave magnets in combination with superconductivity, thus leaving largely unexplored other types of unconventional magnets possessing magnetic order with  higher angular momenta. The interplay between  these unconventional magnets and superconductivity is very likely to offer far more exciting opportunities for novel phenomena. Furthermore, on the experimental front, several questions are still open, such as the identification of unconventional magnets by conductance measurements or their impact on the Josephson effect, to name just two examples. Nevertheless, we believe that the existent theoretical predictions of novel superconducting phenomena will motivate experiments hopefully in the very near future.

\begin{acknowledgements}
 We thank Y. Asano, P. Burset, D. Chakraborty, P.-H. Fu, J.-X. Hu, S. Ikegaya, M. Junzhang, S. Kashiwaya, J.  Knolle, J. Linder, K. Maeda, T. Mizushima, N. Nagaosa, R. Seoane, P. Sukhachov, X. Qiu, M. Uchida, W. J. Zhao  for useful discussions.  Y. F. and K. Y. acknowledge financial support from the Sumitomo Foundation.  B. L. acknowledges financial support from the National Natural Science Foundation of China (project 12474049).    Y.\ T.\ acknowledges financial support from JSPS with Grants-in-Aid for Scientific Research (KAKENHI Grants Nos. 23K17668, 
24K00583, 24K00556, and 24K00578). J. C. acknowledges  financial support from the Carl Trygger’s Foundation (Grant No. 22: 2093), the    Sweden-Japan Foundation (Grant No. BA24-0003), and the Swedish Research Council  (Vetenskapsr\aa det Grant No.~2021-04121), 
 \end{acknowledgements}


\bibliography{biblio}

\end{document}